\documentclass[12pt,preprint]{aastex}

\newcommand{\T}{\rule{0pt}{2.ex}}
\usepackage{natbib}
\usepackage{amssymb}
\usepackage[figtopcap]{subfigure}

\shorttitle{Cross-Match of 2MASS and SDSS. II.}
\shortauthors{Gei{\ss}ler et al.}

\begin{document}


\title{A Cross-Match of 2MASS and SDSS. II. Peculiar L Dwarfs, Unresolved Binaries, and the 
Space Density of T Dwarf Secondaries}


\author{Kerstin Gei{\ss}ler, Stanimir Metchev}
\affil{Department of Physics and Astronomy, State University of New York, Stony Brook, New York 11794-3800, USA}
\author{J.~Davy Kirkpatrick, G.~Bruce Berriman}
\affil{Infrared Processing and Analysis Center, MC 100-22, California Institute of Technology, Pasadena, California 91225, USA}
\and
\author{Dagny Looper}
\affil{Institute for Astronomy, University of Hawaii, 2680 Woodlawn Drive, Honolulu, Hawai'i 96822, USA}
\email{geissler@astro.sunysb.edu}

\begin{abstract}
We present the completion of a program to cross-correlate the SDSS Data Release 1 and 2MASS 
Point Source Catalog in search for extremely red L and T dwarfs. The program was initiated 
by Metchev and collaborators, who  presented the findings on all newly identified T dwarfs 
in SDSS DR1, and estimated the space density of isolated T0--T8 dwarfs in the solar 
neighborhood.  In the current work we present most of the L dwarf discoveries.  
Our red-sensitive ($z-J\geq2.75$~mag) cross-match proves to be efficient in detecting 
peculiarly red L dwarfs, adding two new ones, including one 
of the reddest known L dwarfs.  
Our search also nets a new peculiarly blue L7 dwarf and, surprisingly, two M8 dwarfs.  We 
further broaden our analysis to detect unresolved binary L or T dwarfs through spectral 
template fitting to all L and T dwarfs presented here and in the earlier work by Metchev 
and collaborators.  We identify nine probable binaries, six of which are new and eight 
harbor likely T dwarf secondaries.  We combine this result with current knowledge of 
the mass ratio distribution and frequency of substellar companions to estimate an overall 
space density of 0.005--0.05~pc$^{-3}$ for individual T0--T8 dwarfs.
\end{abstract}




\section{INTRODUCTION}

The cross-correlation of imaging surveys across a broad wavelength range allows efficient and 
reliable identification of photometrically unusual objects, such as cold brown dwarfs 
\citep[e.g.,][]{knapp2004, chiu2006, lodieu2007, pinfield2008, zhang2010}, high-redshift 
quasars \citep[e.g.,][]{fan2001, mortlock2009}, or white dwarfs with unresolved (sub)stellar 
companions or debris disks \citep[e.g.,][]{rebassa-mansergas2009, steele2009}.  Searches for 
such objects are generally performed in a sequential manner, with candidates selected in one 
survey according to a set of desired color criteria and then confirmed in other surveys at 
similar locations but different wavelengths or epochs.  

As we demonstrated in \citet[][henceforth, Paper~I]{metchev2008}, this method is prone to 
overlooking an unknown fraction of objects of interest.  The reason is that the initial step 
of the candidate selection process is performed on a single survey database, and as such it is 
limited by quality flag considerations that are inevitably used to constrain the number of 
potential candidates to include mostly astrophysically viable sources.  While at high 
signal-to-noise levels (SNR $>25$) quality flags are an efficient way to remove true 
contamination, at low signal-to-noise ratios ``suspect'' flag combinations (e.g., resulting 
from flux interpolation or from de-blending of closely-separated sources) carry a 
disproportionate weight in rejecting possible candidates.  The effect of this process on the 
final object statistics can not be readily quantified.  Consequently, the use of quality flags 
to screen against probable artifacts incurs an unknown level of incompleteness in any search 
for faint objects.

In Paper~I we showed that the loss of viable candidates and the resulting incompleteness can 
be avoided if the initial round of candidate selection is done by using {\it cross-survey} 
colors, such as $z-J$ in the Sloan Digital Sky Survey \citep[SDSS;][]{stoughton2002} and the 
Two-Micron All-Sky Survey \citep[2MASS;][]{skrutskie2006}.  The extra verification obtained 
from joint $z$- and $J$-band identifications in SDSS and 2MASS 
allowed us to do away with all quality-flag constraints in either database.  Undeniably, the 
approach is still prone to missing potentially interesting objects (e.g., very red $z$-band 
drop-outs).  However, because source detection in any imaging survey can be reliably and 
reproducibly defined solely in terms of signal-to-noise thresholds, the incompleteness incurred 
by requiring joint survey identification can be readily determined as a function of source 
flux.  Hence, it is possible to obtain a statistically robust description of the resultant 
population of bona-fide objects down to very faint flux limits.

With many large-area imaging surveys completed over the past decade, multi-wavelength, 
multi-database cross-matching is at the heart of the planned functionality of electronic 
platforms that integrate astronomy data access, such as the Virtual Astronomical 
Observatory\footnote{http://www.usvao.org/}. However, while cross-correlations at scale are 
now feasible with modern computational facilities, the scientific validation of the results 
remains an open problem.  

Along with the cross-survey candidate identification approach piloted in Paper~I, we proposed 
a multi-step validation scheme for the extraction of bona-fide astrophysical objects.  We 
based our investigation on a cross-correlation of the SDSS Data Release 1 
\citep[DR1;][]{abazajian2003} and the 2MASS Point Source Catalog \citep[PSC;][]{cutri2003}. 
The specific science goal was the discovery of all T dwarfs in the overlap of the SDSS~DR1 and 
2MASS footprints.  For the purpose we selected all objects within $\Delta r = 6\farcs0$ of 
each other in the two databases that passed a $z-J\geq2.75$~mag selection criterion, without 
imposing any quality flag constraints on candidate sources in either survey.  (A complementary 
cross-survey identification approach, aimed at the detection of high-proper motion 
$\Delta r > 5\farcs0$ M, L, and T dwarfs between SDSS and 2MASS, was subsequently presented 
in \citealt{sheppard2008}.) Our automated object validation approach, based on a comparison of 
the total number of positional identifications within the matching radius to the number of 
identifications that satisfied the color criterion, allowed us to throw out 99.9\,\% of false 
candidates.  Following an observational spectroscopic campaign, we accounted for all 13 known 
T dwarfs in SDSS~DR1 and discovered two new ones (i.e., previous SDSS DR1 searches had 
unknowingly been 13\% incomplete).   With the candidate identification process 
well-defined---based solely on signal-to-noise ratios and colors---through Monte Carlo 
simulations we were able to accurately estimate the incompleteness of our search to T dwarfs 
of all flux levels.  We consequently produced the first estimate of the space density 
of T dwarfs across all (T0--T8) spectral subtypes.

In the current paper we finalise the investigation commenced in Paper I, and characterize 
all other newly discovered objects, including L and M dwarfs. The $z-J\geq2.75$~mag 
selection criterion was expected to produce objects as early as L3. Numerous candidate L 
dwarfs were indeed recovered, although their tally was subsequently restricted by an 
additional $i-z\geq3.0$~mag color cut on all objects brighter than $i=21.3$~mag. Altogether, 
24 new and candidate L and T dwarfs were presented in Paper~I, in addition to 19 that were 
already known in the SDSS~DR1 footprint.  Nineteen of the new candidates still required 
spectroscopic characterization, and 17 of these are presented here, while two could not 
be observed due to bad weather.  

For details on the design and implementation of the SDSS~DR1/2MASS cross-correlation, the 
criteria for excluding false candidates, and the final yield of the cross-match, we refer 
the reader to Paper I.

\section{OBSERVATIONS \label{sec_observations}}

We observed 17 of the 22 candidate L dwarfs identified in Paper I.  Spectra for an additional 
three were already published in Paper I.  Two candidates remain unobserved to date: 
2MASS J11571680$-$0333279 and 2MASS J21203387$-$0747208 due to clouds during the 
scheduled observing run.

We obtained low-resolution 0.9--2.5~$\micron$ spectra of the L dwarf candidates identified 
in the SDSS~DR1/2MASS cross-match with the SpeX spectrograph \citep{rayner2003} on the NASA 
Infrared Telescope Facility (IRTF) between August 2007 and March 2008. The observations 
were taken in prism mode with the $0\farcs5\times15\farcs0$ or the 
$0\farcs8\times15\farcs0$ slit, resulting in resolutions of $R\sim100-150$. The slit 
orientation was maintained to within 20$^{\circ}$ of the parallactic angle for all targets. 
We employed a standard A--B--B--A nodding sequence along the slit to record object and sky 
spectra. Individual exposure times were 180~s per pointing. Standard stars were used for 
flux calibration and telluric correction.  Flat-field and argon lamps were 
observed immediately after each set of target and standard star observations for use in 
instrumental calibrations.  Observation epochs and instrument settings for each science 
target are given in Table~\ref{tab:spectr_obs}.

All reductions were carried out with the SpeXtool package version 3.2 \citep{cushing2004, 
vacca2003}, using an optimal spectroscopic extraction approach \citep{robertson1986, 
horne1986}.The reduced spectra were smoothed to the instrumental resolution corresponding 
to the chosen slit width, using the Savitzky-Golay smoothing kernel \citep{press1992}.

The low resolution near-infrared (near-IR) spectra of the 17 new ultra-cool dwarfs are 
presented in order of increasing right ascension coordinate in Figure~\ref{fig:seq}.  SpeX 
spectra of M and L dwarf standards \citep{kirkpatrick1991, kirkpatrick1999} are overplotted 
for comparison.

\section{SPECTRAL CLASSIFICATION AND UNRESOLVED BINARITY}

\subsection{Spectral Classification}
\label{sec:classification}

The spectral classification of our targets was done primarily by comparison to near-IR SpeX 
spectra of optical M7--M9 \citep{kirkpatrick1991}, optical L0--L8 \citep{kirkpatrick1999}, 
and near-IR L9--T8 \citep{burgasser2006b} dwarf standards.  We note that the L9 standard,  
2MASSW~J0310+1648, used by \citet{burgasser2006b} to define a near-IR classification scheme 
for L8--T8 dwarfs, has an optical spectral type of L8 \citep{kirkpatrick2000}.  

Low-resolution SpeX spectra of most of the spectral standards were available from the Spex 
Prism Spectral Libraries\footnote{http://web.mit.edu/ajb/www/browndwarfs/spexprism/}.
Spectra of the L4 and L6 \citeauthor{kirkpatrick1999} standards were not available, and 
instead we used 2MASSI~J1104$+$1959 and 2MASS~J1010$-$0406, respectively. Both objects were 
originally classified in the optical and their near-IR colors are close to the average of 
the corresponding spectral types (Tab. \ref{tab:standards}).  All standard M and L dwarf 
spectra cover a spectral range from 0.65~$\micron$ to 2.55~$\micron$ and have a resolution 
of $R\sim120$, similar to that of our spectra.  The full set of spectral standards is listed 
in Table~\ref{tab:standards}.  

We re-iterate that while the \citet{kirkpatrick1991, kirkpatrick1999} M--L dwarf 
classification system is defined in the optical, we used near-IR spectra of the suggested 
standards for comparison to our data. The near-IR spectra of L dwarfs do not follow a 
monotonous sequence as a function of 
effective temperature as they do in the optical \citep{mclean2003}.  The $>$1~$\micron$ 
spectral shapes of L dwarfs reflect the varying depth of the photosphere as affected by the 
wavelength dependence of the combined dust and molecular opacity in the atmosphere 
\citep{ackerman2001}.  For this reason we limited our spectroscopic classification to the 
0.95--1.35~$\micron$ region, avoiding most of the strong H$_2$O absorption between 
1.3--1.5~$\micron$, and ensuring sampling of the temperature-sensitive short wavelength 
continuum.  The classification itself was obtained by using $\chi^2$ minimization to find 
the best-matching standard spectrum.  For comparison, we also computed spectral types based 
on $\chi^2$ minimization over most of the 0.95--2.35~$\micron$ region of our spectra, 
excluding only the low signal-to-noise regions (1.35--1.45~$\micron$ and 
1.80--2.20~$\micron$) in the midst of the near-IR H$_2$O bands.  Spectral types for all of 
our objects are listed in Table~\ref{tab:spt} and alongside the plotted spectra in 
Figure~\ref{fig:seq}.

\subsection{Spectral Binary Fitting}
\label{sec:binarity}

To investigate if binarity may be the underlying reason for some of the peculiar features 
observed in the presented L dwarf sample, we fitted single and composite spectra to the 
data. For a comprehensive description of the fitting procedure we refer the reader to 
\cite{cushing2008} and \cite{burgasser2010}.

A set of 190 single and composite 0.90--2.55~$\micron$ spectra was constructed using the 
chosen L0--L9 and T0--T8 spectral standards as templates. In order to flux-calibrate the 
spectra, all templates were scaled with respect to one another according to the $K_s$ 
magnitude vs. spectral type relation defined by \citet{looper2008a}, before generating the 
composite spectra. The spectra of all composite templates were 
then allowed to freely scale to the spectra of the science targets.
 The best-fit composite spectrum for each 
candidate was determined by calculating a weighted $\chi^{2}$ statistic, with the 
individual wavelength bins (detector pixels) weighted by their spectral bandwidth: 
$w_i=\Delta\lambda_i$ \citep[see][]{burgasser2010, cushing2008}. The fit was performed only 
over the 0.95--1.35~$\mu$m, 1.45--1.8~$\mu$m, and 2.0--2.35~$\mu$m wavelength ranges in 
order to avoid regions of low signal and high telluric absorption. 

As mentioned by \citet{burgasser2010}, the much larger number of composite (171) 
vs.\ single (19) templates implies almost certainly better 
$\chi^{2}$ fits with composite rather than single object templates.  We applied the 
one-sided F test to evaluate the significance of the best-fit composite spectrum using the 
$\eta_{\rm SB}$ statistic, defined by \citet[][Eq.~2]{burgasser2010} as the ratio of the 
$\chi^2$ values of the best-fit single template and the best-fit composite template. The 
number of degrees of freedom is $\nu = N_{\rm eff}-1$ to account for the scaling process 
during fitting, where $N_{\rm eff}$ is the effective number of data points in each 
spectrum.  In a slight departure from the parameterization of \citet{burgasser2010}, we 
define $N_{\rm eff}$ as the number of {\it independent} resolution elements in our spectra, 
equal to the total number of pixels $N$ divided by the number of pixels per slitwidth. Most 
of our spectra are taken with a 5~pix-wide ($0\farcs8$) slit, and a few with a 
3~pix-wide ($0\farcs5$) slit (Table~\ref{tab:spectr_obs}), and $N_{\rm eff}$ is either 
59 or 99, correspondingly.

To rule out the null hypothesis, i.e., that a candidate is a single object, at the 
99\% confidence level (CL), we require $\eta_{SB}>1.85$ for the 5~pix-slitwidth 
spectra or $\eta_{SB}>1.61$ for the higher-resolution 3~pix-slitwidth spectra. In order to 
test the reliability of the best-match composite fit, we varied the initial flux-calibration 
of the templates within the RMS scatter of the $K_s$ magnitude vs.\ spectral type relation 
\citep[$\pm$0.33~mag;][]{looper2008a}. We repeated the $\chi^{2}$ minimization 1000 times 
for each of the binary candidates, randomly varying the flux-calibration of the primary and 
the secondary independently within the range permited by the scatter in the relation. 

Considering that four of the nine L dwarf standards that are used as spectroscopic 
templates are actually binary systems themselves \citep[the L2, L3, L5, and L7 standards 
from][]{kirkpatrick1999}, the individual spectral types of the components should not 
be taken as definitive. Nevertheless, a superior $\chi^{2}$ fit of a composite template to 
any of the candidates suggests that the object is probably an unresolved binary system.

\section{RESULTS}
\label{sec_res}

Fifteen of the new candidates are confirmed either as L dwarfs, or as unresolved binaries 
with L composite spectral types, and two as M dwarfs. 
Among the new L-type objects, three 
are identified as unresolved binaries (L+L or L+T), two are peculiarly red, and one is 
peculiarly blue.  In our selection of objects with peculiar colors, we have followed the 
construct of \citet{faherty2009}, requiring that an object has a $J-K_s$ color that is either 
2$\sigma$ or 0.4~mag away from the mean for its spectral subtype. 

An additional search for unresolved binarity among the L and T dwarfs 
reported in Paper~I reveals that two of those L dwarfs and four of the T dwarfs are also 
likely binaries.  One of the T dwarf binaries has been spatially resolved into two 
components by \citet{burgasser2006b}, two others have been suggested as probable binaries from 
spectral template fitting by \citet{burgasser2010}, and the remaining one is new.

We first discuss the probable binary systems, then the peculiarly blue or red L dwarfs, and 
eventually select ordinary L and M dwarfs discovered in the cross-match.

\subsection{Candidate Binaries }
\label{sec_bin}

Three of our 15 L dwarfs are significantly better fitted by a composite than by 
a single template at the $>$99\% confidence level. Best-fit single and composite spectra 
for these binary candidates are shown in Fig.~\ref{fig:bin}. Six additional unresolved 
binaries from the Paper~I sample are presented in Fig.~\ref{fig:addbin}. 
Table~\ref{tab:fit} lists the two most likely spectral template combinations for each of 
the unresolved binary candidates, along with the fraction of Monte Carlo outcomes in which 
the given combinations were the best fit ones.  The individual systems are discussed in the 
following.

\subsubsection{New Candidate Unresolved Binaries}
\label{sec_new_bin}

\paragraph{2MASS J0735+2720.}
The 0.95--1.35~$\mu$m continuum of this object is best-fit by an L1 standard 
(Fig.~\ref{fig:seq1}).  However, with a $z-J$ color of $3.03\pm0.17$~mag, 2MASS J0735+2720 
is red compared to the average for the L1 spectral subtype (Table~\ref{tab:standards}). The 
$H$ band peak is flat for an L1 dwarf and the $K$ band peak is shifted towards the blue. 
Spectral fitting suggests that the spectrum of 2MASS J0735+2720 is a composite of an L1 and 
an L4 dwarf.

\paragraph{2MASS J1423+6154.} 
If single, this dwarf would be classified as an L4 (Fig.~\ref{fig:seq3}).  However, the 
2MASS colors and the SED are slightly blue in $J-H$ and $J-K_s$. Spectral fitting shows 
that the spectrum of 2MASS~J1423+6154 is best reproduced by an L2\,+\,T5 composite 
spectrum, with an L2\,+\,T4 composite almost as likely.

\paragraph{2MASS J1737+5953.}
If single, this dwarf would be classified as a blue L9 (Fig.~\ref{fig:seq6}), since its 
$J-K_{s}$ color is more than 2\,$\sigma$ (or 0.4\,mag) bluer than the 
average of the L9 subtype. The spectrum shows a marked methane absorption feature 
at 1.6~$\mu$m and the $K$ band peak is rounded, with its maximum shifted slightly towards 
the red. The best-fit L5\,+\,T5 composite template reproduces all of these spectral 
features.  We note that 2MASS~J1737+5953 is the only binary candidate for which we cannot 
formally claim binarity at the 99\% confidence level ($\eta_{SB}=1.55$, while the 99\% 
threshold is $\eta_{SB}>1.61$ for its 3 pix-slitwidth spectrum).  However, the 1.6~$\mu$m 
methane absorption leaves little ambiguity about the presence of an unresolved mid-T component.

\subsubsection{Additional Binary Candidates from Paper I}

The L and T dwarfs reported in Paper I were not checked for unresolved binarity.  We 
do so here, and find six more probable binaries (Fig.~\ref{fig:addbin}).

\paragraph{2MASS~J0052+0012}
was reported as a moderately blue L2 dwarf in Paper~I, and was designated as peculiar.  
In fact, the object is only moderately blue in $J-K_s$ and not a $>2\sigma$ outlier.  
Hence, the peculiar designation is unwarranted under the presently adopted definition, even 
if the color may be suggestive of unusual properties. The best-fit composite spectra indeed 
indicate that is it likely a combination of an L4/L2 dwarf and a T3 dwarf.

\paragraph{SDSS~J1731+5310}  
was an already known L6 dwarf in the SDSS~DR1 footprint, first 
announced by \citet{chiu2006}.  The near-IR spectrum of SDSS~J1731+5310 is best fit by a 
combination of an L5 and an L8 template.

\paragraph{SDSSp J0926+5847} 
is a T4.5 dwarf originally announced by \citet{geballe2002}.  Our binary template fitting 
indicates two components: a T3 or a T4 dwarf and a T6 dwarf.  \citet{burgasser2006c} report 
a marginal elongation in high angular resolution 1.1~$\micron$ and 1.7~$\micron$ imaging of 
this object with NICMOS on the {\it Hubble Space Telescope}, from which they infer that 
SDSSp~J0926+5847p is a binary with near-equal flux T4:\,+\,T4: components. This result is 
consistent with our inference of a T3/T4 and a T6 component.

\paragraph{SDSS J1214+6316} 
is a T3.5 dwarf originally discovered by \citet{chiu2006}.  It has not been identified 
as a potential binary before.  Our fitting indicates that it is a probable composite 
of a T2 and a T6 component.

\paragraph{2MASS J1324+6358} 
is a peculiar T2 dwarf reported independently in \citet{looper2007} and in Paper I.  
\citet{burgasser2010} suggest it is a possible L8\,+\,T3.5 binary.  Our fitting indicates 
that the most likely component spectral types are L9 and T2. Given the ambiguity among the 
optical spectra of L8 and L9 dwarfs, the two sets of findings are mutually consistent.

\paragraph{SDSS J1516+0259}
is a T0\,$\pm$1.5 dwarf discovered by \citet{knapp2004}. \citet{burgasser2010} list 
SDSS J1516+0259 as a highly probable binary candidate, with spectral types of L$7.5\pm1.1$ 
and T$2.5\pm2.2$. Our spectral fitting, based on a more limited number of spectral templates 
with integer-valued subtypes, yields spectral types of L9 and T0 for the two components.


\subsection{A Peculiarly Blue Single L dwarf: 2MASS J1542$-$0045}
\label{sec_blue}

The 0.95--1.35~$\micron$ spectrum of 2MASS~J1542$-$0045 is most adequately fit by a single 
L7 template (Fig.\,\ref{fig:seq4}).  No pairwise combinations of L or T dwarfs produce a 
significantly better match to the overall near-IR SED.  Yet, the SED is suppressed redwards 
of 1.4~$\mu$m, and is responsible for its very blue 2MASS colors compared to other L7 
dwarfs. The $K$ band flux peak appears rounded with the peak slightly shifted towards 
redder wavelengths. The 2.3~$\mu$m CO absorption is weaker than in the comparison 
standards, while the H$_{2}$O absorption at 1.4~$\mu$m is enhanced. 2MASS~J1542$-$0045 also 
shows a deeper 0.99~$\mu$m FeH absorption line than in the L7 standard (see 
Fig.~\ref{fig:veryblue}). 

A comparison to the known blue dwarf SDSS~J112118.57$+$433246.5 \citep[L7.5,][]{chiu2006} 
reveals that both dwarfs have a similar spectral shape (Fig.~\ref{fig:veryblue}). 
2MASS~J1542$-$0045 is marginally bluer than SDSS J1121+4332 in $J-K_s$ color, but its 
0.99\,$\mu$m FeH absorption is shallower. Indeed, 2MASS~J1542$-$0045 shows all signs 
generally associated with the near-IR spectra of blue L dwarfs: an enhanced 1.4~$\mu$m 
H$_{2}$O absorption, weak CO absorption and, of course, an unusually blue SED 
\citep{burgasser2008}. Optical spectroscopy of 2MASS~J1542$-$0045 is required to confirm its L7 designation.


\subsection{Peculiarly Red Single L Dwarfs}
\label{sec_red}

Two peculiarly red, likely single L dwarfs are presented here: 2MASS J09175418+6028065 
(\S\ref{sec_0917}) and 2MASS J16154255+4953211 (\S\ref{sec_1615}).  An additional pair of 
peculiarly red L dwarfs were identified within the larger sample of SDSS DR1 L dwarfs: 
2MASS J01262109+1428057 and 2MASS J01075242+0041563.  The former was already discussed as 
a red, moderately low gravity L2  dwarf in Paper I.  The latter was first identified and 
classified in the near-IR as an L5.5 dwarf by 
\citet[][SDSSp J010752.33+004156.1]{hawley2002} and in the optical as an L8 by 
\citet{hawley2002}.  We note the very red $J-K_s$ color of this object, and also designate 
it as a peculiar L8.

\subsubsection{2MASS J0917+6028}
\label{sec_0917}

This dwarf was already noted for its unusually red $z-J$ color in Paper~I, although no 
spectrum was available at that time.  At $z-J=3.48\pm0.32$~mag, it: (1) has the reddest 
$z-J$ color of any of the L and T dwarfs found in the entire SDSS~DR1/2MASS cross-match, 
(2) is one of the reddest known L dwarf, and (3) is as red as mid-T dwarfs (see 
Table~\ref{tab:standards}).

The spectrum of 2MASS~J0917+6028 (Fig.\,\ref{fig:seq1}) cannot be fit by any of the L or T 
spectral standards, although we note that the lack of methane absorption precludes it from 
being a T dwarf.  A conspicuous emission feature near 2.17~$\micron$ is a likely result of 
the incomplete removal of broadened Bracket~$\gamma$ absorption in the telluric standard.  
{\it Spitzer Space Telescope} 3.6--8.0~$\micron$ photometry presented in Paper~I places it 
among other mid-L dwarfs, although slightly redder in [4.5$\micron$]$-$[5.8$\micron$] color 
(see Figure~4 in Paper~I).

We tentatively classify 2MASS~J0917+6028 as an L5 dwarf.  
The overall spectrum is much redder than that of any of the standards, even though 
curiously the 2MASS PSC colors do not reflect it. Near-IR colors computed from the 
spectrum are in disagreement by up to 2~$\sigma$ with the 2MASS colors, and classify 
2MASS~J0917+6028 as red outlier with a $J-K_{s}$ color more than 0.4\,mag redder than the 
average. 

Considering its red SED shape, 2MASS~J0917$+$6028 is likely dusty and could be young.  
Better agreement is achieved with the spectrum of G196--3B \citep{allers2007}, a 
60--300~Myr L2 dwarf.  The shallow K~I absorption at 1.17~$\micron$ and 
1.25~$\micron$ suggests low surface gravity.  Figure~\ref{fig:Hpeak} compares 
2MASS~J0917$+$6028 to G196--3B and Fig.\,\ref{fig:red} compares it to the even younger 
(1--50~Myr) L0 dwarf 2MASS~J0141$-$4633 \citep{kirkpatrick2006}, both of which appear 
to have stronger K~I absorption features.

Even so, 2MASS~J0917+6028 lacks the peaked $H$-band spectrum characteristic of 
low-gravity L dwarfs.  The low SNR of 5--15 between 0.95\,-\,1.3~$\micron$ prevents a 
reliable assessment of the strength of the gravity sensitive features. 
Finally, we note the $\sim$2.17~$\micron$ bump in our spectrum, suggesting that 
hydrogen absorption features may not have been adequately removed from the telluric 
standard during post-procesing.  This likely affects the $H$ band continuum, and may 
have affected its spectral shape.

Besides low gravity, high metallicity may be a possible explanation for a redder SED. 
\citet{looper2008b} noted that theoretical models with higher metallicity result in 
overall redder SED's than solar metallicity models.  Higher metallicity would not 
result in a peaked $H$-band spectrum.  However, in the absence of low gravity we would 
expect stronger alkali line absorption, unlike what is observed.

\subsubsection{2MASS J1615+4953}
\label{sec_1615}

This dwarf has already been discovered by \citet{cruz2007}, where it is classified as an 
L4 from an optical spectrum. Here we tentatively classify it as an L6 (Fig.\,\ref{fig:seq5}). 
However, its 0.95--1.35~$\mu$m flux is suppressed compared to the standard, making an 
assessment of the spectral type difficult. 

The SED of 2MASS~J1615+4953 is red throughout the near-IR. The $K$ band peak is slightly 
shifted to the red, and the $H$ band peak has a slightly pointed triangular shape. Alkali 
absorption lines are subdued in the 1.1--1.3~$\mu$m region.  All of these characteristics 
suggest that 2MASS~J1615+4953 may be moderately young.  The optical spectrum of 
2MASS J1615+4953 also displays low gravity signatures \citep{cruz2007}.

A comparison to G196--3B \citep{allers2007}, a known young L2 dwarf (Fig.\,\ref{fig:Hpeak}), 
reveals the similar spectral shapes of the two objects. Figure\,\ref{fig:red} compares 
2MASS~J1615+4953 to the young L0 dwarf 2MASS~J0141$-$4633 and 2MASS~J0917+6028. Both 
2MASS~J1615+4953 and 2MASS~J0917+6028 exibit similar SED's, with 2MASS~J1615+4953 emiting 
stronger at $>$1.55~$\mu$m.


\subsection{Notes on Select Ordinary L dwarfs}

\paragraph{2MASS~J0229$-$0053} has been classified as an L2 dwarf (Fig.\,\ref{fig:seq1}).  
While its SDSS / 2MASS $i-z$ and $z-J$ colors are very red, the spectrum is fully 
consistent with an L2 standard.  The apparently peaked $H$ band continuum may be an 
artifact of the low signal to noise of the observation (SNR $\sim$\,12).  The alkali line 
absorption strengths are comparable to a standard L2 dwarf, and the overall SED differs 
significantly from that of the lower-gravity L2 dwarf G196--3B (Fig.\,\ref{fig:Hpeak}).

\paragraph{2MASS~J1308+6103} has been classified as an L2 (Fig.\,\ref{fig:seq3}). The dwarf 
has a slightly red $z-J$ color ($\simeq$\,1\,$\sigma$ above the average) for its spectral 
type, but the 0.95--1.35~$\mu$m continuum is consistent with L2.
 
\paragraph{2MASS~1414+0107} has been classified as an L4 (Fig.\,\ref{fig:seq3}). 
The spectrum displays deeper water absorption between 1.35~$\mu$m and 1.5\,$\mu$m, and 
again at 1.9\,$\mu$m than the comparison spectra.

\paragraph{2MASS~J1534+0426} has been given only a tentative classification, L0: 
(Fig.\,\ref{fig:seq4}), because the spectrum has a low signal-to-noise ratio (SNR $\sim$ 10). 

\paragraph{2MASS~J1716+2945} has been classified as an L3 (Fig.\,\ref{fig:seq5}). The blue 
2MASS $J-K_s$ color ($\gtrsim$\,1.5\,$\sigma$ below the average) is inconsistent with the 
spectrum, which is well fit by the standard.

\paragraph{2MASS~J2116$-$0729} has been classified as an L6 (Fig.\,\ref{fig:seq6}). 
However, the flux of 2MASS~J2116$-$0729 is enhanced longward of 1.3~$\mu$m, resulting in a
slightly redder $J-K_s$ ($\sim$\,1.5\,$\sigma$) color than for the average L6 dwarf. The 
dwarf shows no signs generally associated with low gravity, leaving higher metallicity 
and/or enhanced atmospheric dust content as a possible explanation for the resulting red SED. 

\subsection{M dwarfs}
\label{sec:mdwarf}

Two of the candidate L dwarfs, 2MASS J0926+5230 (Fig.\,\ref{fig:seq2}) and 2MASS~J1551+0151 
(Fig.\,\ref{fig:seq5}) turn out to be late-M dwarfs. Their spectra are well fit by an M8 
standard and show no unusual signatures of redness in the near-IR.
This is despite the fact that the recorded SDSS/2MASS $z-J$ 
colors ($2.77\pm0.18$~mag and $2.88\pm0.28$~mag, respectively) of the M dwarfs are much 
redder than the average for the M8 spectral type ($2.07\pm0.18$~mag).  

Given the applied color cuts the detection of M and early L dwarfs (L0--L2) is 
unexpected, but most likely explained by statistical fluctuations in the SDSS $z$ and 2MASS 
$J$ band photometry.  With $z$ and $J$ magnitudes fainter than 19.4~mag and 16.5~mag, 
respectively, all of the ordinary early L dwarfs and the two M8 dwarfs are low SNR detections 
in both SDSS and 2MASS.  Their magnitudes are uncertain (errors $>$0.10~mag), and so the $z-J$ 
colors have larger than typical errorbars ($\gtrsim$0.15~mag). Given the likely thousands of 
late-M and early-L dwarfs in SDSS~DR1, it is conceivable that a few of them would have 
$\sim$4$\sigma$ discrepant $z-J$ colors simply due to statistical variance in the number of 
recorded photons.  A simple comparison of the reported 2MASS colors of our targets with ones 
synthetically generated from the SpeX spectra confirms this notion: the low-SNR 2MASS 
photometry is rather broadly scattered around the one-to-one correspondence line with the much 
higher SNR synthetic colors (Fig.~\ref{fig:color}).

\section{Discussion}
\label{sec_dis}

\subsection{The Fraction of Binary or Peculiar L and T Dwarfs}
\label{sec_dis_part1}
The SDSS DR1 and 2MASS PSC cross-match returned a total of 26 L dwarfs: 8 previously known, 
3 presented in Paper I, and 15 presented here (see Paper~I for a complete list). For 22 of 
the 26 L dwarfs near-IR spectra are available either from the present data, or from Paper~I, 
or from the SpeX Prism Spectral Libraries. Six of these 22 L dwarfs have discrepant $J-K_s$ 
colors (based on synthetic photometry over the SpeX spectra), which is $>$2~$\sigma$ or 
0.4\,mag away from the mean for their spectral subtype.  This number includes one blue L 
dwarf, four red L dwarfs and one candidate unresolved binary (a blue color outlier; 
Table~\ref{tab:spt}). The remaining four unresolved L dwarf binary candidates have $J-K_s$ 
colors within the 2$\sigma$ range for their respective composite spectral types. We also 
checked the 15 known T dwarfs in SDSS~DR1 for binarity, and found four probable binaries, one 
of which is a red color outlier (Table~\ref{tab:spt}). After removing candidate unresolved 
binaries, none of the remaining T dwarfs have peculiar $J-K_s$ colors or SEDs.

Among the 278 L dwarfs in the proper motion study of \citet{faherty2009}, 11 (4.0\%) and 22 
(7.9\%) were classified as blue and red outliers, respectively, based on their $J-K_s$ colors. 
All of these objects were considered regardless of binarity, i.e., their spectral types are 
potentially composite. In our sample of 22 L-composite dwarfs for which we have available 
spectra, we would thus expect one blue and two red dwarfs.  The factor of two higher rate of 
discovery of red L-composite dwarfs is not unusual given the color bias of our SDSS/2MASS 
cross-correlation, and attests 
to the efficiency of the $z-J$ color selection in identifying peculiarly red objects.  The 
factor of two higher fraction of blue color outliers (in $J-K_s$) is intriguing, although only 
marginally discrepant with expectations.  It is possible that the $z-J\geq2.75$~mag criterion 
enhances sensitivity to mid-L plus mid-T dwarf binaries, whose $z-J$ colors are artificially 
reddened by the T component's $J$-band flux.  The $J$-band flux peak of the T dwarf in these 
systems would also make the composite $J-K_s$ color peculiarly blue, such as in the L5+T5 
candidate binary 2MASS~J1737+5934 (\S~\ref{sec_new_bin}).

The number of L and T dwarf binary candidates identified in our analysis approximately agrees 
with statistical expectations from published results.  \citet{goldman2008} compile data on 
high angular resolution imaging observations of L and T dwarfs and find that 24 out of 130 
(18.5\%$^{+3.8}_{-2.9}$) L0--L9.5 dwarfs and 8 out of 38 (21\%$^{+8}_{-5}$) T0--T8 dwarfs are 
resolved binaries \citep[see also][]{burgasser2007a}.  The corresponding expectations are for 
four L-composite type and three T-composite type binaries in our sample.  The observed 
numbers are five and four, respectively, again marginally higher than the expectations.  Our 
inferred frequencies of L and T dwarf binaries in our flux-limited samples of 22 L and 15 T 
dwarfs, for which near-IR spectra were available, are $23^{+11}_{-7}$\% and $27^{+13}_{-8}$\%, 
respectively, conditional on the confirmation of all candidate unresolved binaries.

\subsection{The Space Density of Isolated T Dwarfs and T Dwarf
Secondaries}

One of the main results of Paper I was a determination of the space density of T dwarfs in the 
solar neighborhood.  In that paper we announced the discovery of two T dwarfs in addition to 
the 13 already known in the SDSS~DR1 catalog.   Based on a Monte Carlo analysis of the SDSS 
and 2MASS detection limits, we estimated a space density of $0.007\pm0.003$~pc$^{-3}$ (95\% 
confidence interval) for T0--T8 dwarfs within $\approx$90~pc of the Sun.

We do not report any new T dwarfs here, but we do report altogether nine candidate unresolved 
binaries, in eight of which one or both of the components are T dwarfs (\S\ref{sec_bin}). It 
is pertinent to discuss whether the T dwarf space density estimate of Paper I needs revision.

As far as composite spectral types of unresolved T dwarf systems are concerned, a revision is 
not warranted.  In Paper~I we simulated the space density of unresolved T dwarf systems, 
regardless of whether they were individual objects or close binaries with a T-composite 
spectral type. Since none of the newly reported objects have T-composite spectral types, the 
space density on T dwarf systems remains the same.

Given our analysis of unresolved binarity, we are now in a position to determine the space 
density of T dwarfs regardless of their multiplicity, i.e., to estimate the number of 
individual T dwarfs among both isolated objects and multiples.  We note that our SDSS/2MASS 
cross-match was not ideally designed to answer this question, since unresolved T dwarf 
companions to much earlier-type stars would not be recovered. Any binary system containing a 
T secondary and a primary earlier than L3 will have a composite spectral type $<$L3, and hence 
will most likely be excluded by the $z-J\geq2.75$~mag criterion. E.g., a close binary like 
SCR~1845$-$6357A/B \citep[M8.5/T6;][]{biller2006} would not have been picked up.  Nevertheless, 
the frequency of unresolved T dwarf companions can be estimated from the current knowledge of 
binary and individual (sub)stellar populations, and constrained from the present analysis.

In Paper I unresolved T dwarf binarity was simulated by simply doubling the flux of an 
inidividual dwarf at any given T subtype, i.e., by assuming a mass ratio distribution that is 
a delta function at $q\equiv M_2/M_1=1$. This is an adequate approximation for very low mass 
binaries, whose mass ratio distribution is sharply peaked near unity \citep{burgasser2007a}. 
A more comprehensive estimate of the frequency of T dwarf companions requires assumptions of 
the stellar mass function, the frequency and mass distribution of low-mass substellar 
companions to more massive brown dwarfs and stars, and the star formation history of the Milky 
Way galaxy. We attempt only an approximate estimate here.

Using a \citet{kroupa2002} multi-part power-law mass function for single stars, binarity 
rates of objects with B to T spectral types from the literature ($\sim$80\% for AB stars, 
\citealt{shatsky2002, kouwenhoven2005}; $\sim$55\% for FGK stars, \citealt{duquennoy1991}; 
25\%--42\% for M stars, \citealt{leinert1997, fischer1992}; $\sim$20\% for late-M to T 
dwarfs, \citep[][and references therein]{burgasser2007a}, and a mass ratio distribution for 
both stellar and substellar companions of the form 
$\Gamma(q) = d N/d q\propto q^\beta$ ($\beta\approx-0.4$ for B--K stars, \citealt{shatsky2002, 
kouwenhoven2007, metchev2009}; $\beta\approx0.3$ for M0--M6 stars, based on results from the 
9~pc M dwarf multiplicity survey of \citealt{delfosse2004}; and $\beta\sim5$ for late-M, L, 
and T dwarfs, \citealt{burgasser2007a}), we find that $\sim$13\% of all companions in binary 
systems are substellar ($M_2<0.072M_\sun$) and that their space density is 
$\sim$0.024\,pc$^{-3}$. Adopting a uniform age distribution over 0--10~Gyr and assuming that 
any object with an effective temperature between 500~K and 1400~K is a T dwarf, an application 
of the substellar evolutionary models of \citet{baraffe2003} indicates that $\approx$85\% of 
all $>$0.01~M$_\odot$ brown dwarf companions are T dwarfs. Hence the estimated space density 
of T dwarf {\it secondaries} based on the above assumptions is 
$\rho_{\rm T\,comp}\sim$0.02~pc$^{-3}$.

Not all of the above parameters are well constrained, and our estimate can vary substantially 
in either direction given the parameter uncertainties.  The greatest weight is carried by the 
adopted parameters for the population of very low mass (VLM) primaries.  Seventy percent of 
all T dwarf companions are estimated to orbit a $<$0.1~M$_\odot$ primary, and varying either 
the VLM star IMF power law index \citep[$\alpha_0=0.3\pm0.7$;][]{kroupa2002} or multiplicity 
fraction \citep[$20\pm10\%$;][]{burgasser2007a} within their empirical uncertainty ranges can 
affect the estimate for $\rho_{\rm T\,comp}$ in either direction by up to 50\% or 35\%, 
respectively.  The empirical evidence for the space density of isolated T dwarfs indeed 
indicates that $\alpha_0$ is at most 0 \citep[Paper~I;][]{reyle2010} or perhaps $-0.5$ 
\citep{pinfield2008, burningham2010}.  These result in 
$\rho_{\rm T\,comp}=0.017$~pc$^{-3}$ or 0.014~pc$^{-3}$, respectively.

Variations in the power law index $\beta_{\rm VLM}$ of the mass ratio distribution $\Gamma(q)$ 
of VLM stars have much less of an impact ($<$5\%), since the distribution is so strongly 
peaked near unity.  We note, however, that the difference between $\beta_{\rm VLM}\sim5$ and 
the power law index $\beta_{\rm M}=0.3$ of the mass ratio distribution of $>$0.1~M$_\odot$ 
field M dwarfs is large compared to that between $\beta_{\rm M}$ and the power law index 
$\beta_{\rm BK}$ for higher mass, B--K type binaries.  The $\beta_{\rm M}=0.3$ value was 
obtained as our own fit ($\chi^2=1.2$) to the  \citet{delfosse2004} mass ratio distribution 
of the 9~pc M-dwarf binary sample, and has a 68\% confidence interval of $[-0.4, 3.2]$.  
\citet{delfosse2004} find that their combined radial velocity and high angular resolution 
survey is nearly 100\% complete to stellar companions with periods up to $10^5$ years and 
75\% complete to brown dwarf companions with periods up to $10^3$ years, so no significant 
incompleteness corrections are needed.  Allowing $\beta_{\rm M}$ to increase up to the upper 
limit of its 68\% confidence interval decreases the space density of T dwarfs by 25\%, all 
else being fixed. Changes in the power law index $\beta_{\rm BK}$ of the mass ratio 
distribution of higher-mass binaries have a $<$5\% impact on the space density of T dwarf 
companions, since B--K stars are estimated to harbor only $\sim$2\% of all T dwarf companions 
in the present framework of assumptions.

Finally, systematic uncertainties in the substellar evolutionary models impact our 
temperature-based definition of a T dwarf: a 500~K $<T_{\rm eff}<1400$~K object.  A 30\% 
decrease in the equivalent mass predictions for this temperature range for all 0--10~Gyr-aged 
substellar objects decreases $\rho_{\rm T\,comp}$ by 35\%.  A similar increase in the 
predicted masses increases $\rho_{\rm T\,comp}$ by 10\%.

Overall, our analytical estimate the space density of T dwarf companions lies in the 
0.005--0.04~pc$^{-3}$ range.  Values near 0.02~pc$^{-3}$ are favored for the nominal values of 
the adopted \citet{kroupa2002} IMF, binary fraction, and mass ratio distribution parameters.  
Values near 0.015~pc$^{-3}$ are obtained for flat or slightly increasing forms of the 
substellar portion of the mass function, as motivated by the population of T dwarfs in the 
solar neighborhood.  

The above estimate is marginally higher than the 0.004--0.01~pc$^{-3}$ space density (95\% 
confidence interval) of unresolved T0--T8 dwarf systems from Paper I.  Our analysis does 
double-count certain types of widely-separated T dwarf companions.  If sufficiently bright, 
these could be identifiable as isolated objects (e.g., Gl~337C/D, Gl~570D, etc) in 
seeing-limited surveys such as SDSS or 2MASS, and would contribute towards the isolated T 
dwarf space density.  At the same time, we count these secondaries as part of the companions 
that are generally undetected or unresolved by SDSS or 2MASS.
However, the effect of the double counting turns out to be negligible.  The 
majority of known widely-separated T dwarf companions detected by seeing-limited imaging are 
associated with primaries with A--K spectral types, i.e, more massive than 0.5~M$_\odot$. As we 
already discussed, we estimate that less than 2\% of all T dwarf companions orbit 
$>$0.5~M$_\odot$ stars. We therefore conclude that T dwarfs exist at least as frequently, and 
perhaps even twice as frequently in binary systems with a higher-mass companion than as 
isolated objects.

Our now completed analysis of the SDSS DR1 and 2MASS PSC cross-match can also place an 
independent empirical lower limit on the frequency of unresolved T dwarf secondaries relative 
to isolated  and composite T dwarf systems.  This can be obtained from the ratio of the number 
of L+T or T+T probable binaries (eight, if all are confirmed) found in the cross-match to the 
total number of T dwarfs in the SDSS~DR1 footprint (15; Paper~I).  We note that unresolved 
equal flux binary systems are over-represented by a factor of $2^{3/2}=2.8$ (less if non-equal 
flux) in a flux-limited survey.  Given eight candidate and confirmed unresolved binaries with 
T dwarf secondaries in our flux-limited sample, the actual relative frequency of unresolved T 
dwarf secondaries to T-composite systems in a volume-limited sample would be 
$\sim\frac{8}{2.8} : 15 \approx 20\%$.  This is a lower limit, since it only includes T 
companions to $\approx$L3 dwarfs or later.  Additional unresolved T companions might also 
exist to the four L dwarfs (out of 26 total recovered in our cross-match), for which we do not 
have SpeX spectra to determine unresolved binarity (Section \ref{sec_dis_part1}). 
Statistically, one of these would be expected to be an unresolved binary, with 
approximately equal probabilities of the secondary being an L or a T dwarf.

In summary, the frequency of T dwarf companions is at the least greater than $\approx$20\%, 
and probably a factor of two higher than that of isolated T0--T8 dwarfs.  Our broadest 
estimate for the space density of T type companions is thus between 0.001~pc$^{-3}$ and 
0.04~pc$^{-3}$, with the overall space density of T0--T8 objects, whether in binaries or in 
isolation, between 0.005~pc$^{-3}$ and 0.05~pc$^{-3}$.

\section{Summary}

With this paper we have concluded a pilot undertaking, first described in \citet{metchev2008}, 
to cross-correlate the SDSS~DR1 and 2MASS PSC databases over their 2099~deg$^2$ common area in 
search for rare objects: very red L and T dwarfs.  The principal scientific results from the 
completed project are:
\begin{itemize}
\item the discovery of two additional T dwarfs in an already thoroughly perused region of the 
sky, both in SDSS and in 2MASS \citep{metchev2008};
\item thus, a completion of the T dwarf sample in the SDSS~DR1 footprint to within the 
combined SDSS/2MASS flux limits, and the first estimate of the space density of isolated T 
dwarfs \citep[spectral types T0--T8;][]{metchev2008};
\item thus, an empirical constraint on the field substellar mass function \citep{metchev2008}, 
enabling predictions for the yield of future sensitive wide-area IR surveys;
\item the identification of eight probable T dwarf companions in spatially unresolved L+T or 
T+T binary systems, five of which are new (this paper);
\item hence, the first estimate of the space density of T dwarf secondaries, and a combined 
estimate of the space density of isolated and secondary companion T dwarfs (this paper);
\item the discovery of an L dwarf (2MASS J0917$+$6028) with one of the reddest known optical 
minus near-IR ($z-J$) colors: a potential new laboratory for studying low-gravity or 
dusty substellar atmospheres, such as those of young extrasolar giant planets (this paper).
\end{itemize}

The above results were obtained by cross-correlating only the 2099~deg$^2$ overlap of the 
SDSS~DR1 and 2MASS footprints.  With SDSS-II recently complete and SDSS-III well under way, 
the 5.6 times larger area (11,663~deg$^2$) of SDSS Data Release 7 \citep{abazajian2009} 
offers proportionately richer prospects.  Combined with the opportunities presented by the 
on-going UKIRT Infrared Deep Sky Survey \citep[UKIDSS;][]{lawrence2007} and {\it Wide-Field 
Infrared Survey Explorer} \citep[{\it WISE};][]{duval2004}, the potential of cross-survey 
science has grown tremendously.

The completed project represents a successful demonstration of the feasibility and 
scientific merits of database cross-correlation and validation at 
scale. It has produced several valuable lessons on the computational logistics of 
cross-correlations, on the required understanding of the physical and statistical 
properties both of the target sources and of possible contaminants, and on the need for 
judicious use of database quality flags for object discrimination. We anticipate that these 
lessons will be beneficial to larger, multi-wavelength cross-survey comparisons in the 
future.



\acknowledgments

This research has benefitted from the M, L, and T dwarf compendium housed at 
DwarfArchives.org and maintained by Chris Gelino, Davy Kirkpatrick, and 
Adam Burgasser. This research has benefitted from the SpeX Prism Spectral Libraries, 
maintained by Adam Burgasser at http://www.browndwarfs.org/spexprism.
This research has benefitted from the Ultracool Dwarf Catalog housed at 
http://www.iac.es/galeria/ege/catalogo$\_$espectral/index.html. This research has 
made use of data obtained from or software provided by the US National Virtual 
Observatory, which is sponsored by the National Science Foundation.

{\it Facilities: \facility{IRTF (SpeX)}}.

\bibliographystyle{apj}
\bibliography{xmatch}

\clearpage



\begin{figure}
\caption{Comparison of the spectra of the discovered dwarfs (red) to the M and L 
dwarf standards (black). The new object spectra have been normalized to the average 
flux in the 1.2\,-\,1.25\,$\mu$m region.  In each panel the science target spectrum is 
reproduced multiple times with constant offsets in between.  The spectra of the standards 
are normalized to minimum $\chi^{2}$ deviations from the respective science spectrum over 
the 0.95\,-\,1.35\,$\mu$m range.}
\label{fig:seq}
\subfigure[2MASS J0229$-$0053 (L2), 2MASS J0735+2720 (L1) and 2MASS J0917+6028 (L5).]{
\label{fig:seq1}
\includegraphics[scale=0.28]{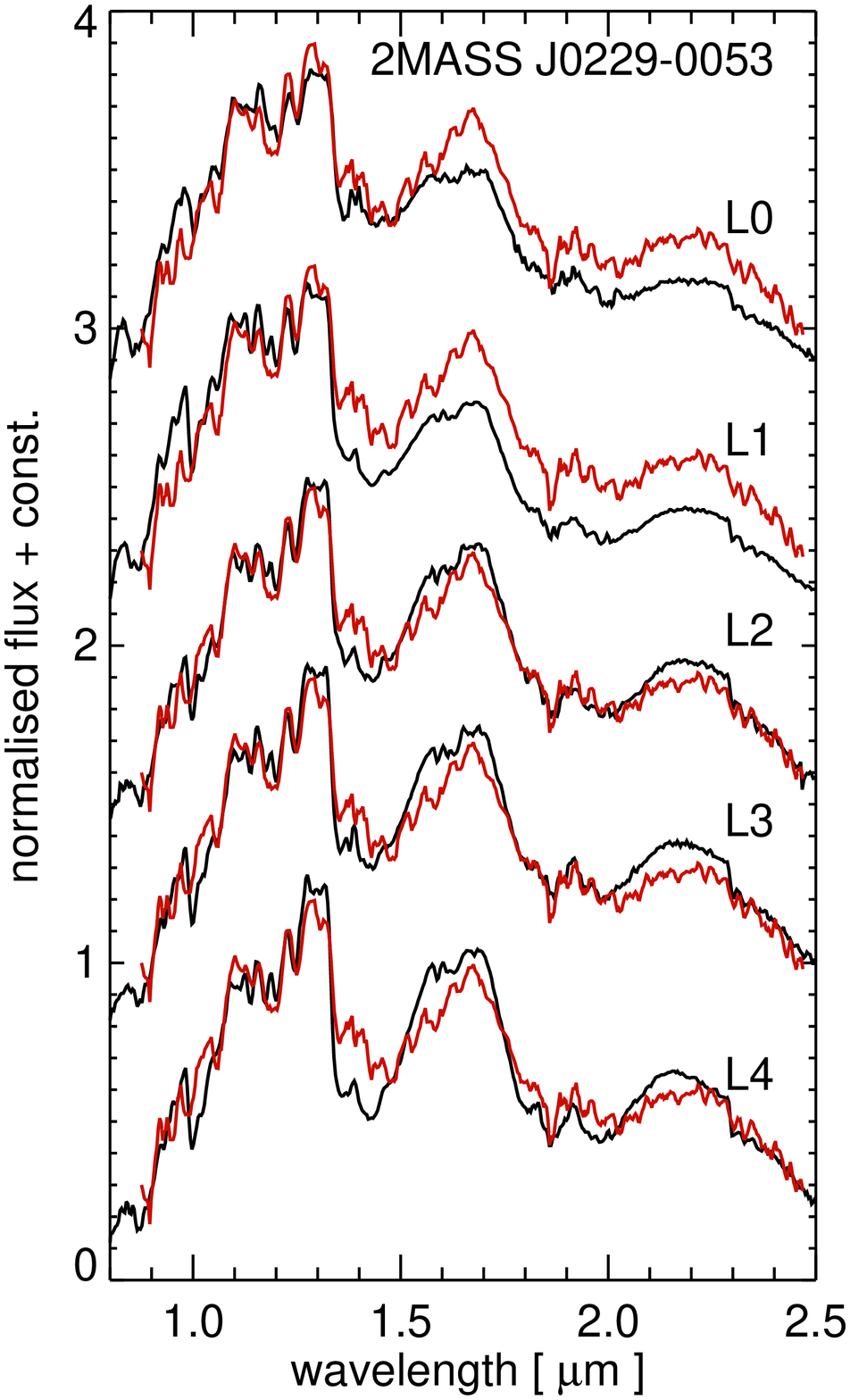}
\includegraphics[scale=0.28]{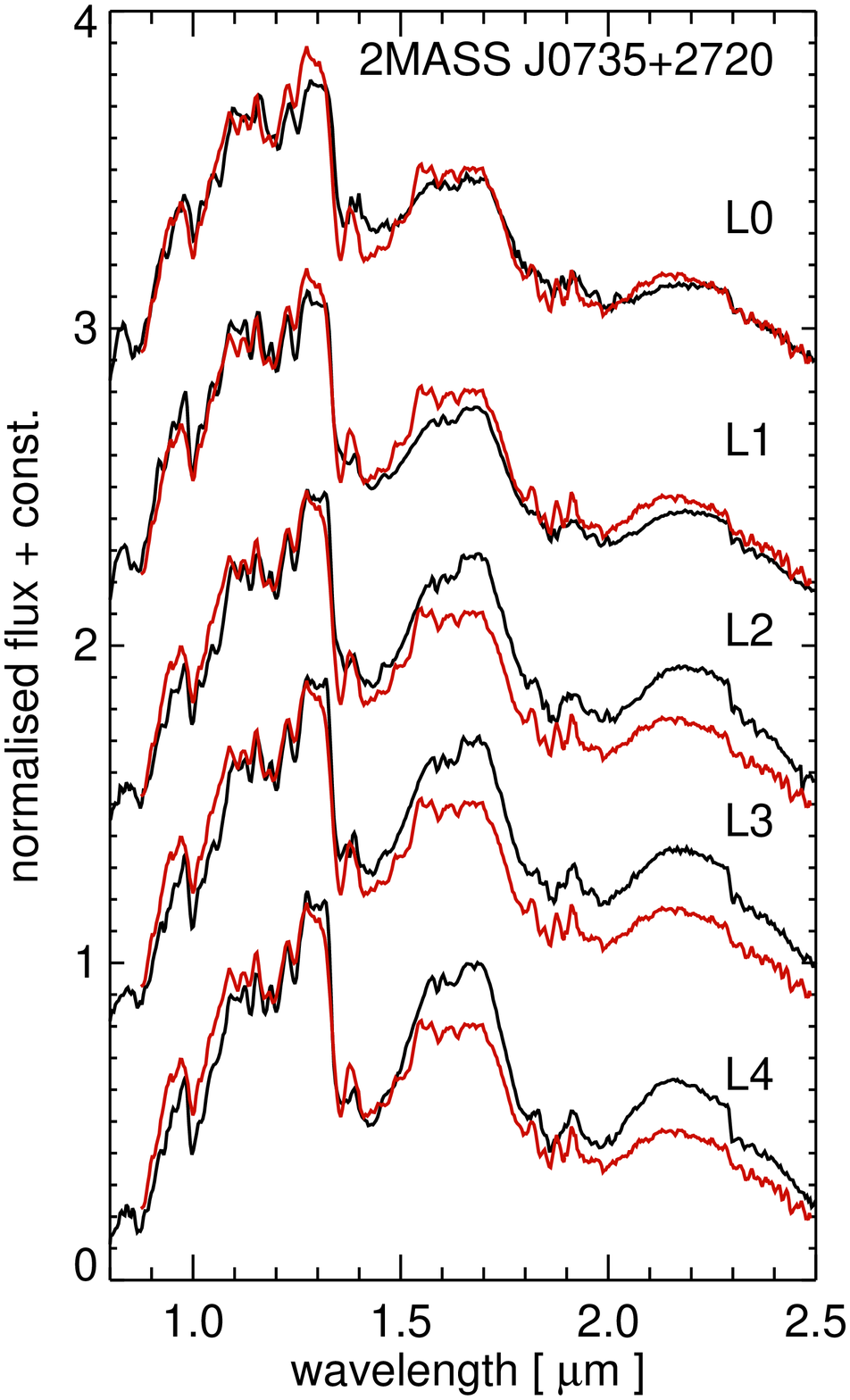}
\includegraphics[scale=0.28]{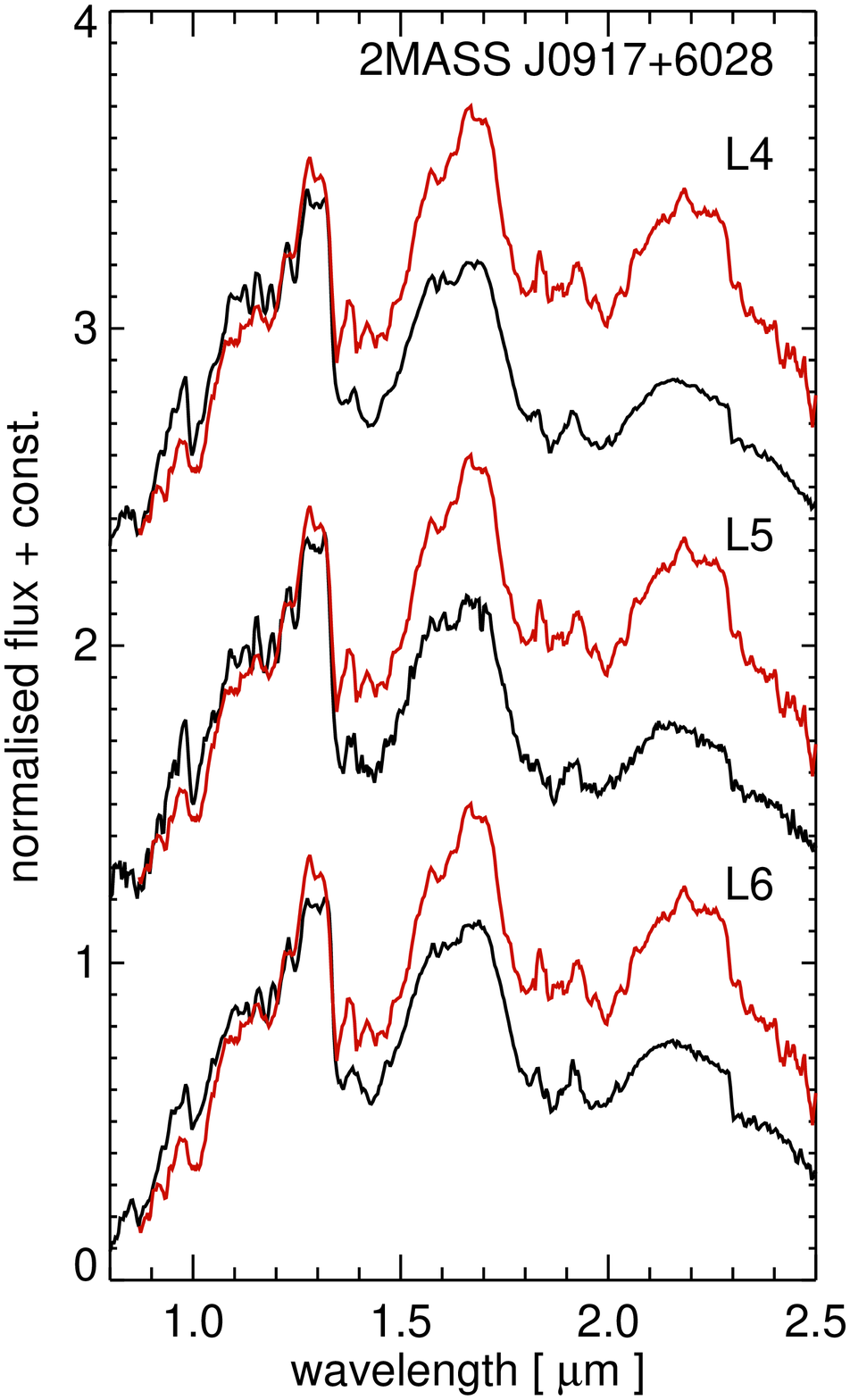}
}
\subfigure[2MASS J0926+5230 (M8), 2MASS J1119+0552 (L4) and 2MASS J1217$-$0237 (L4).]{
\label{fig:seq2}
\includegraphics[scale=0.28]{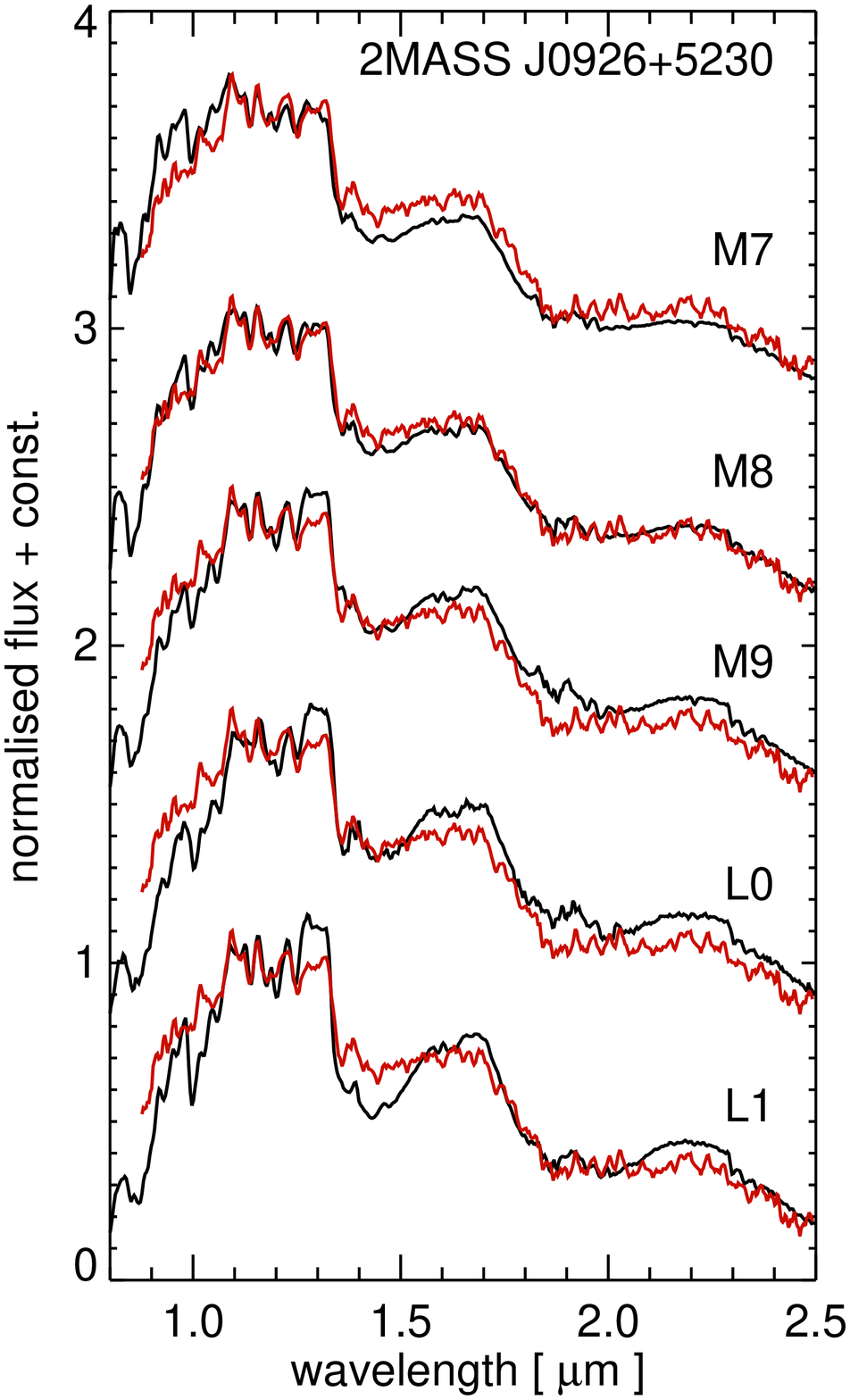}
\includegraphics[scale=0.28]{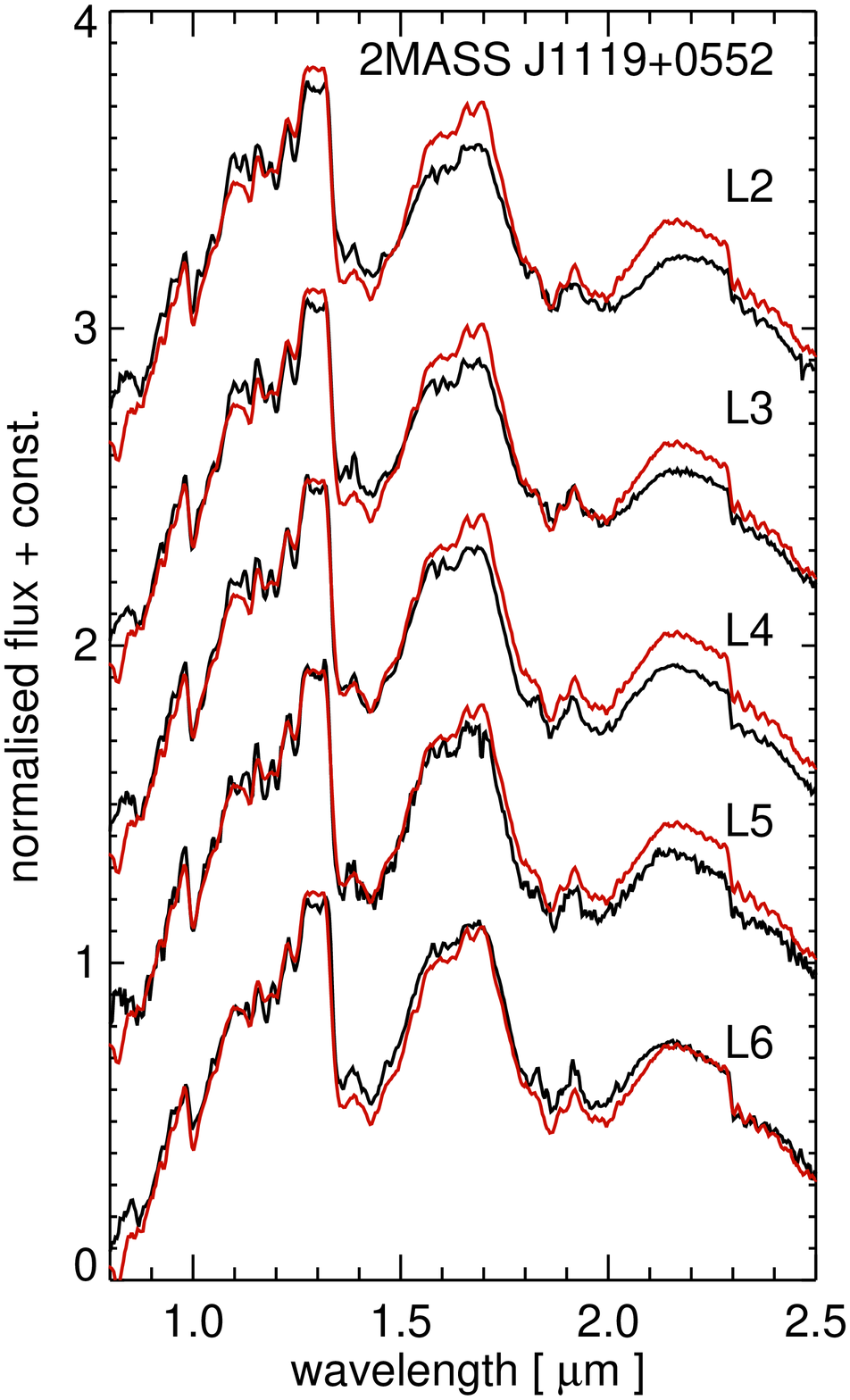}
\includegraphics[scale=0.28]{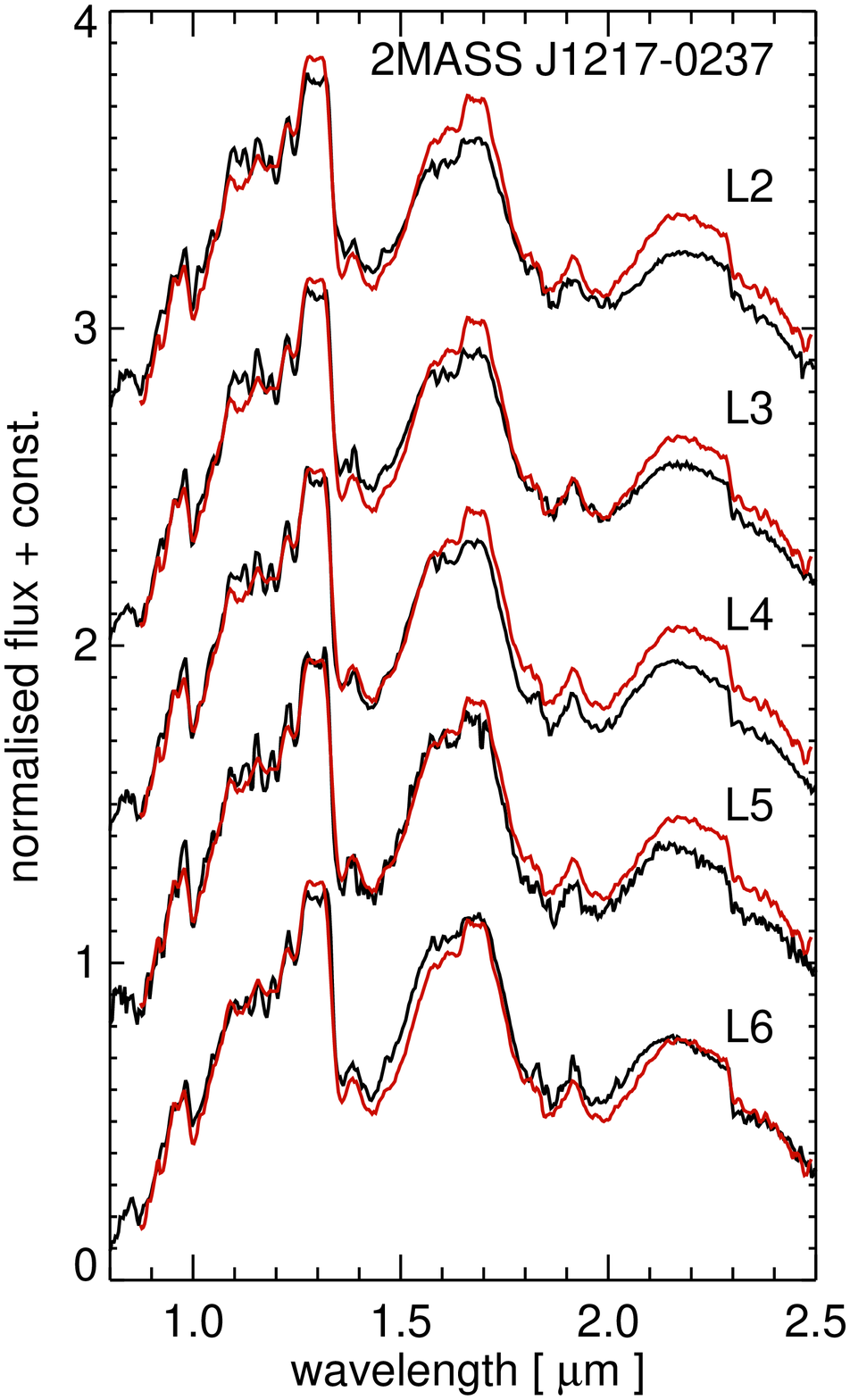}
}
\end{figure}
\addtocounter{figure}{-1}
\begin{figure}
\caption{ continued.}
\addtocounter{subfigure}{2}
\subfigure[2MASS J1308+6103 (L2), 2MASS J1414+0107 (L4) and 2MASS J1423+6154 (L4).]{
\label{fig:seq3}
\includegraphics[scale=0.28]{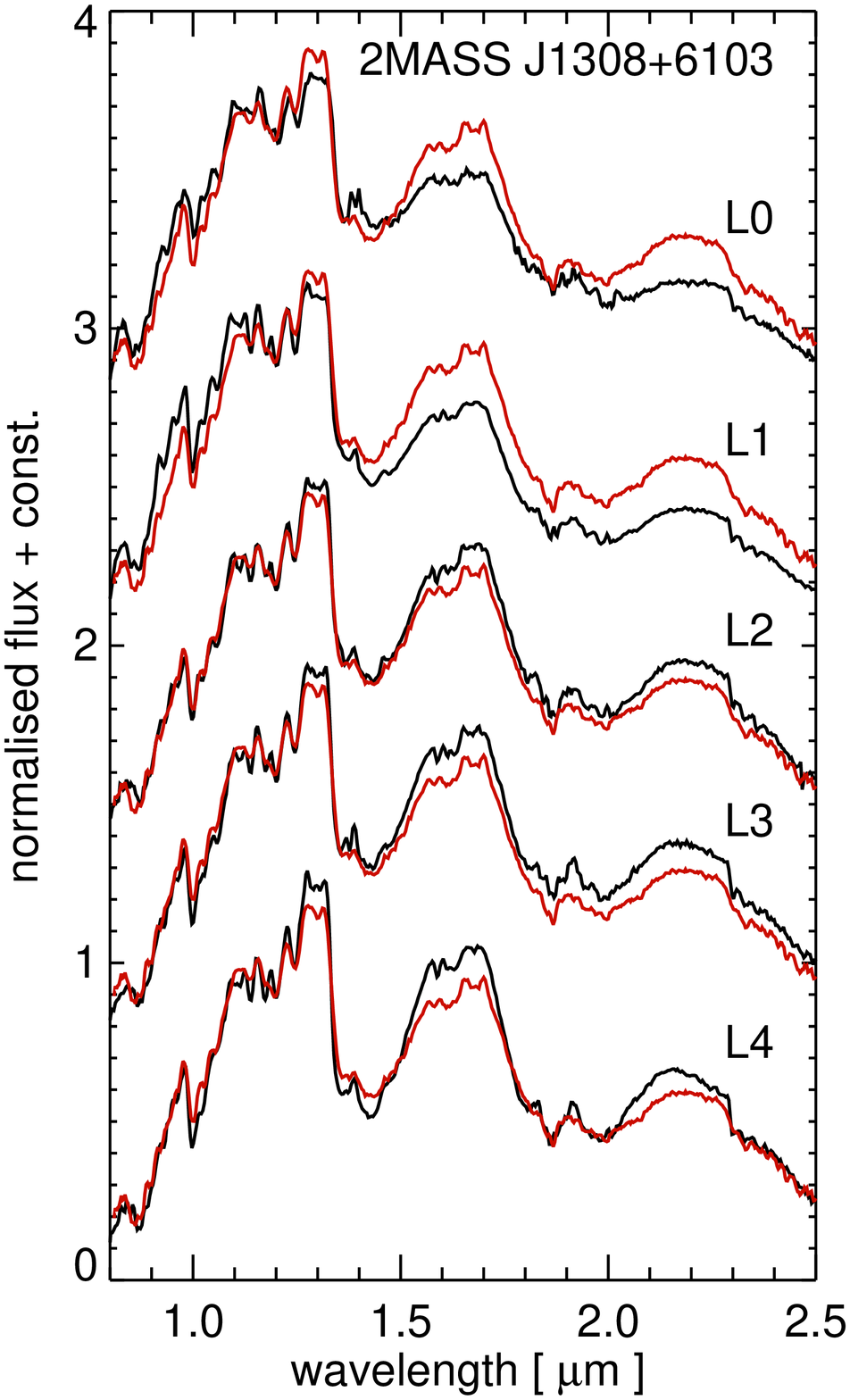}
\includegraphics[scale=0.28]{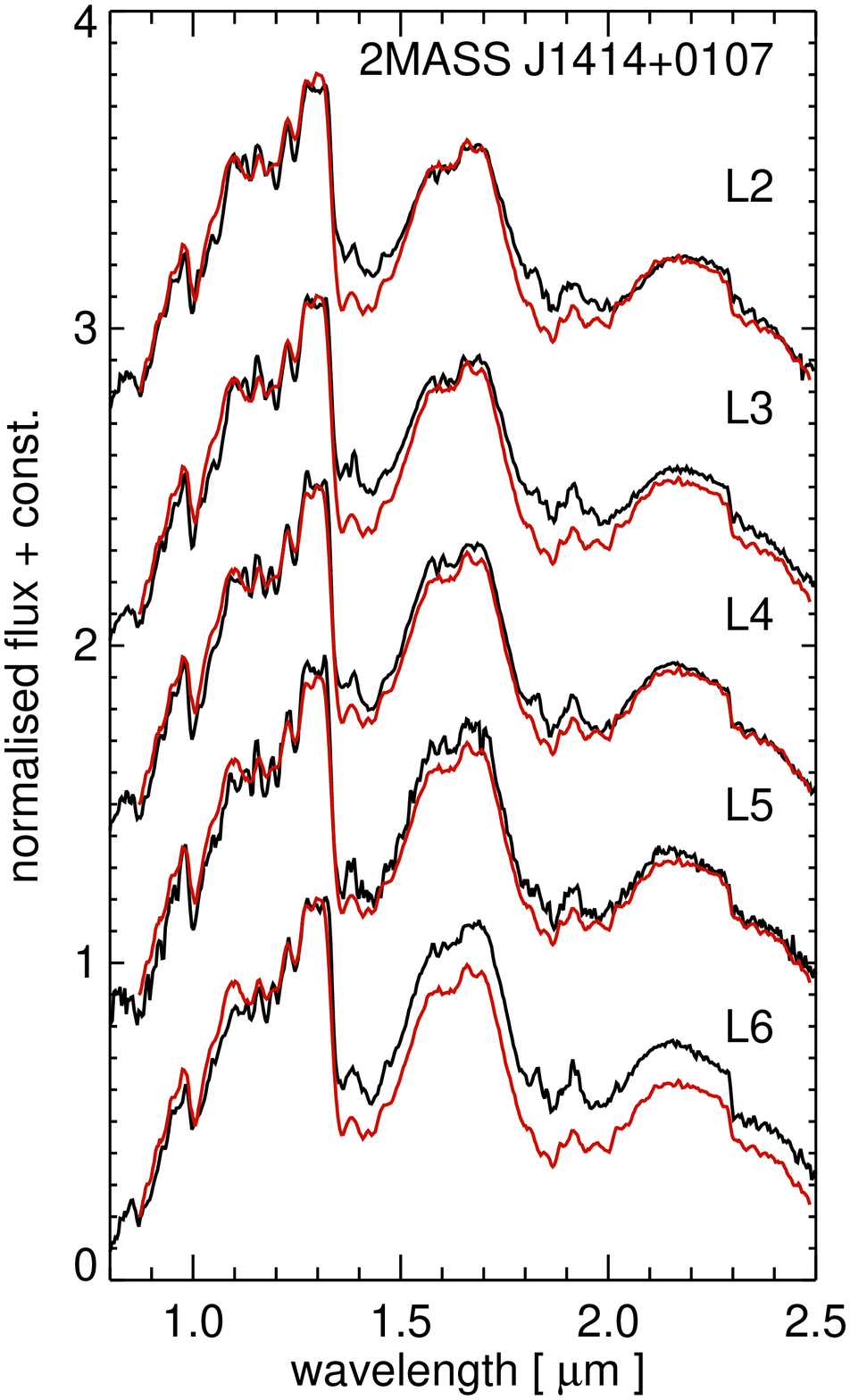}
\includegraphics[scale=0.28]{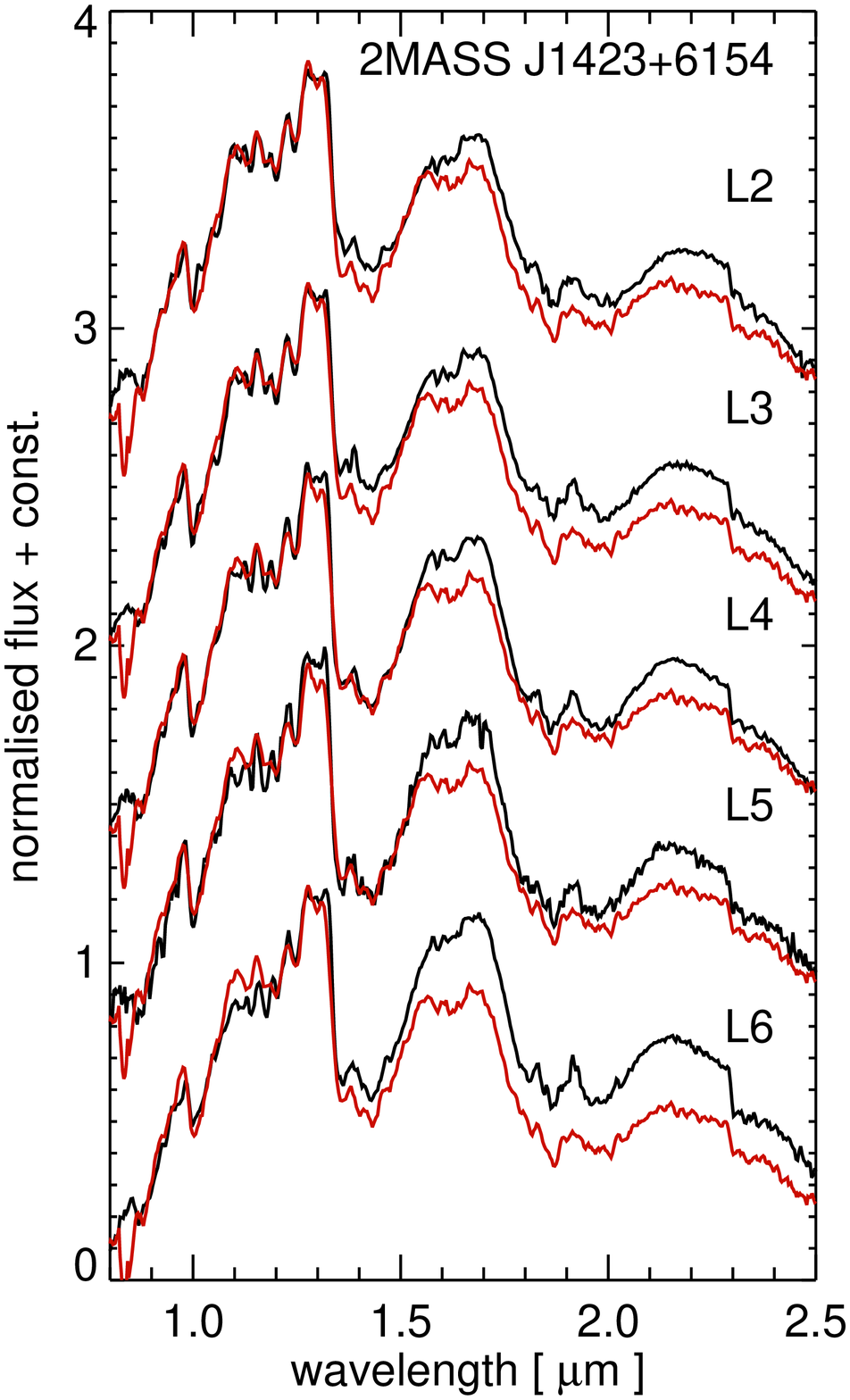}
}
\subfigure[2MASS J1534+0426 (L0), 2MASS J1542$-$0045 (L7) and 2MASS J1542+5522 (L4).]{
\label{fig:seq4}
\includegraphics[scale=0.28]{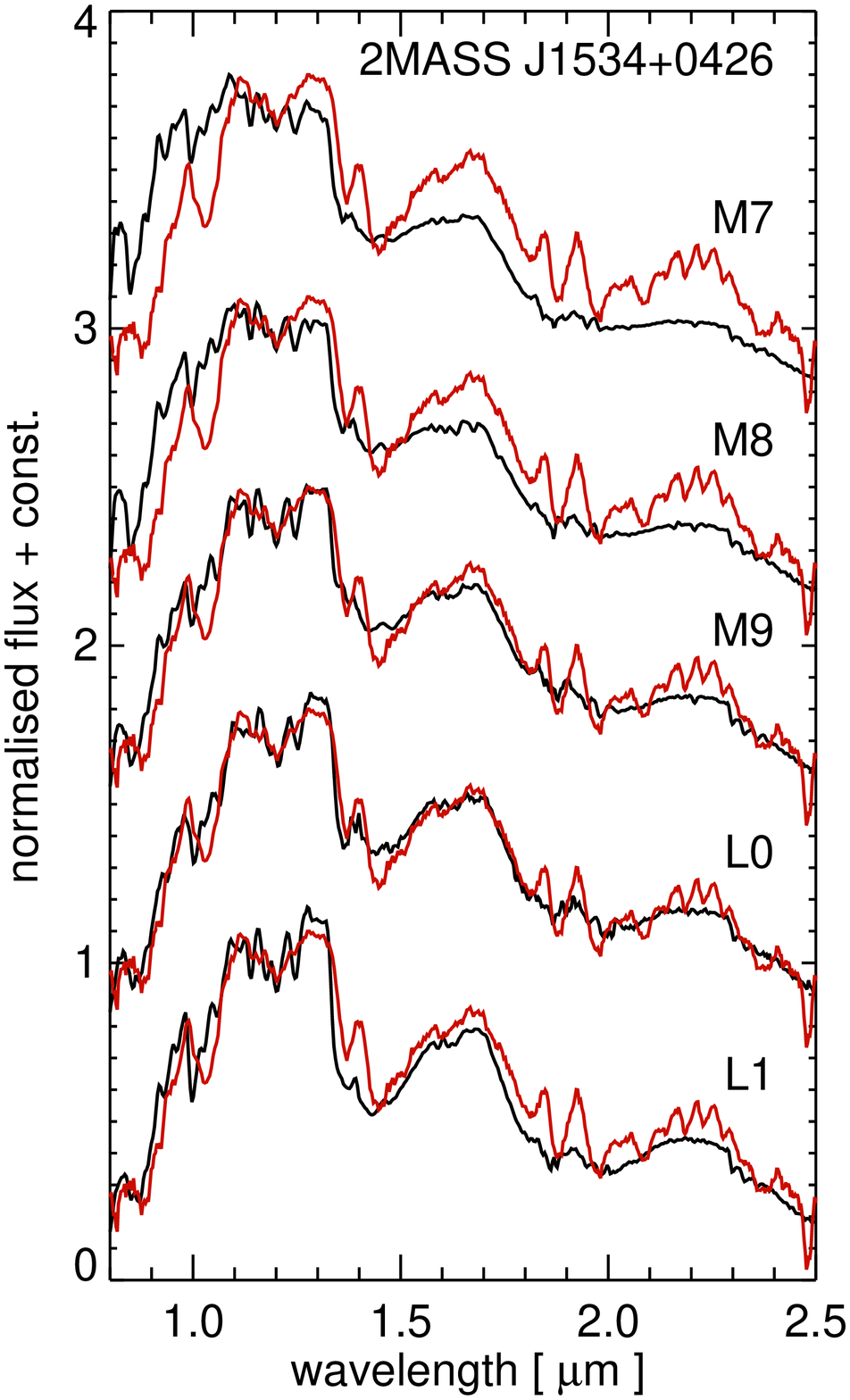}
\includegraphics[scale=0.28]{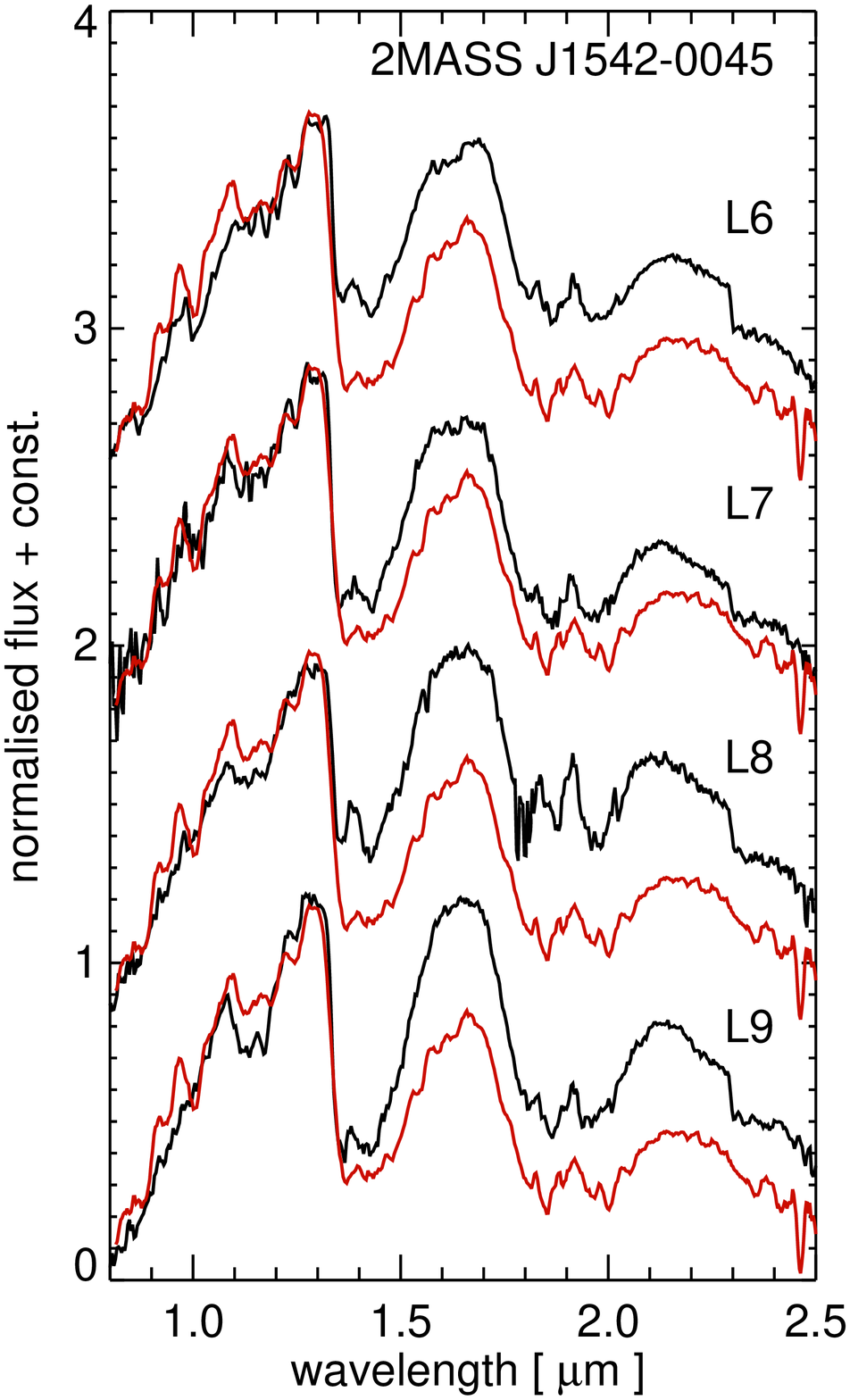}
\includegraphics[scale=0.28]{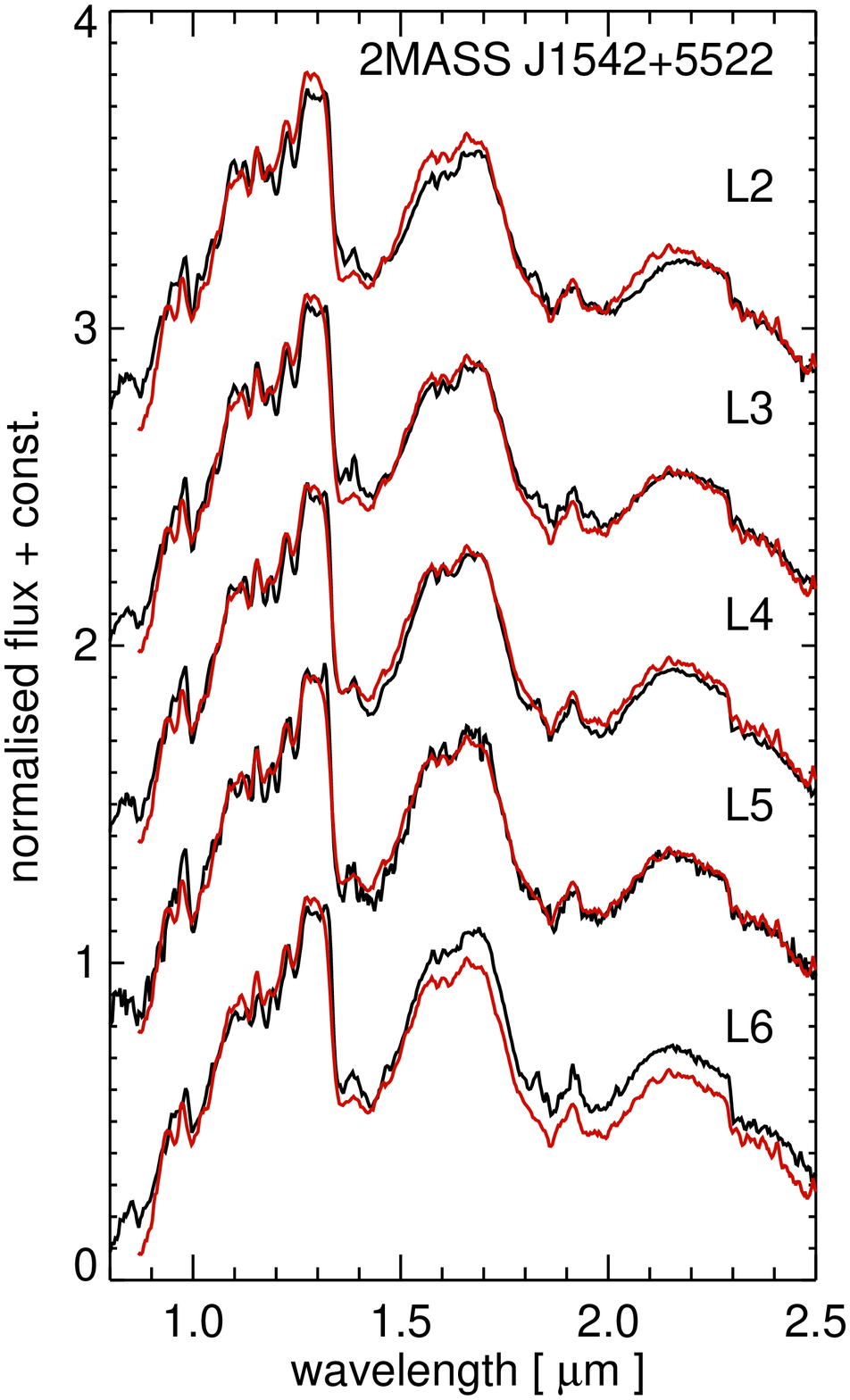}
}
\end{figure}
\addtocounter{figure}{-1}
\begin{figure}
\caption{ continued.}
\addtocounter{subfigure}{4}
\subfigure[2MASS J1551+0151 (M8), 2MASS J1615+4953 (L6) and 2MASS J1716+2945 (L3).]{
\label{fig:seq5}
\includegraphics[scale=0.28]{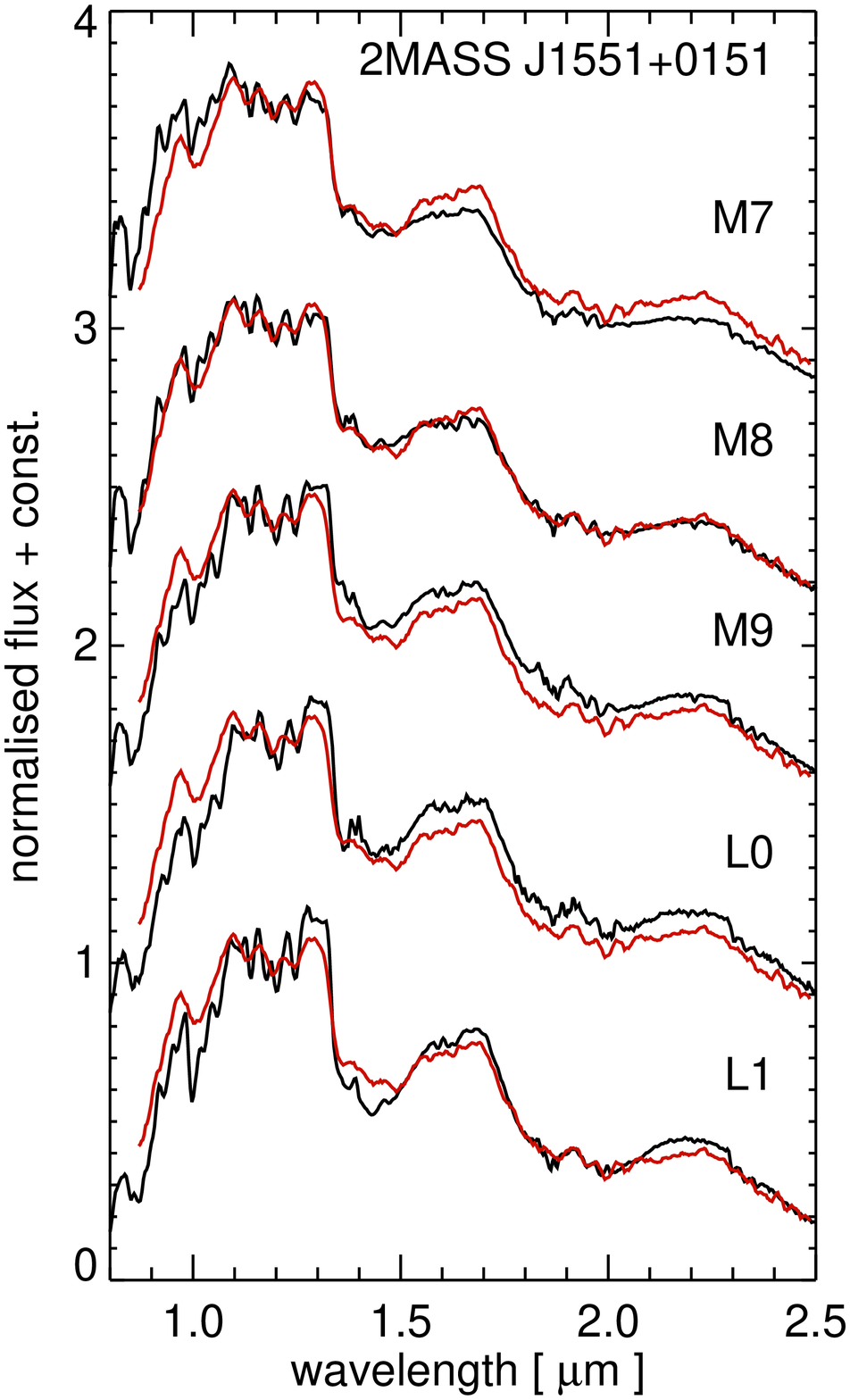}
\includegraphics[scale=0.28]{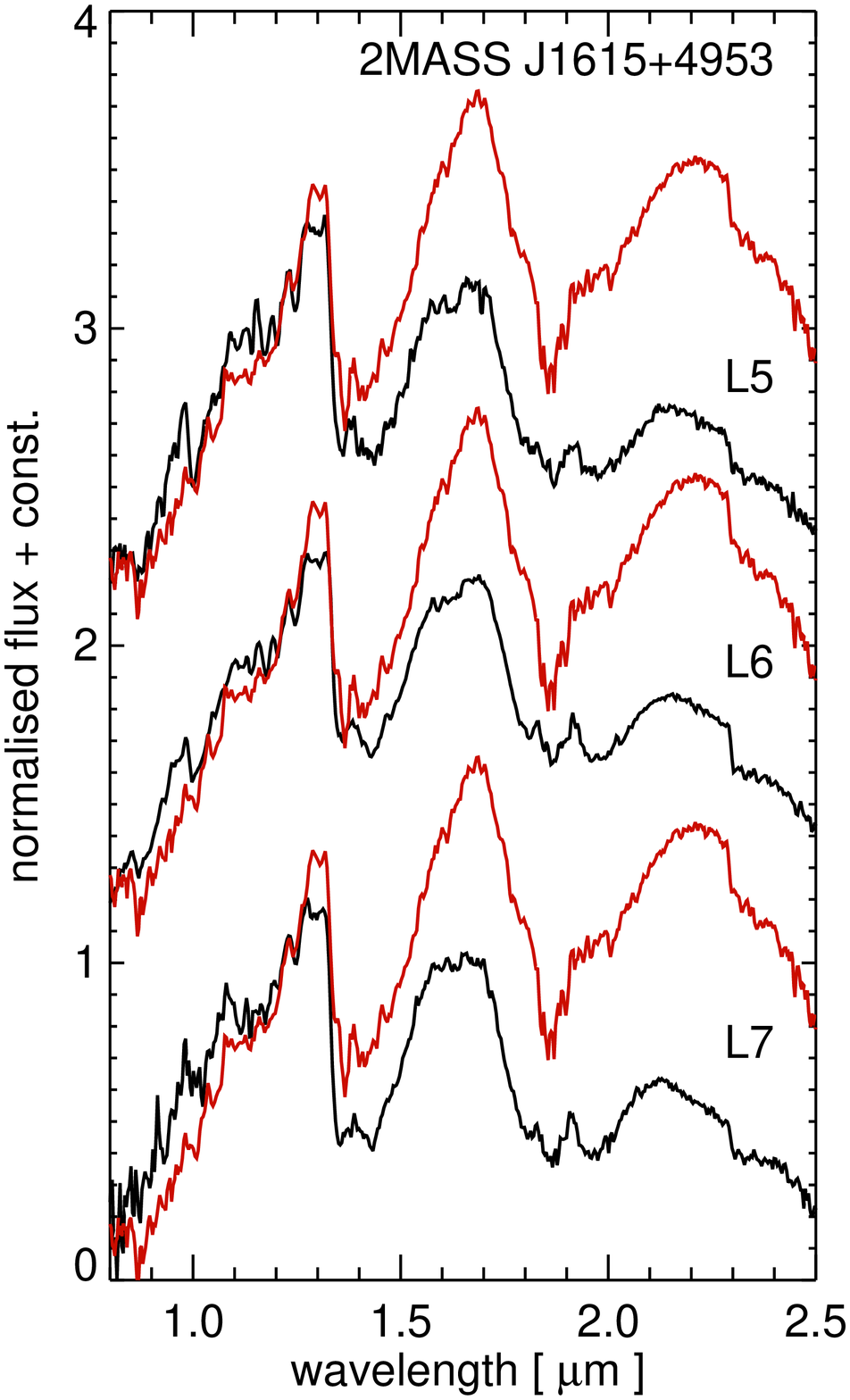}
\includegraphics[scale=0.28]{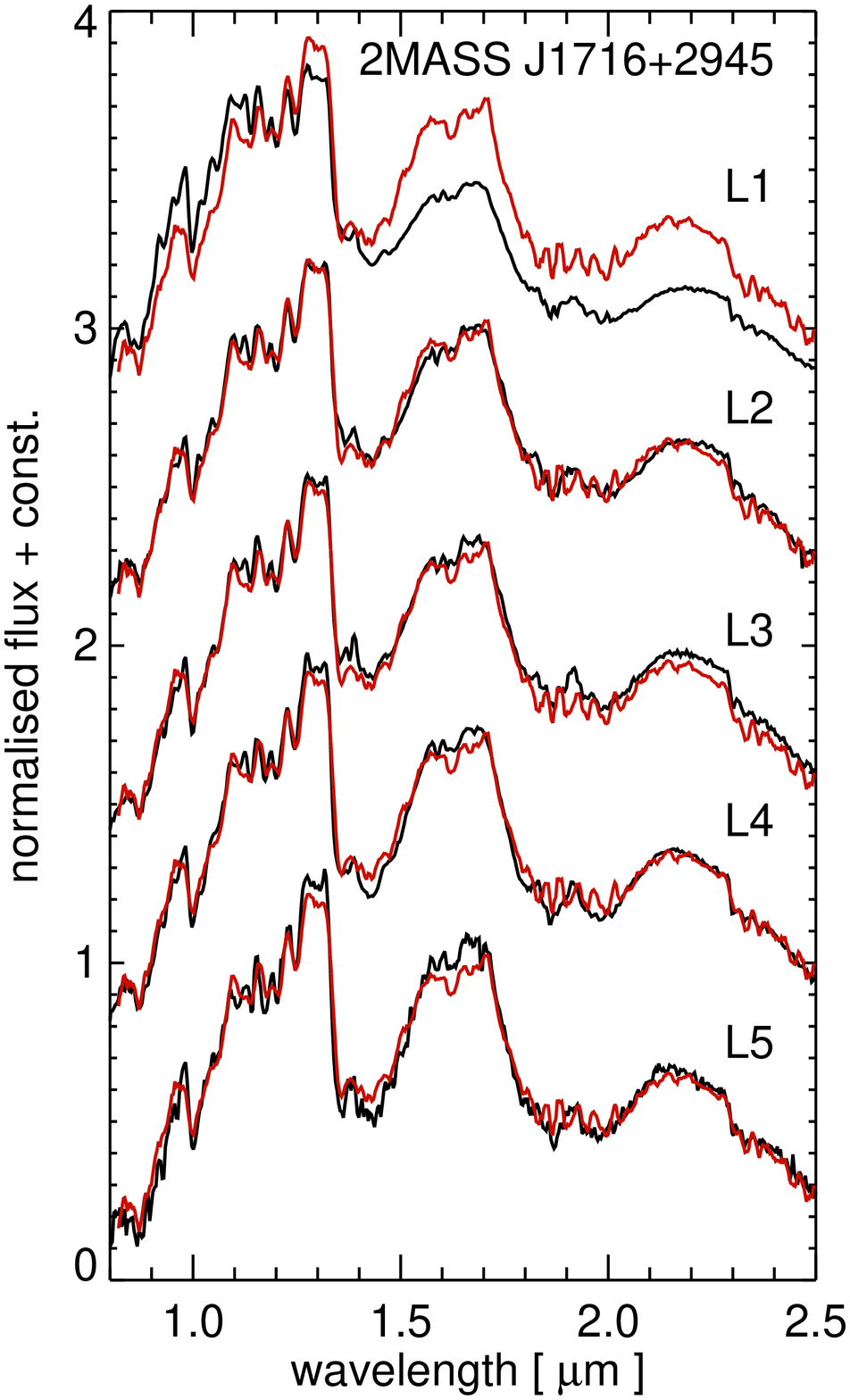}
}
\subfigure[2MASS J1737+5953 (L9) and 2MASS J2116$-$0729 (L6).]{
\label{fig:seq6}
\includegraphics[scale=0.28]{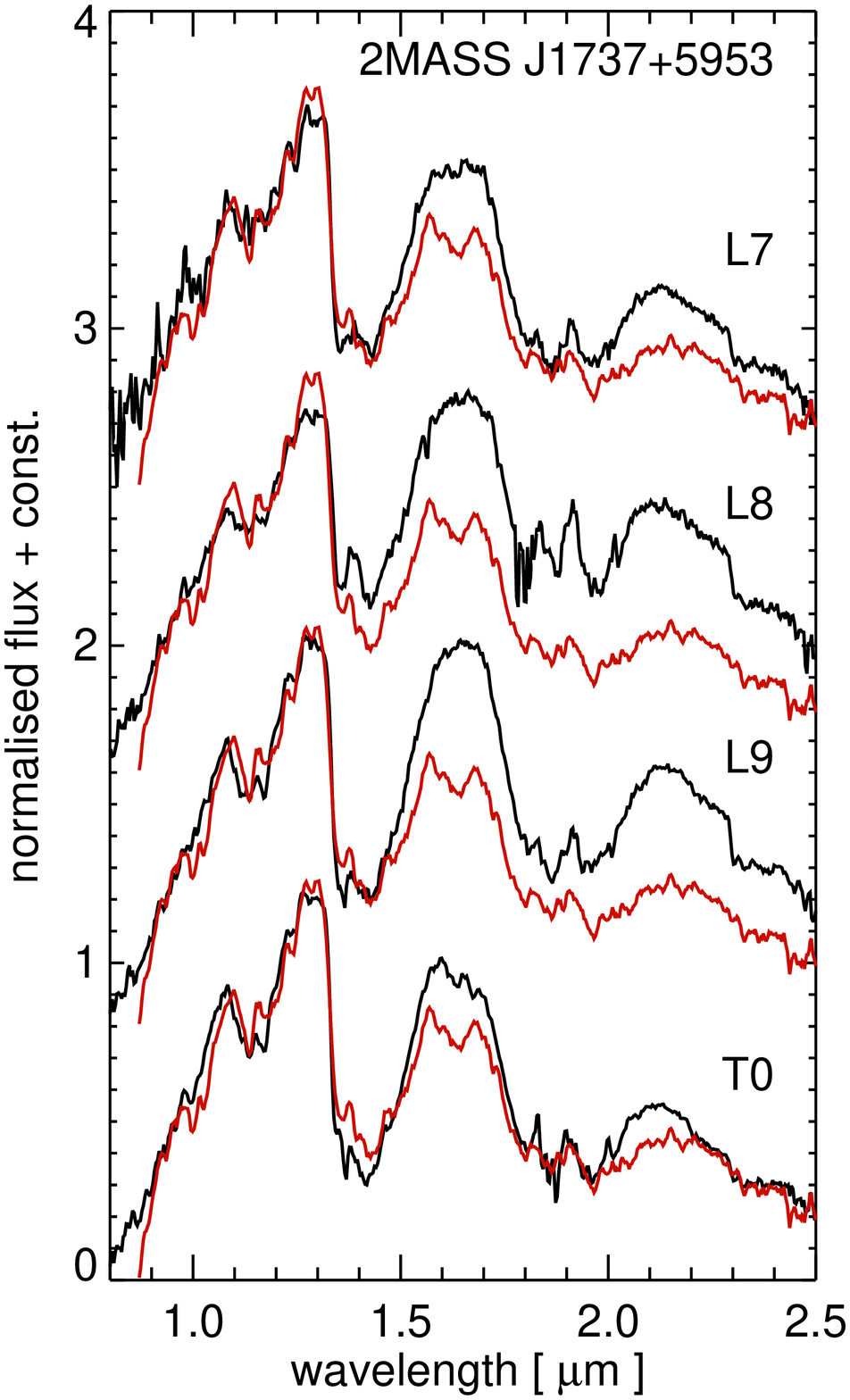}
\includegraphics[scale=0.28]{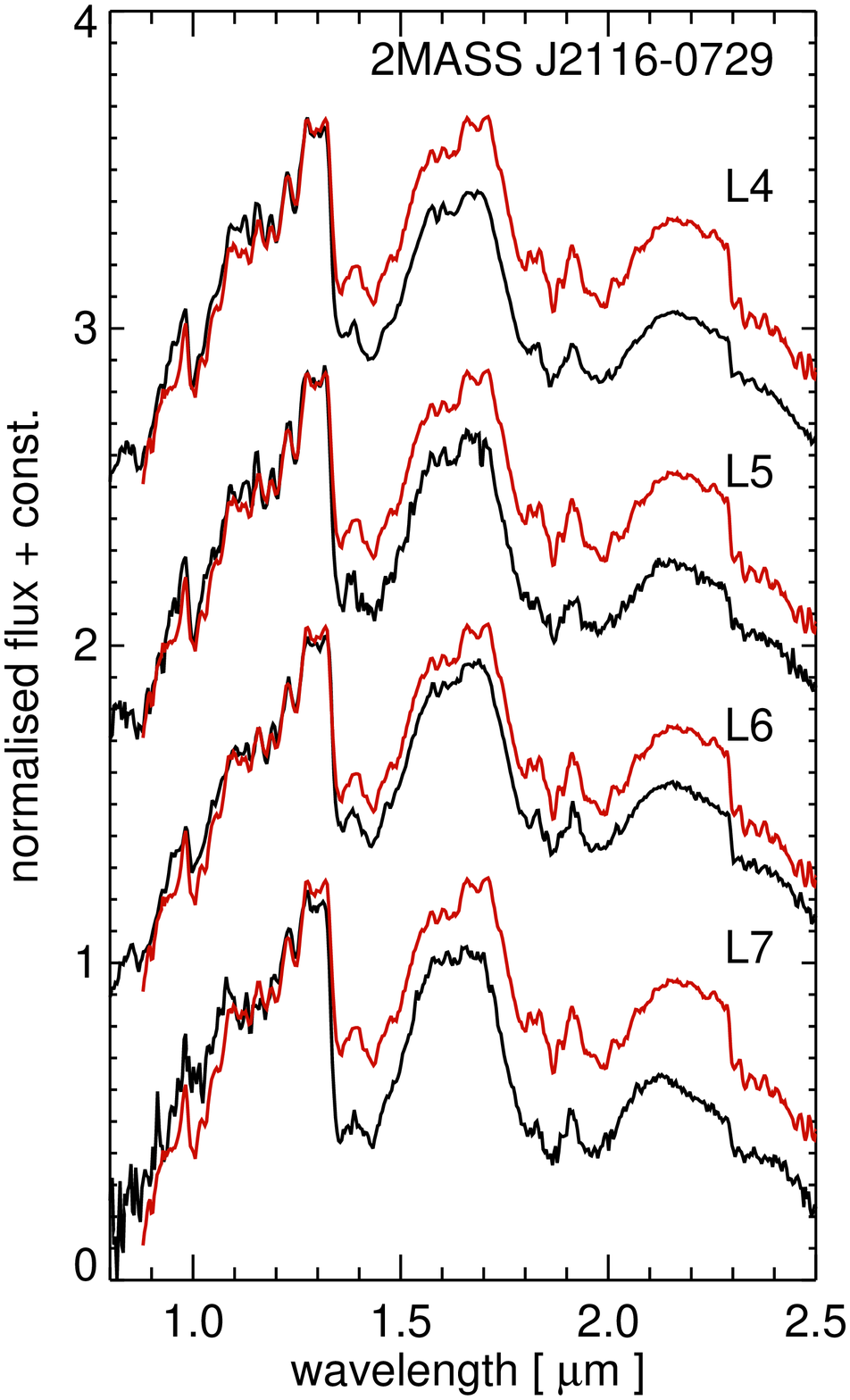}
}
\end{figure}

\begin{figure}
\caption{\label{fig:bin}Probable binary candidates from the SDSS DR1 and 2MASS PSC 
cross-match L dwarf sample. The first two columns compare the target spectrum (black) to 
best $\chi^{2}$ fit single and composite templates (green) over the complete 
0.95--2.35~$\micron$ range. Thereby the left column shows the best fit single template, 
while the middle column shows the best fit composite spectra, composed of the stated single
templates (blue and red). Lastly, the right column shows the best 
0.95--1.35~$\micron$ $\chi^{2}$ fit.  
The binary candidate spectra (black) are normalized to the average flux in the 
1.2\,-\,1.25\,$\mu$m region, while the best-fit single and composite templates are 
normalized to minimum $\chi^{2}$ deviations.}
\includegraphics[angle=90,scale=0.19]{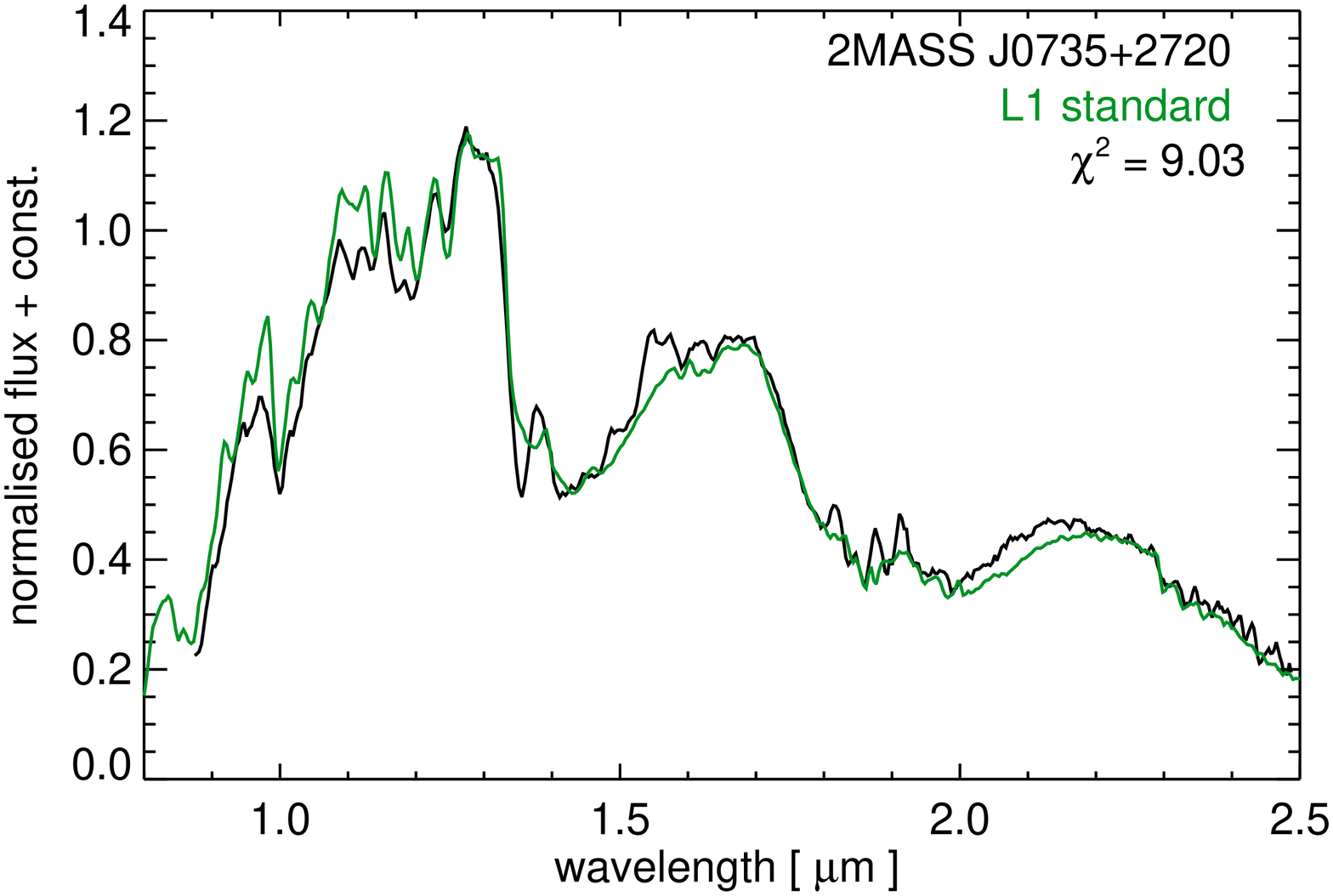}
\includegraphics[angle=90,scale=0.19]{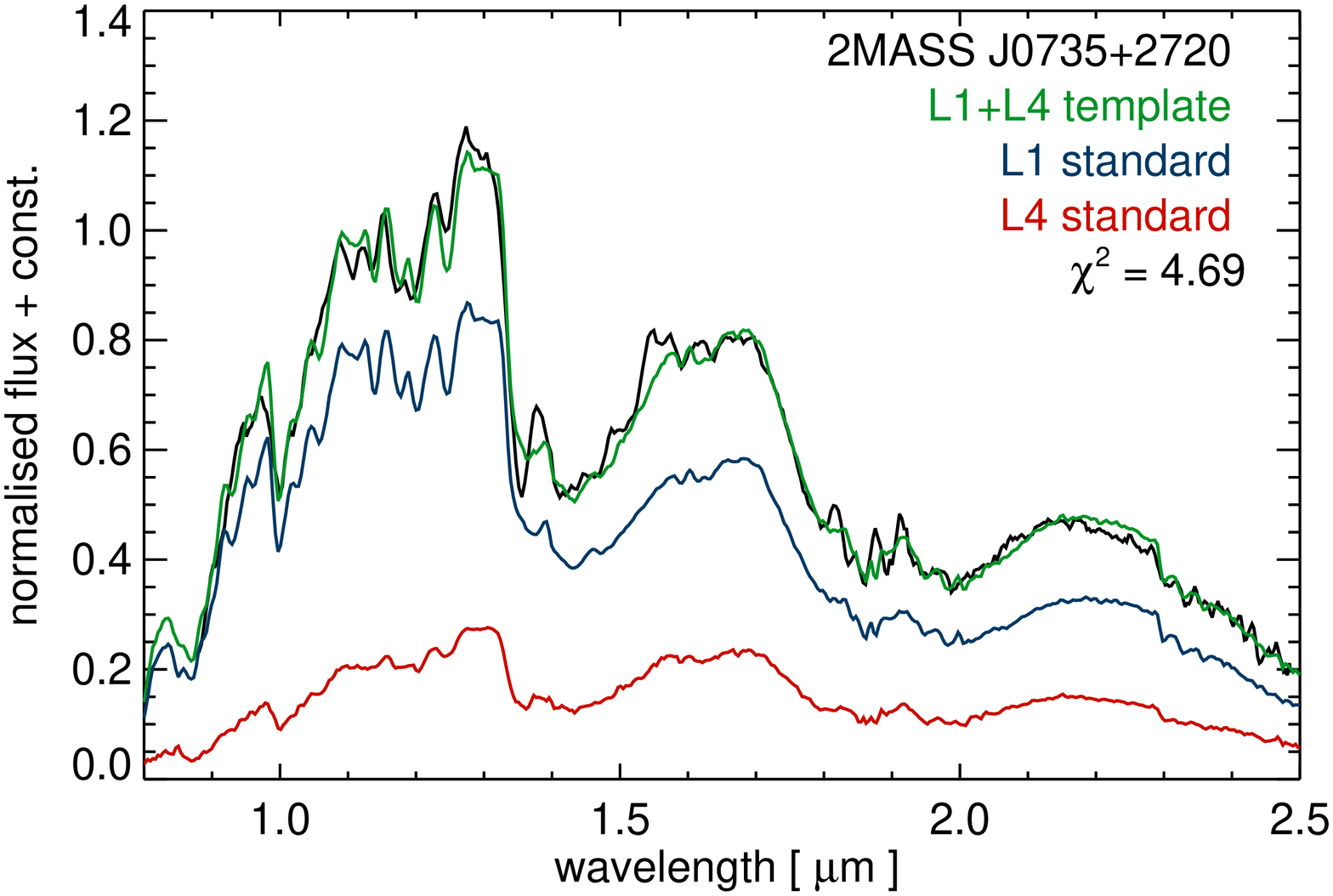}
\includegraphics[angle=90,scale=0.19]{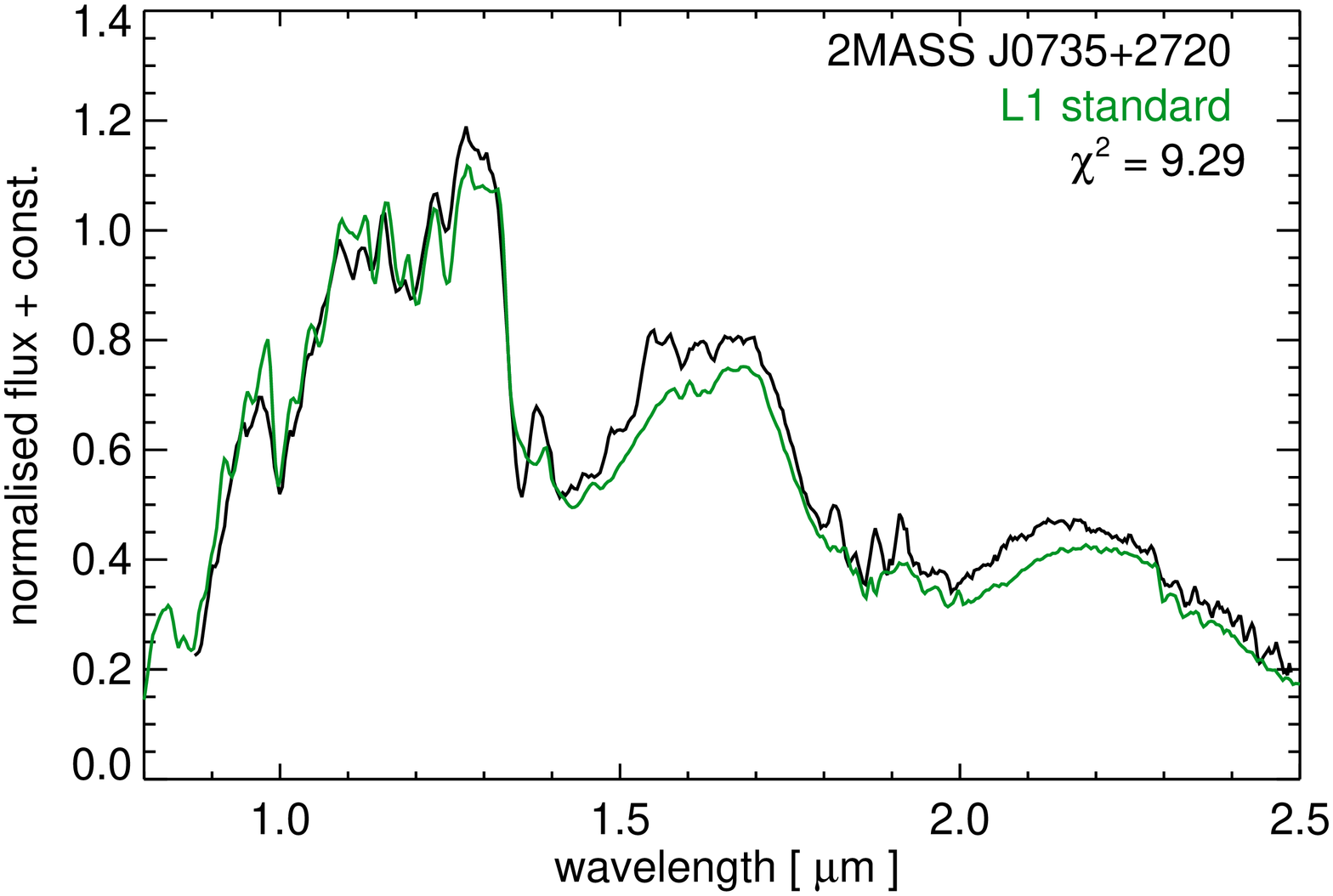} \\
\includegraphics[angle=90,scale=0.19]{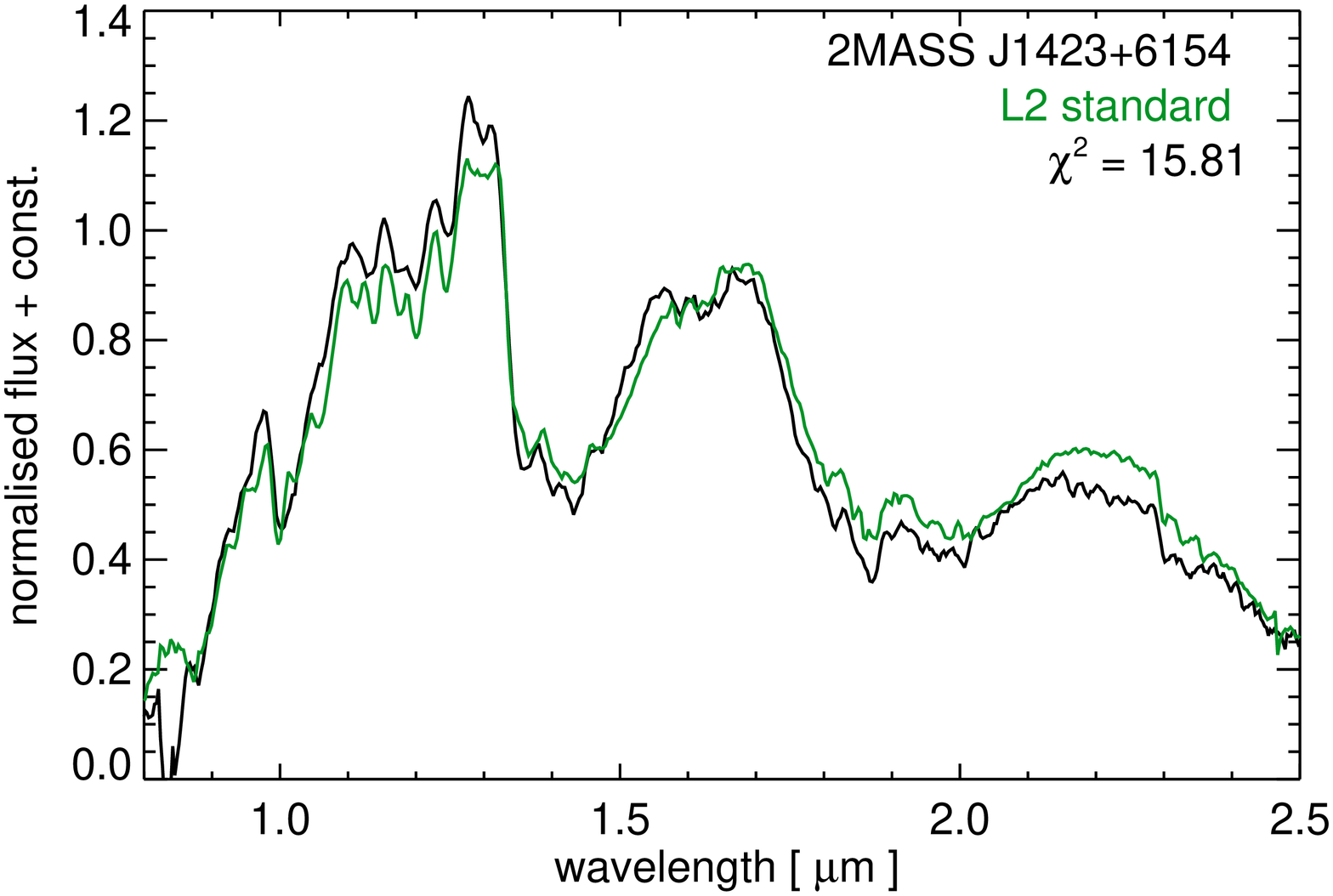}
\includegraphics[angle=90,scale=0.19]{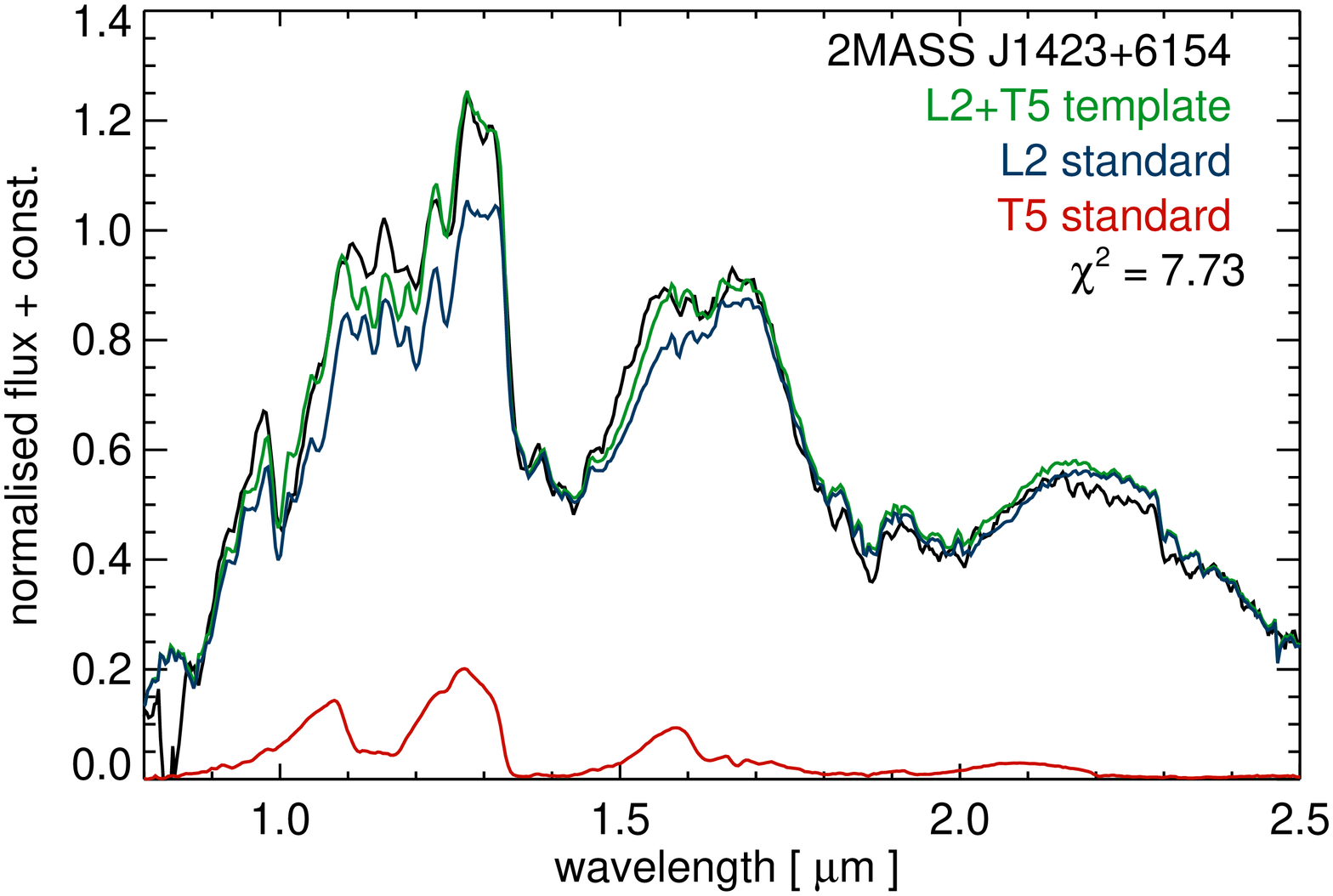}
\includegraphics[angle=90,scale=0.19]{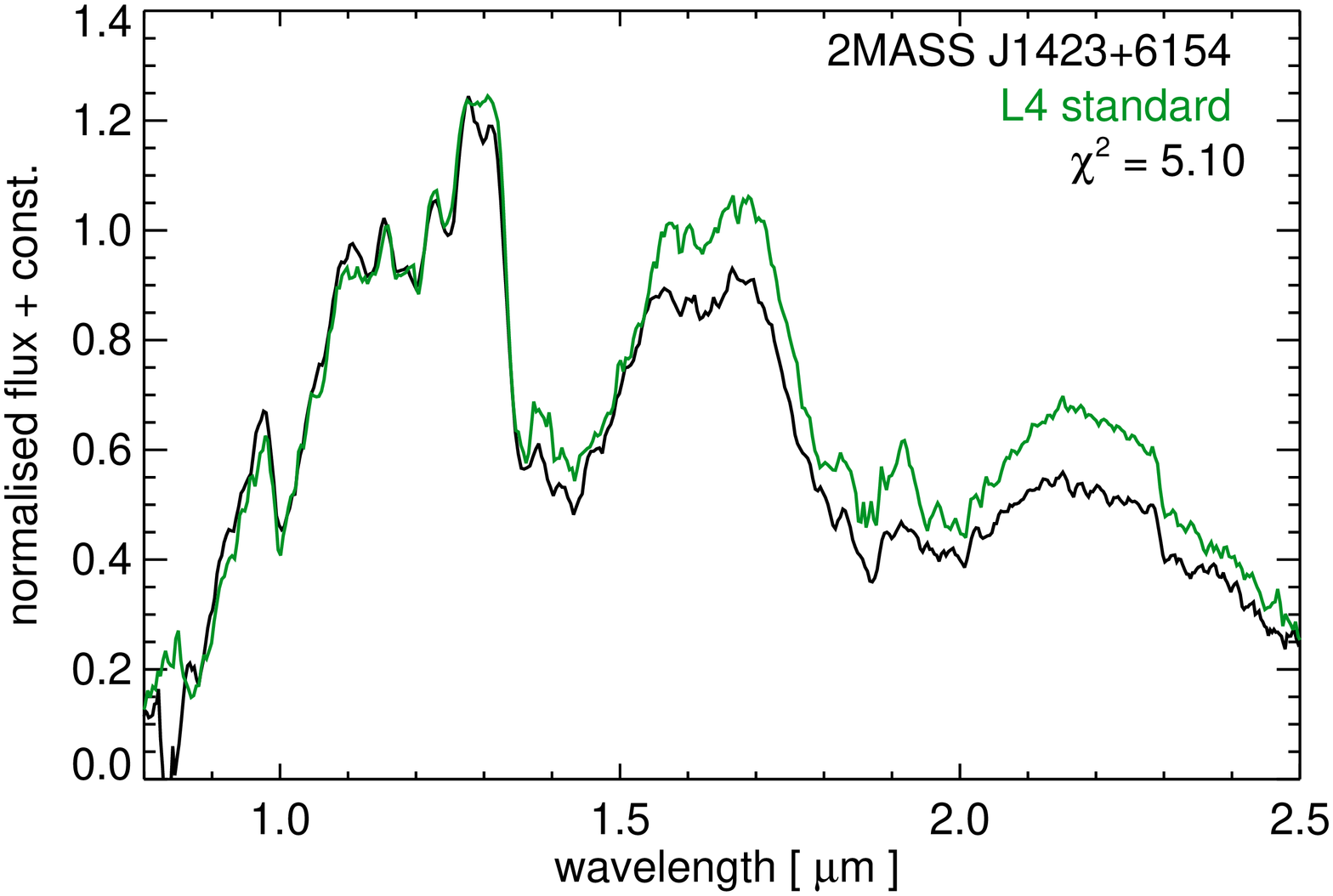} \\
\includegraphics[angle=90,scale=0.19]{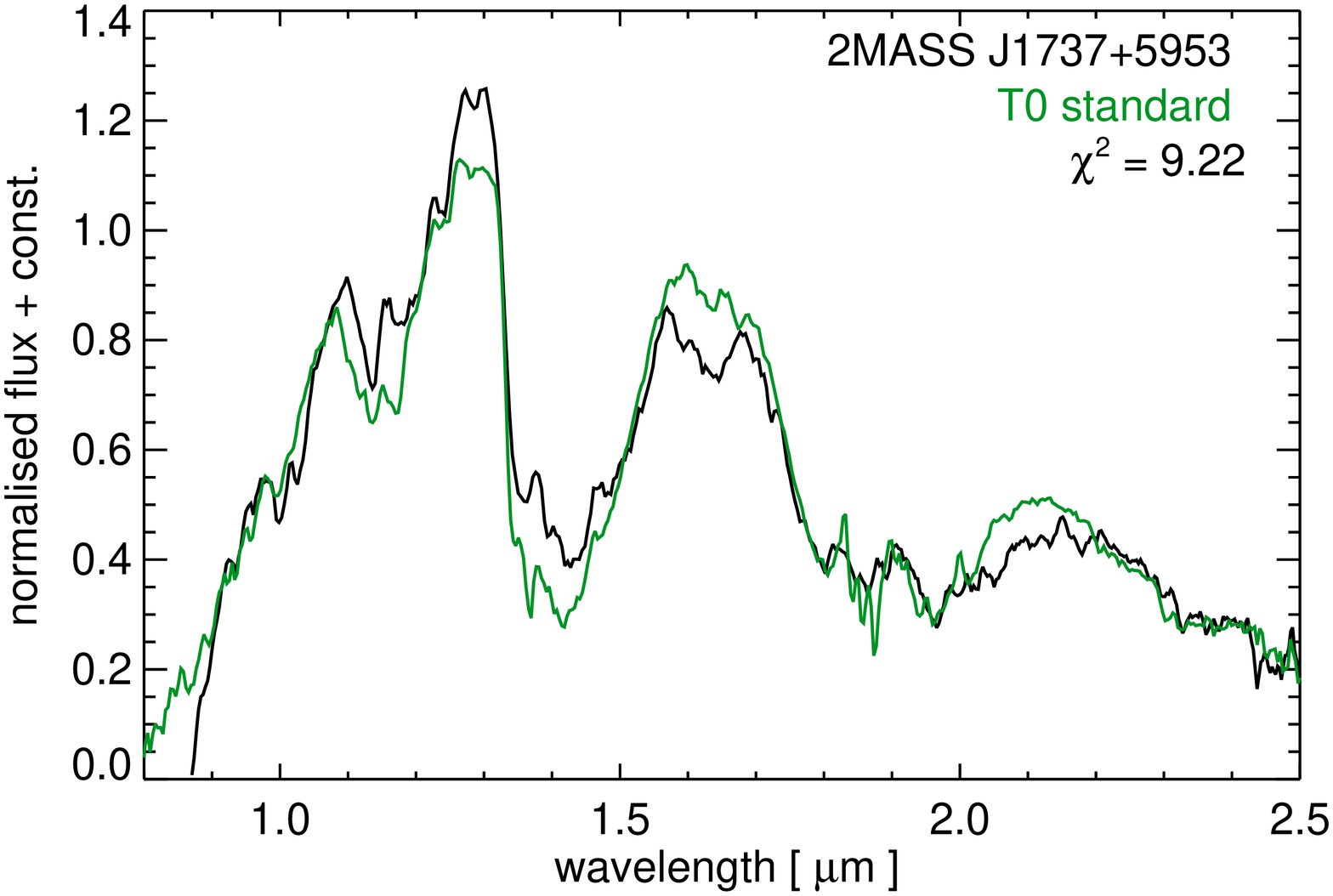}
\includegraphics[angle=90,scale=0.19]{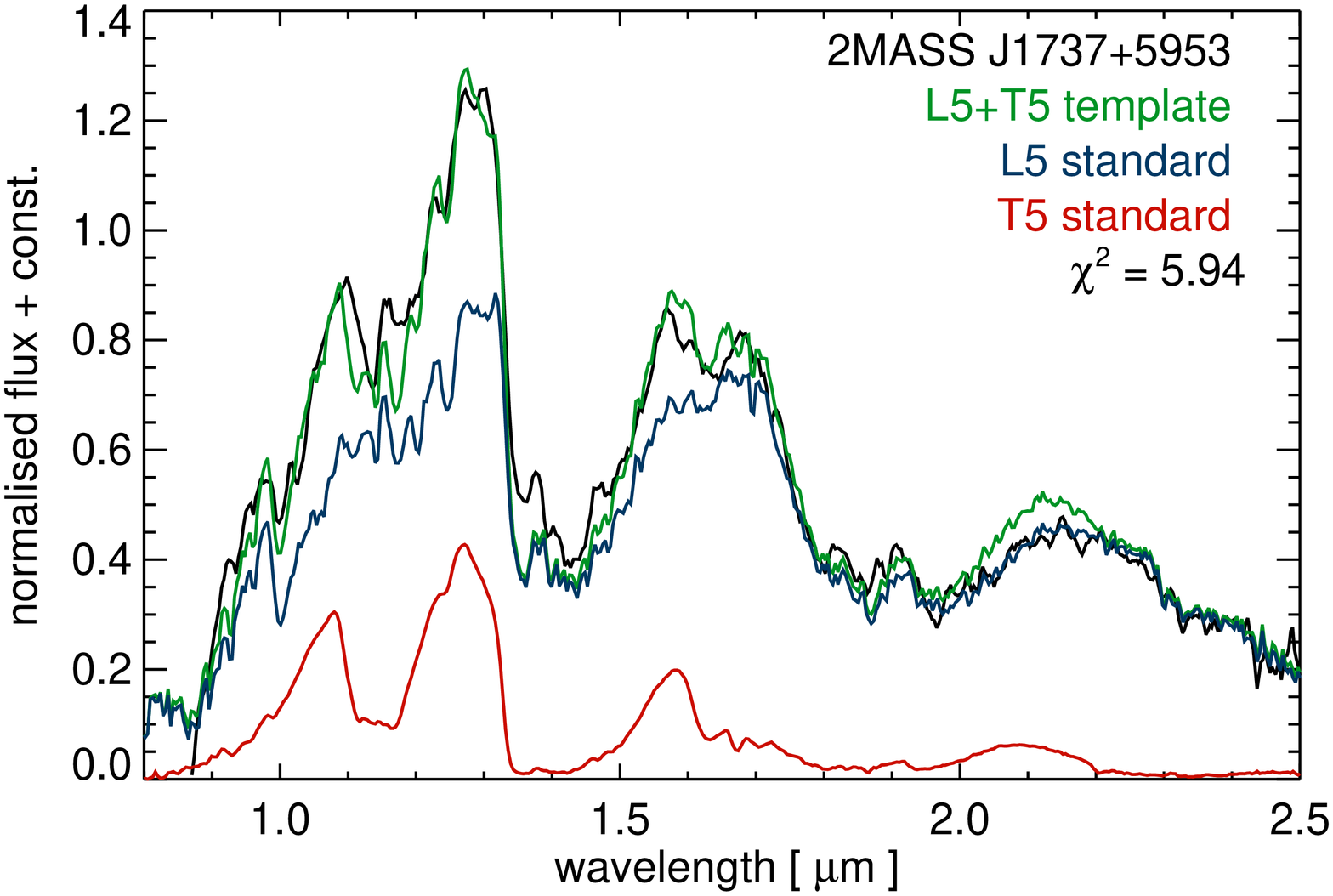}
\includegraphics[angle=90,scale=0.19]{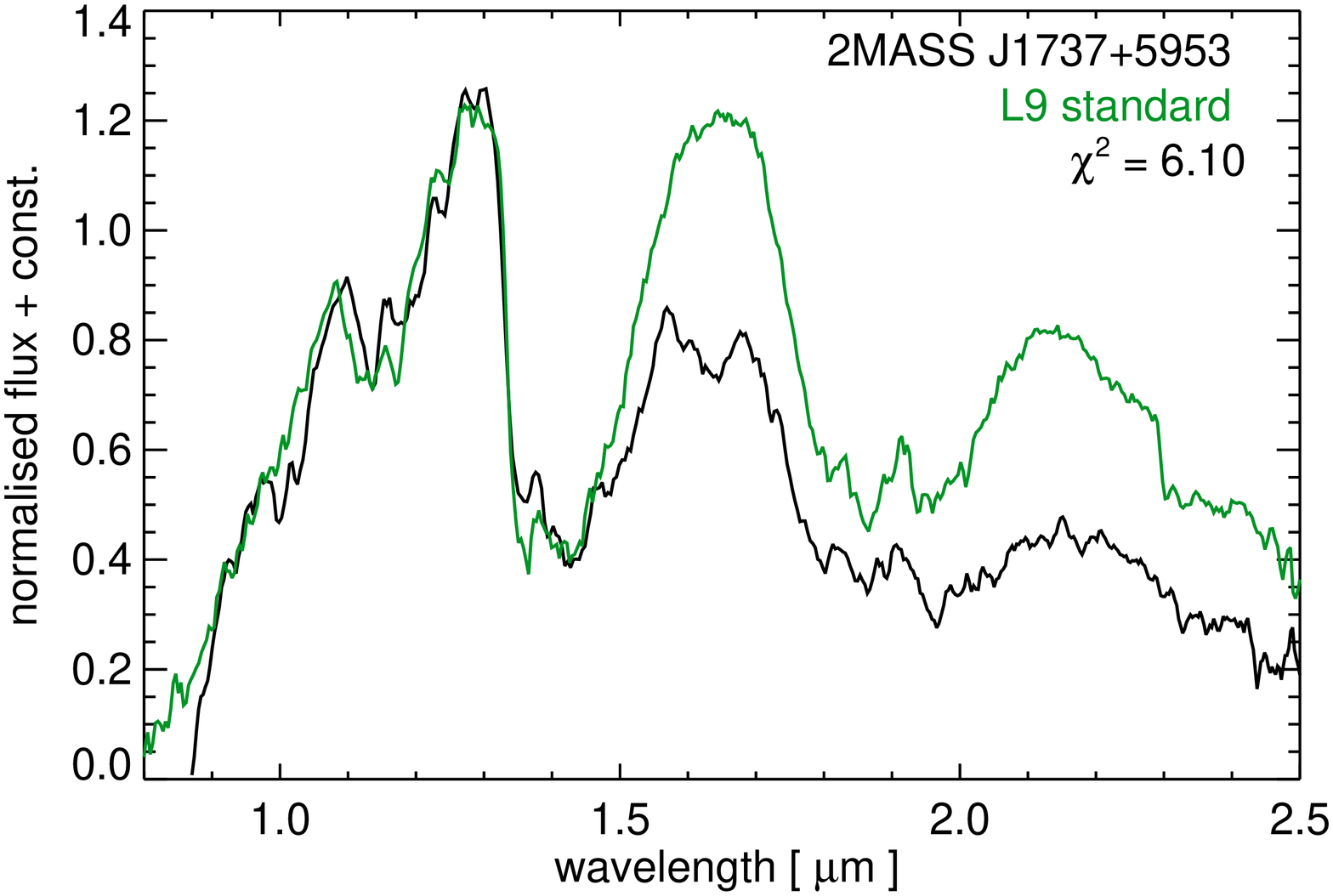} \\
\end{figure}
\begin{figure}
\caption{\label{fig:addbin}Additional binary candidates from Paper~I. The left column 
shows the best-fit single template (green) compared to the targets spectrum (black), while 
the rigth column shows the best-fit composite template (green), along with the associated 
primary and secondary (blue and red, respectively). 
The binary candidate spectra are normalized to the average flux in the 1.2\,-\,1.25\,$\mu$m 
region and the $\chi^{2}$ fit is performed over the wavelength range of 0.95--2.35~$\micron$.  
The best-fit single and composite templates are normalized to minimum $\chi^{2}$ deviations.}
\includegraphics[angle=90,scale=0.29]{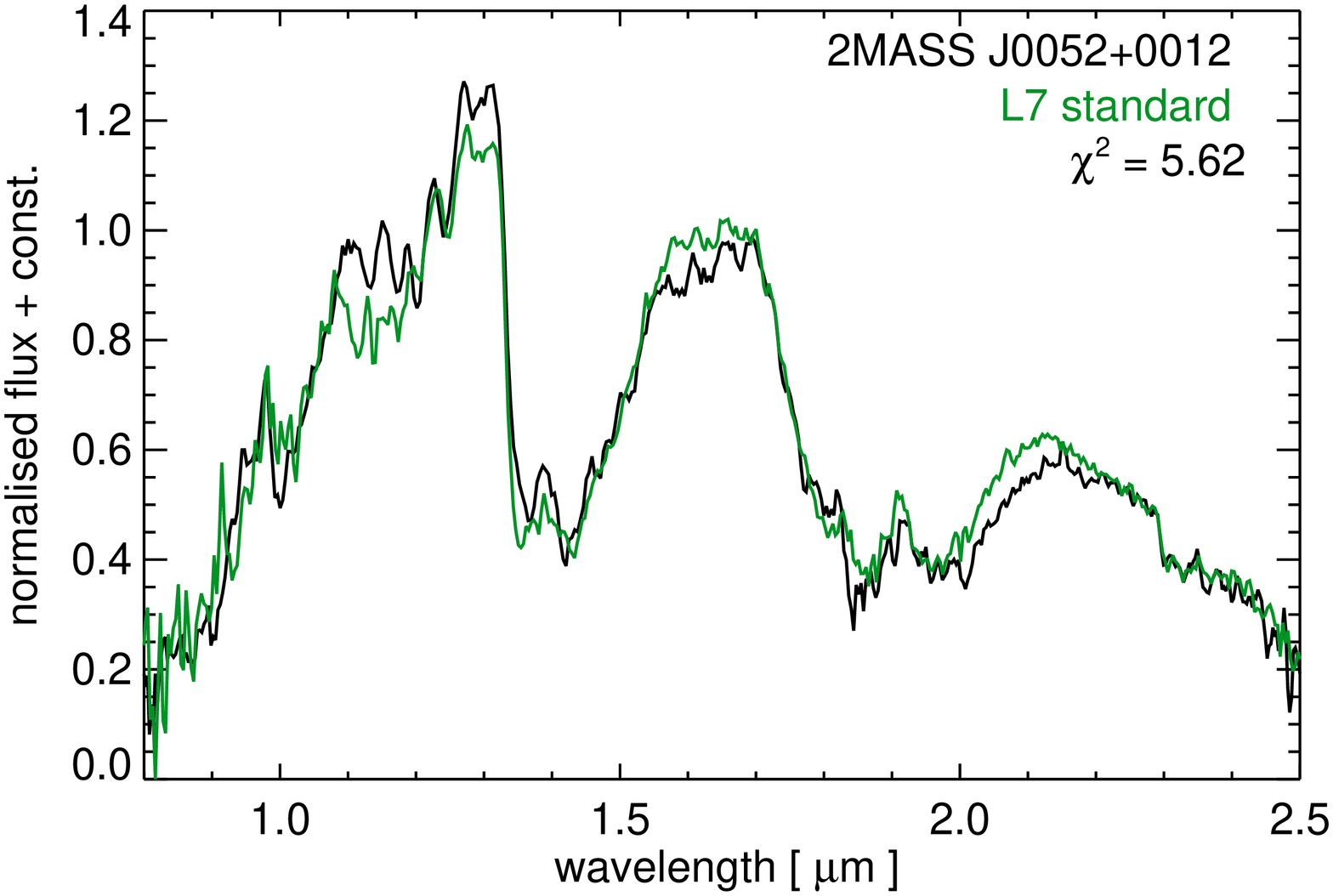}
\includegraphics[angle=90,scale=0.29]{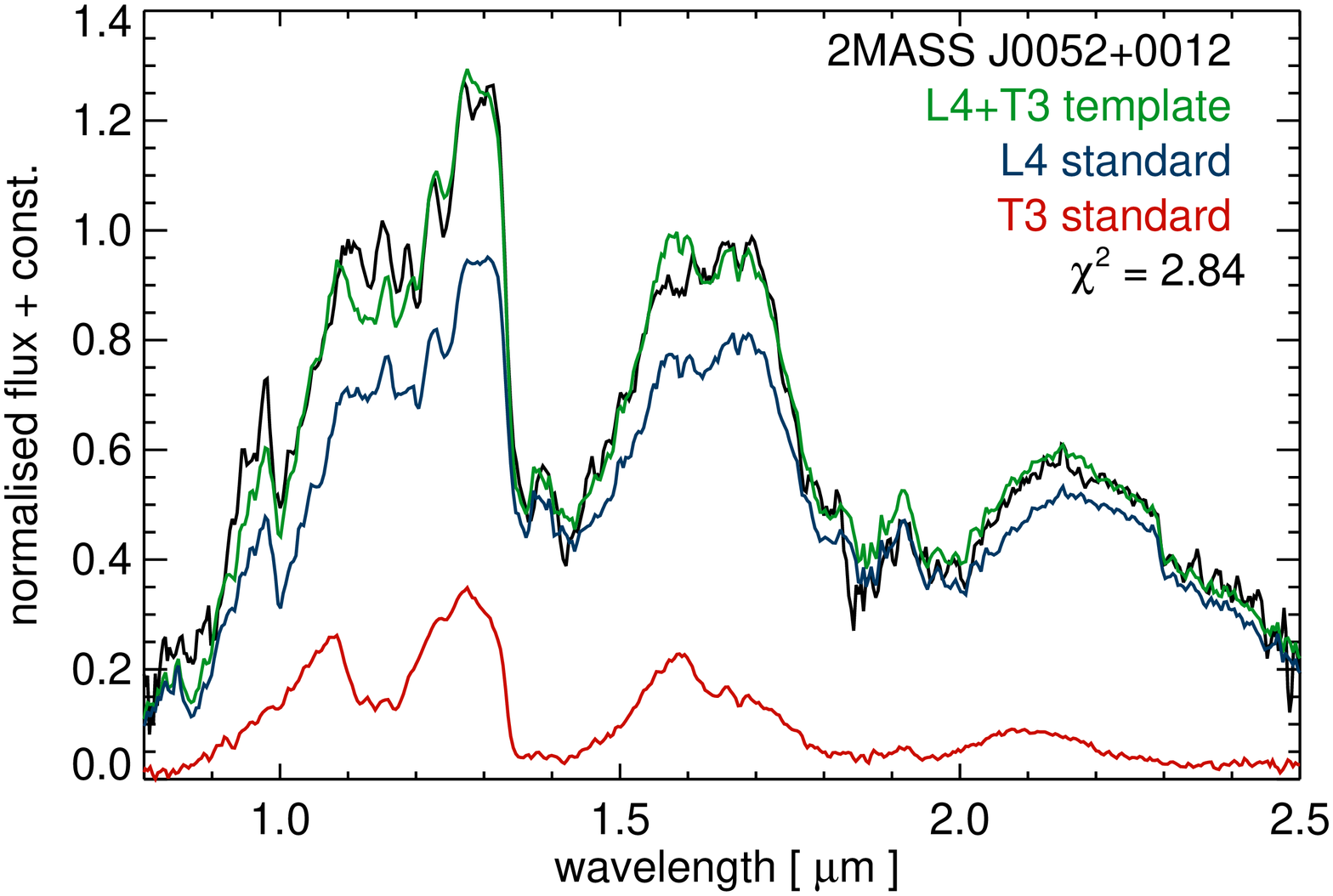} \\
\includegraphics[angle=90,scale=0.29]{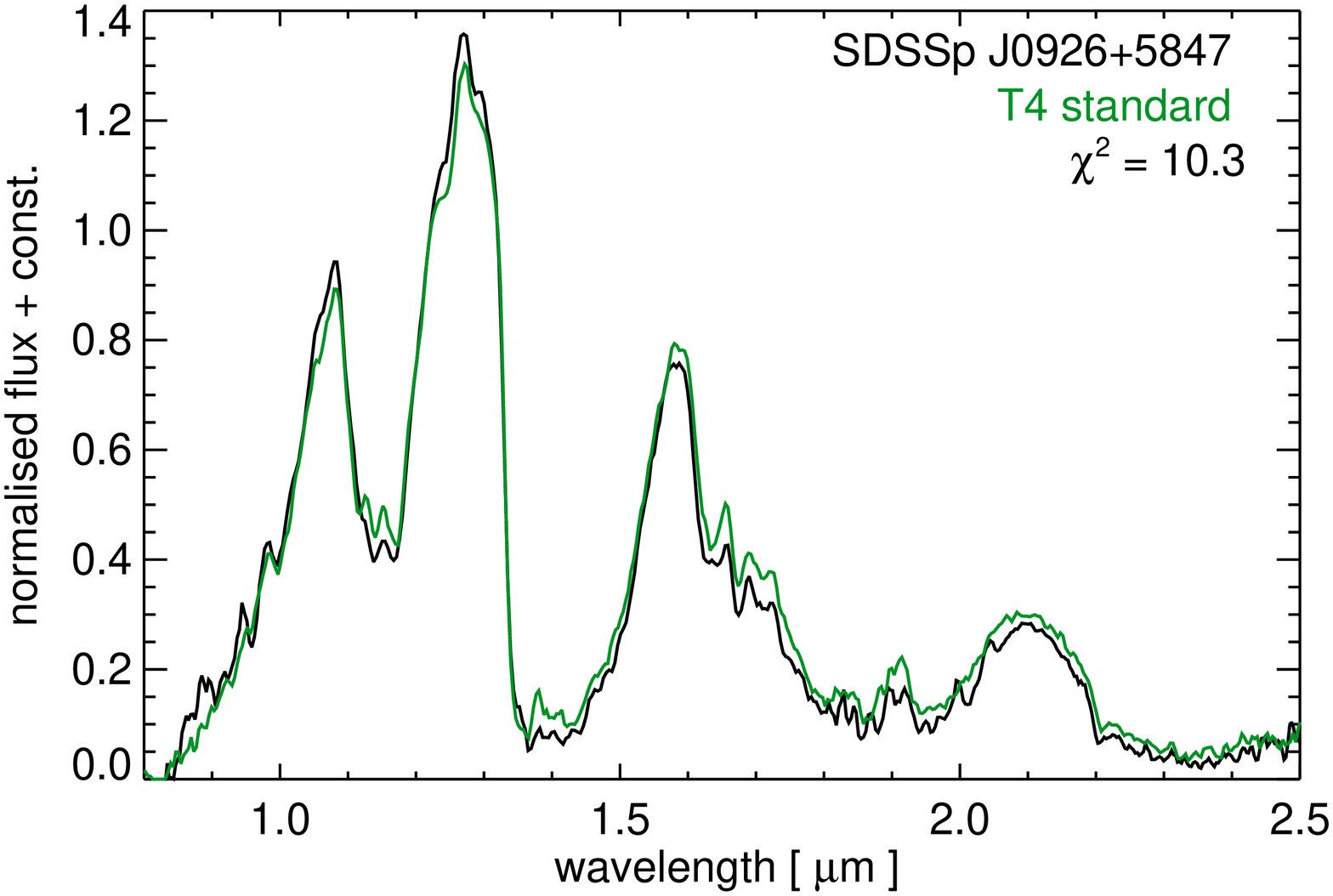}
\includegraphics[angle=90,scale=0.29]{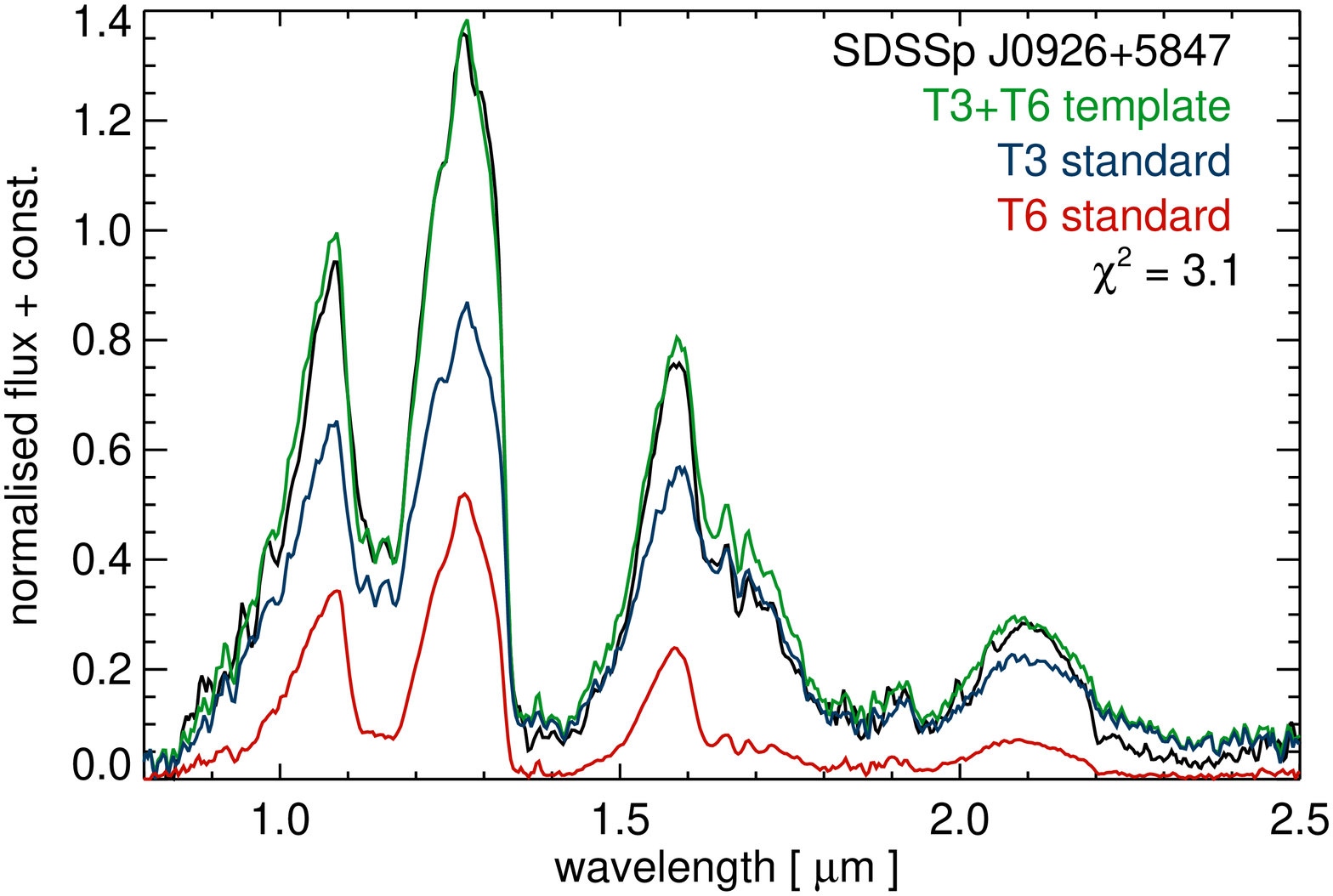} \\
\includegraphics[angle=90,scale=0.29]{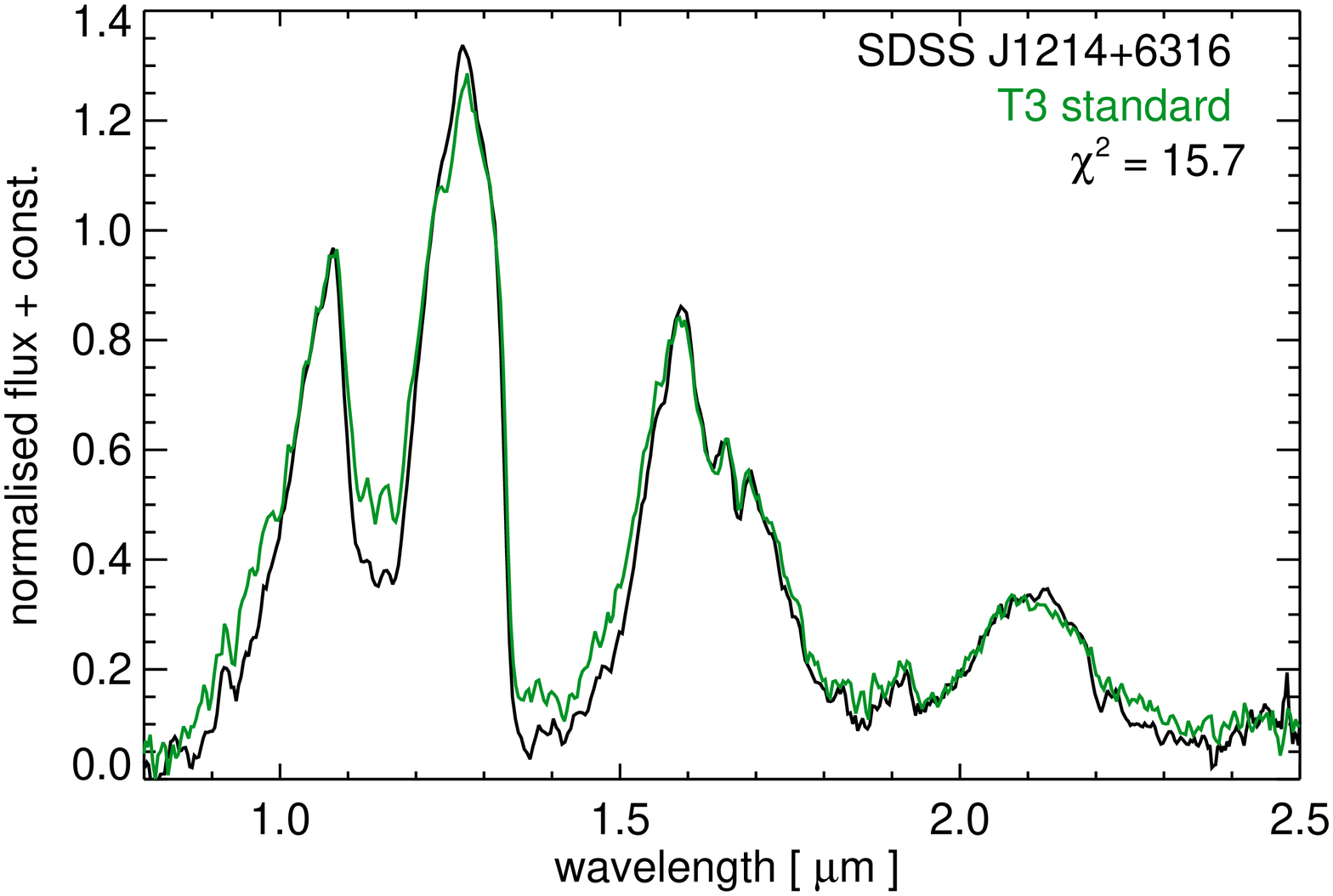}
\includegraphics[angle=90,scale=0.29]{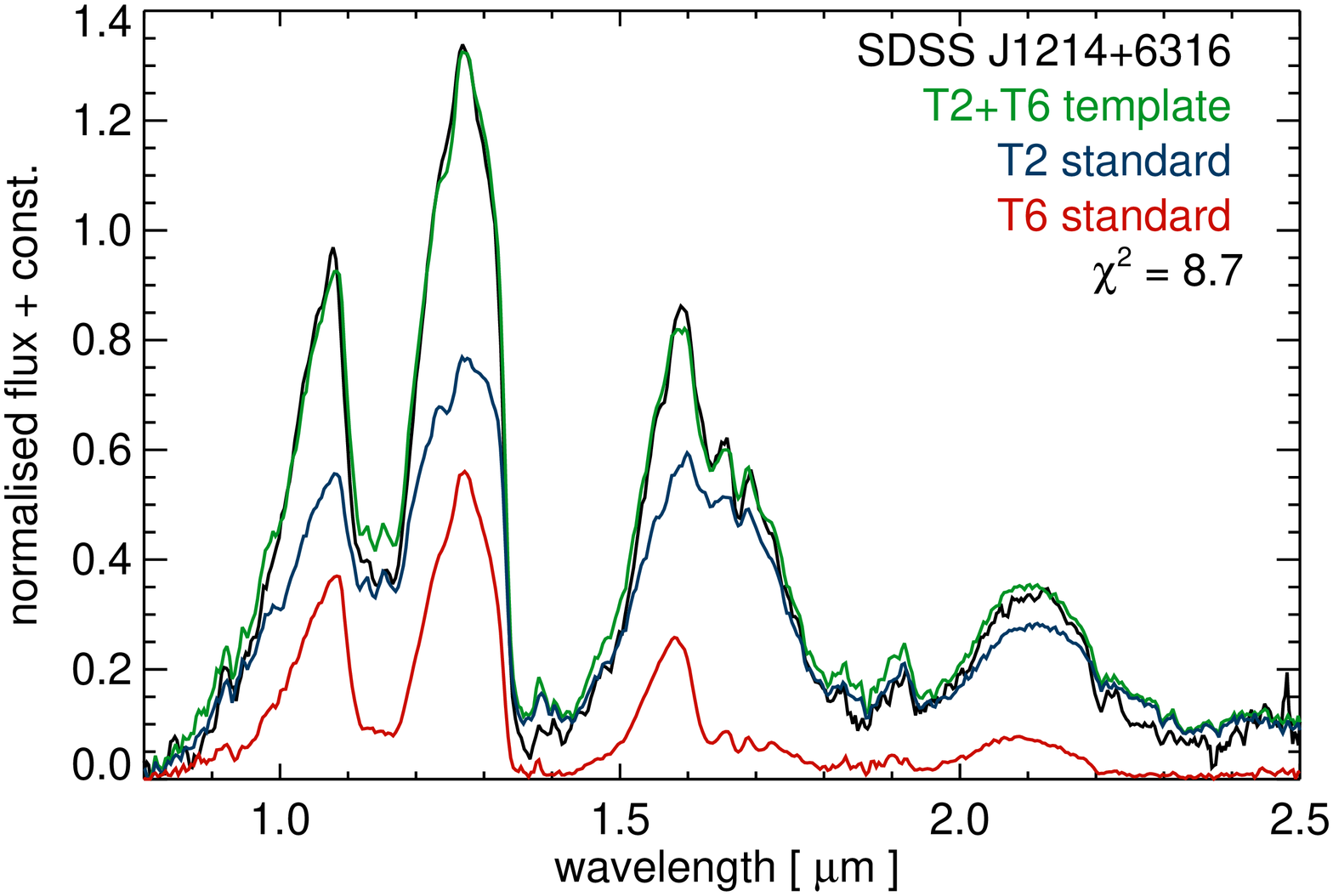} \\
\end{figure}
\addtocounter{figure}{-1}
\begin{figure}
\caption{continued.}
\includegraphics[angle=90,scale=0.3]{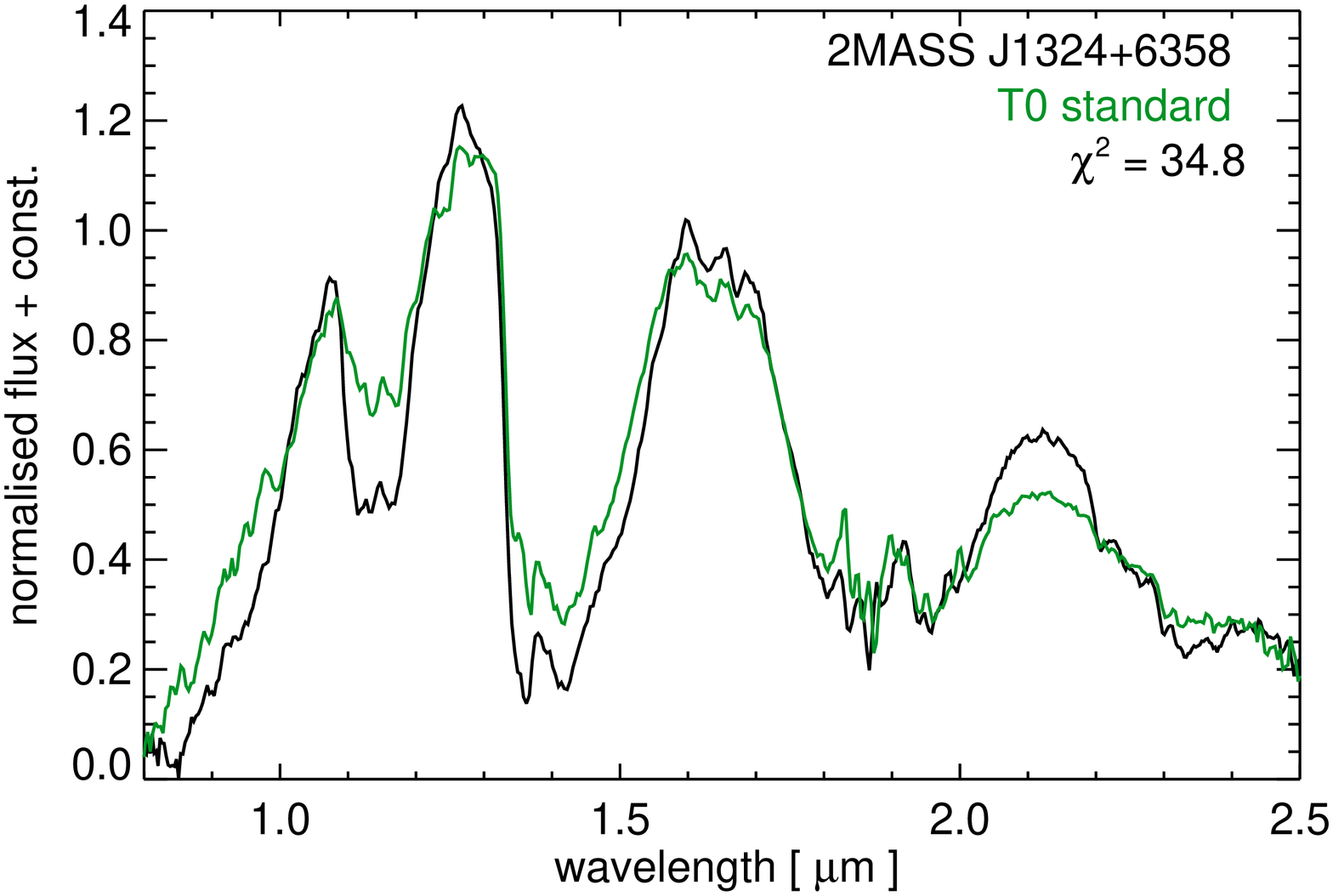}
\includegraphics[angle=90,scale=0.3]{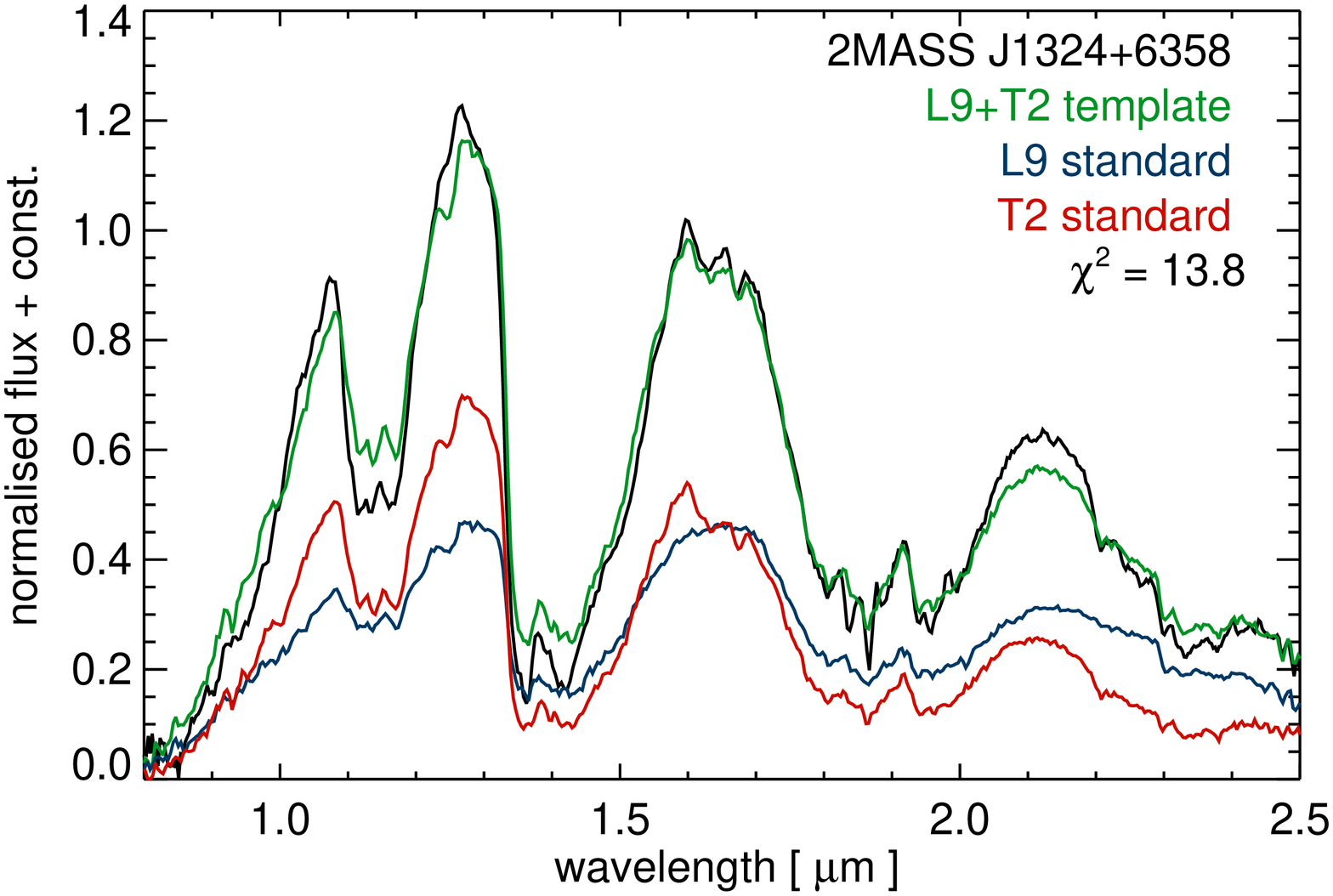} \\
\includegraphics[angle=90,scale=0.3]{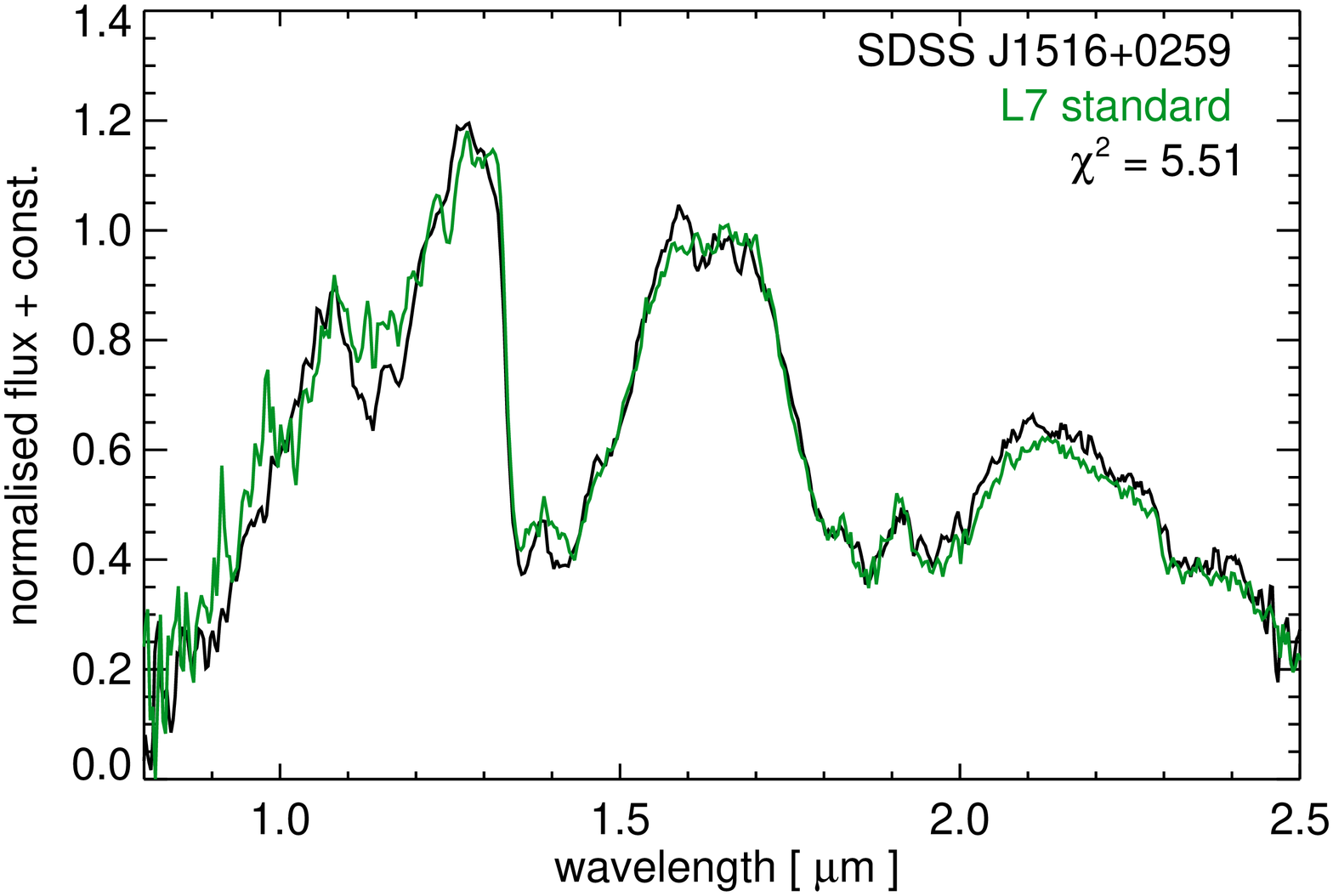}
\includegraphics[angle=90,scale=0.3]{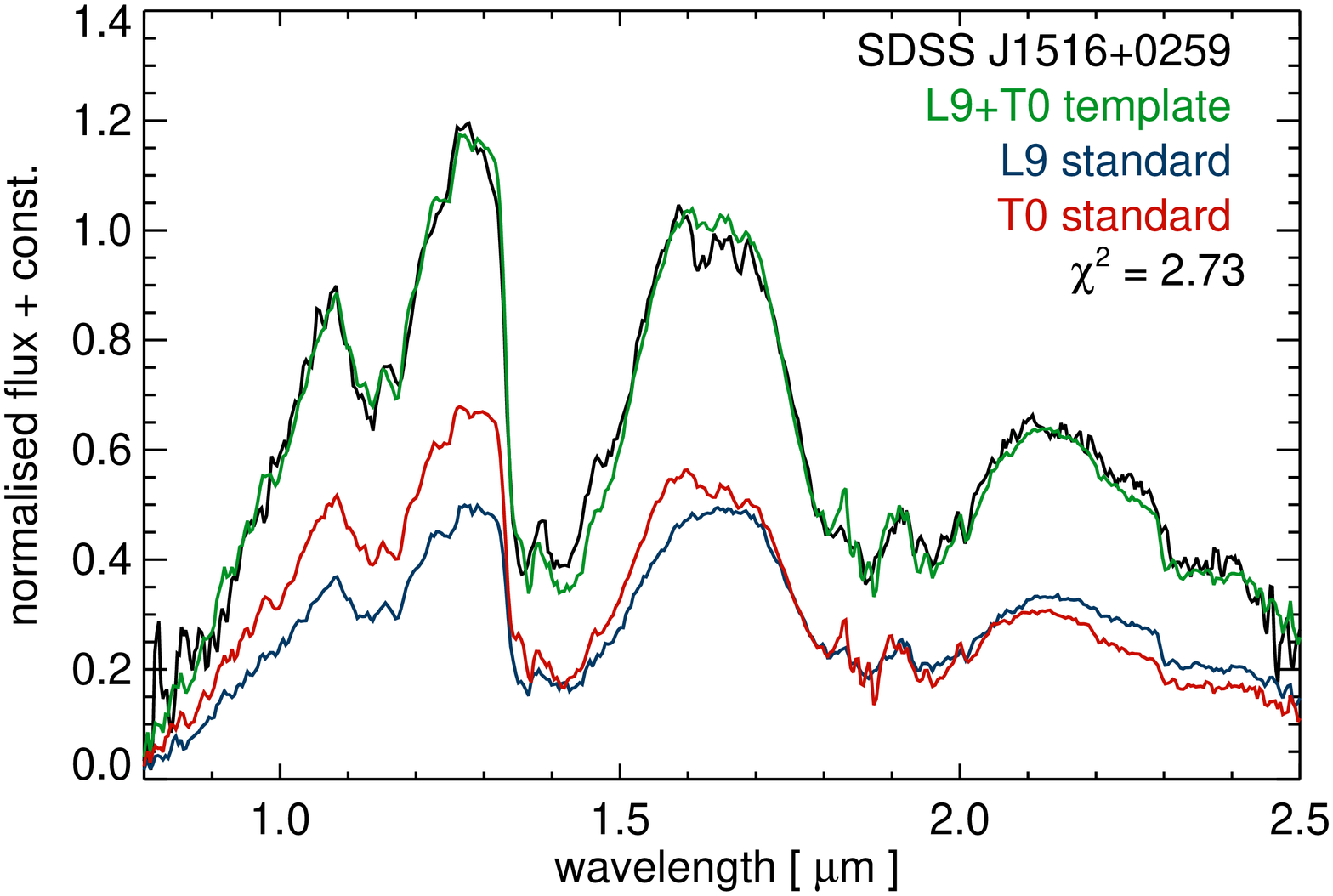} \\
\includegraphics[angle=90,scale=0.3]{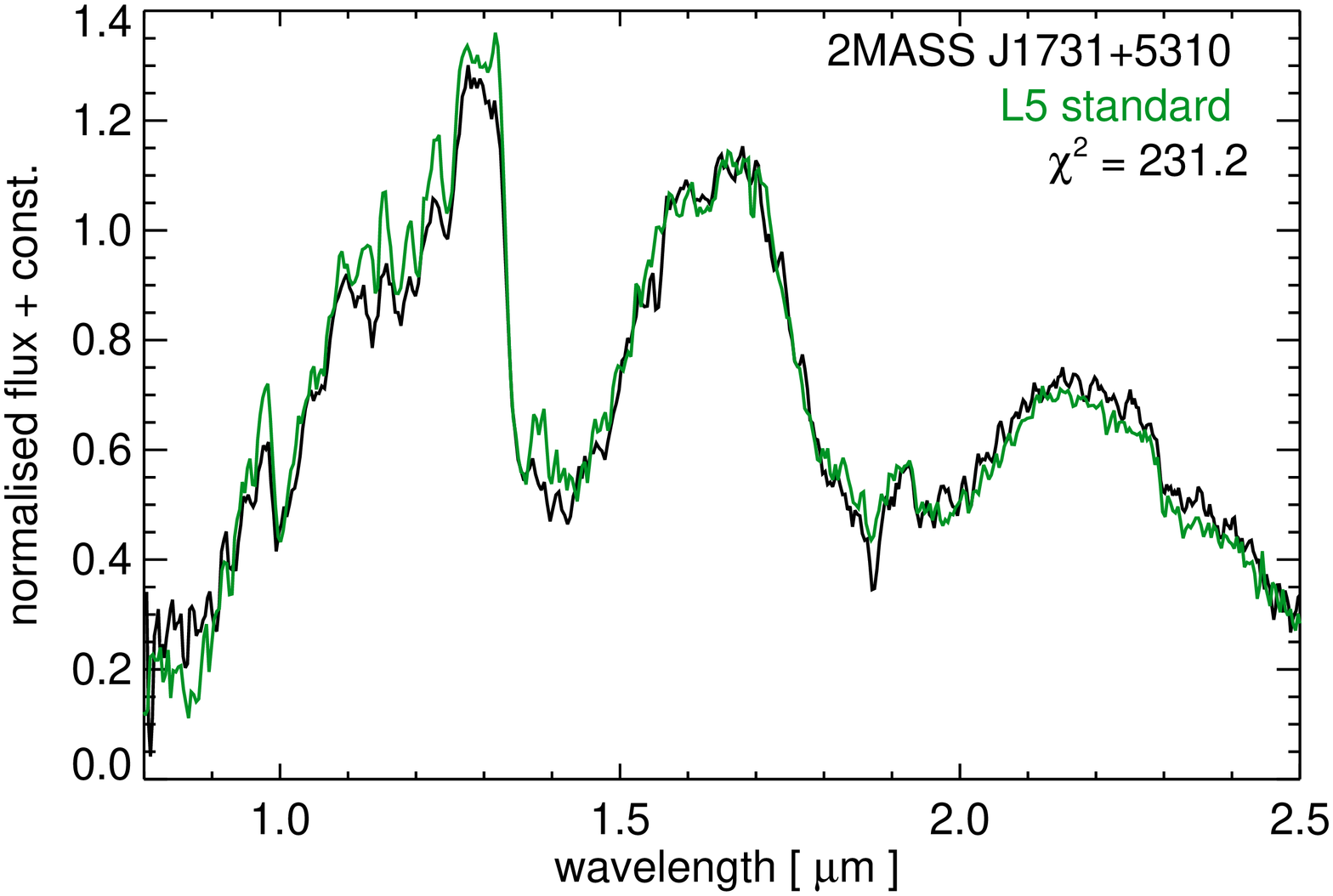}
\includegraphics[angle=90,scale=0.3]{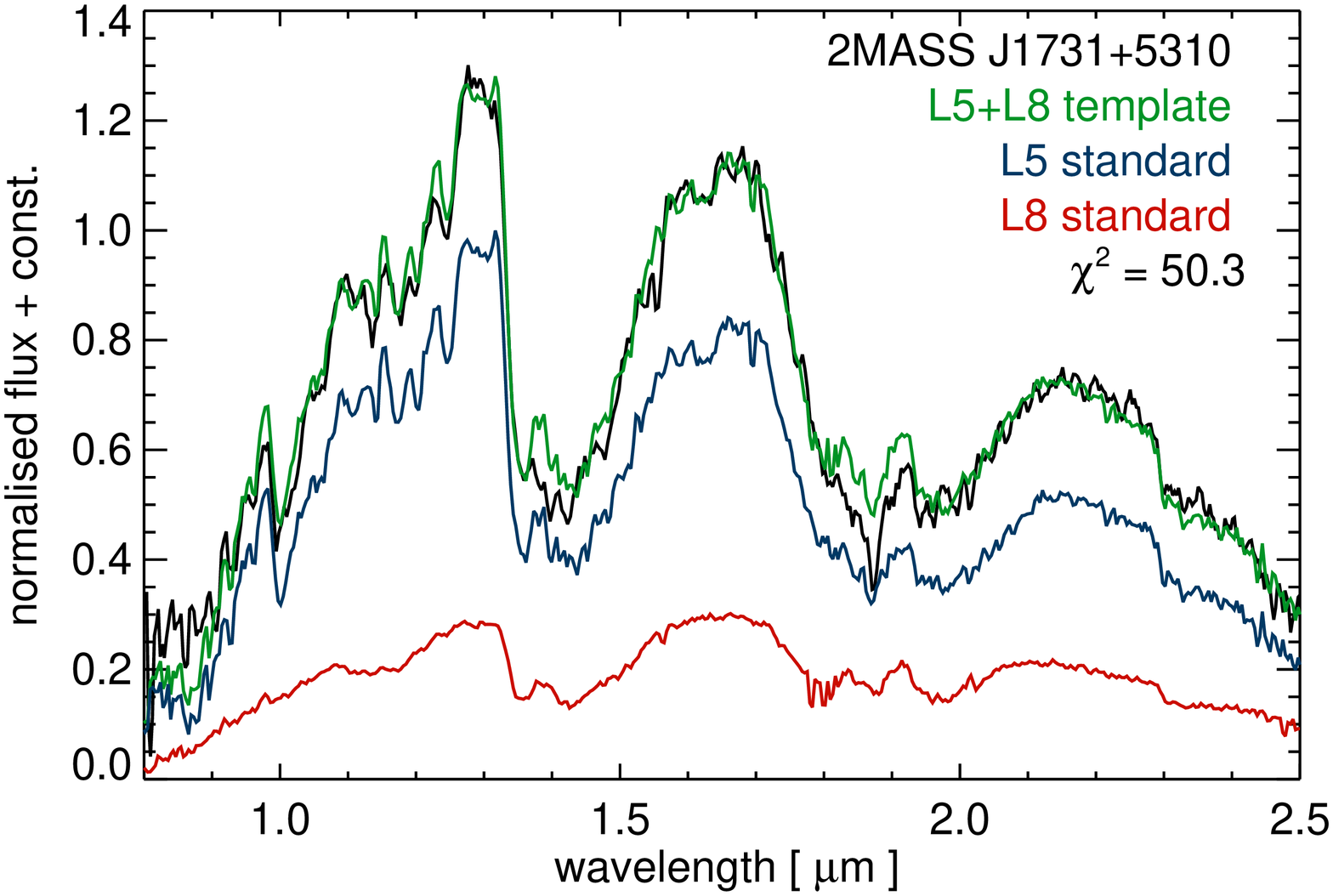} \\
\end{figure}
\begin{figure}
\caption{\label{fig:veryblue} Comparison of 2MASS J1542$-$0045 (solid spectra) to the known 
blue dwarf SDSS J112118.57+433246.5 (L7.5), and to the L7 (2MASS~J0205$-$1159) and the L8 
(2MASS J1632+1904) standards from \citet{kirkpatrick1999}, which are shown as dotted lines. 
All spectra have been normalized to the average flux in the 1.2\,-\,1.25\,$\mu$m region.}
\center
\includegraphics[scale=0.4]{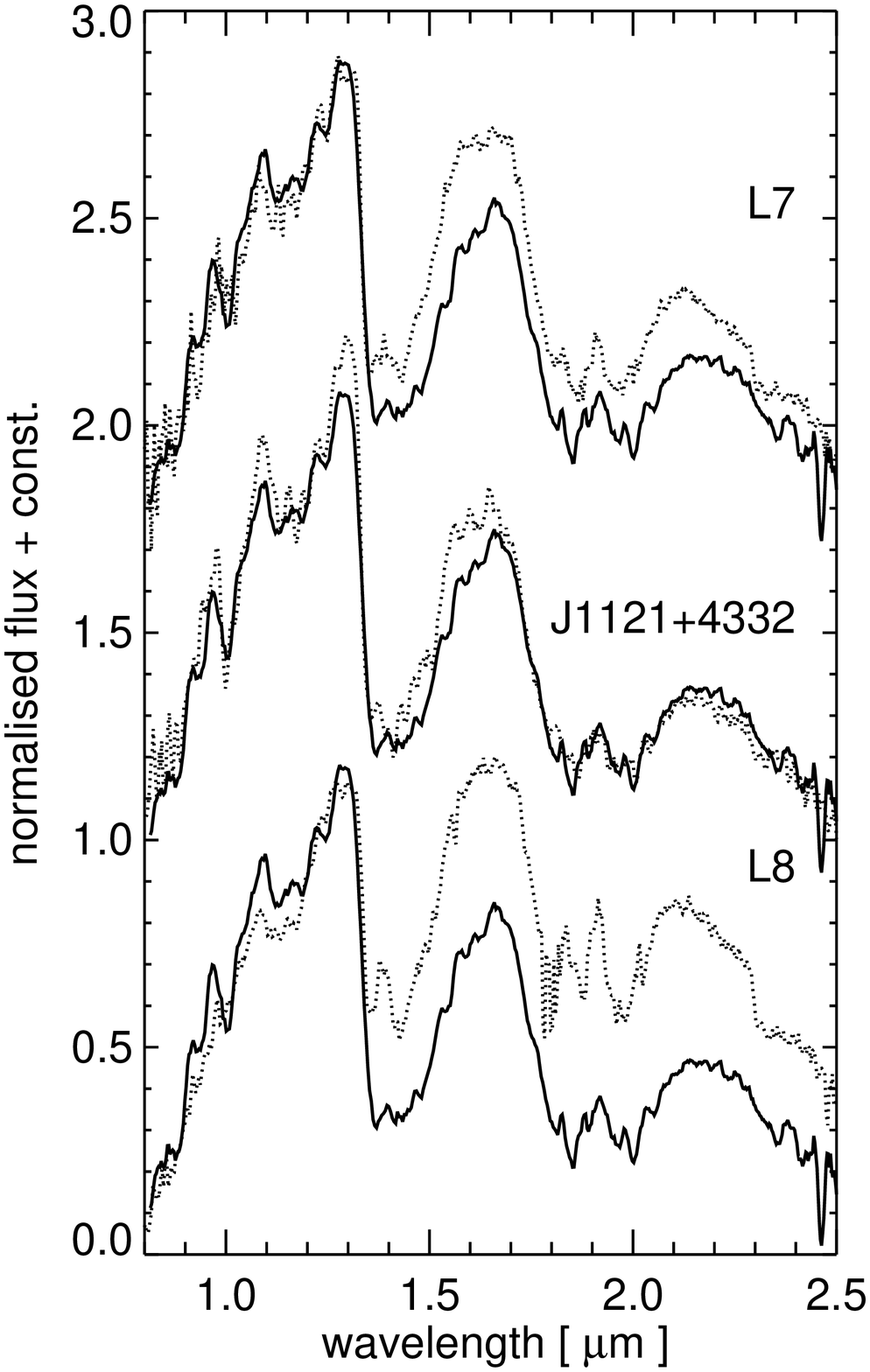}
\end{figure}
\begin{figure}
\caption{\label{fig:Hpeak} Comparison of 2MASS J0229$-$0053, 2MASS J0917+6028 and 
2MASS J1615+4953 (red) to an ordinary L2 dwarf and to the young L2 dwarf G196--3B 
\citep[dotted lines;][]{allers2007}. All spectra have been normalized to the average flux 
in the 1.2\,-\,1.25\,$\mu$m region.}
\center
\includegraphics[scale=0.25]{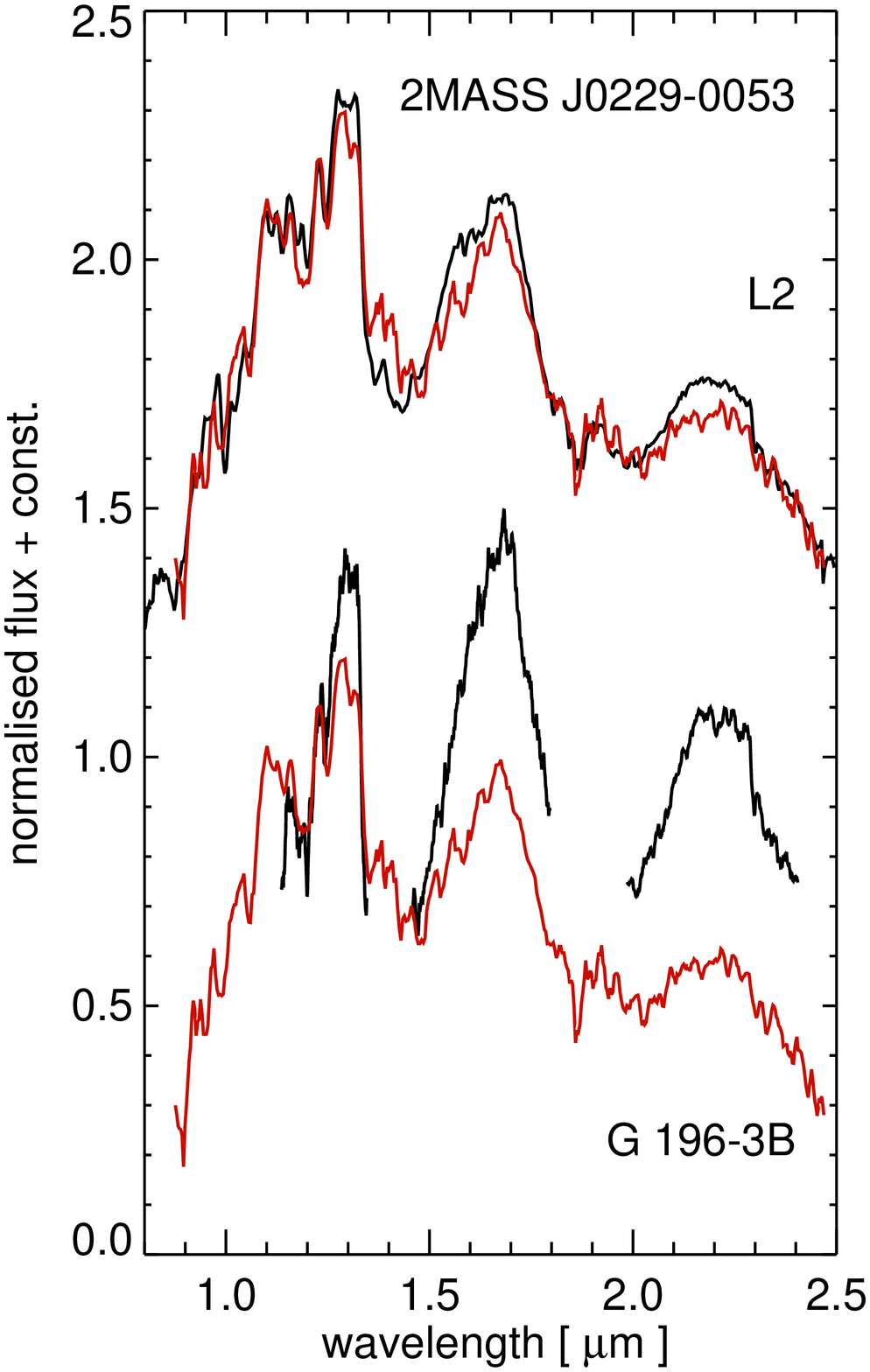}
\includegraphics[scale=0.25]{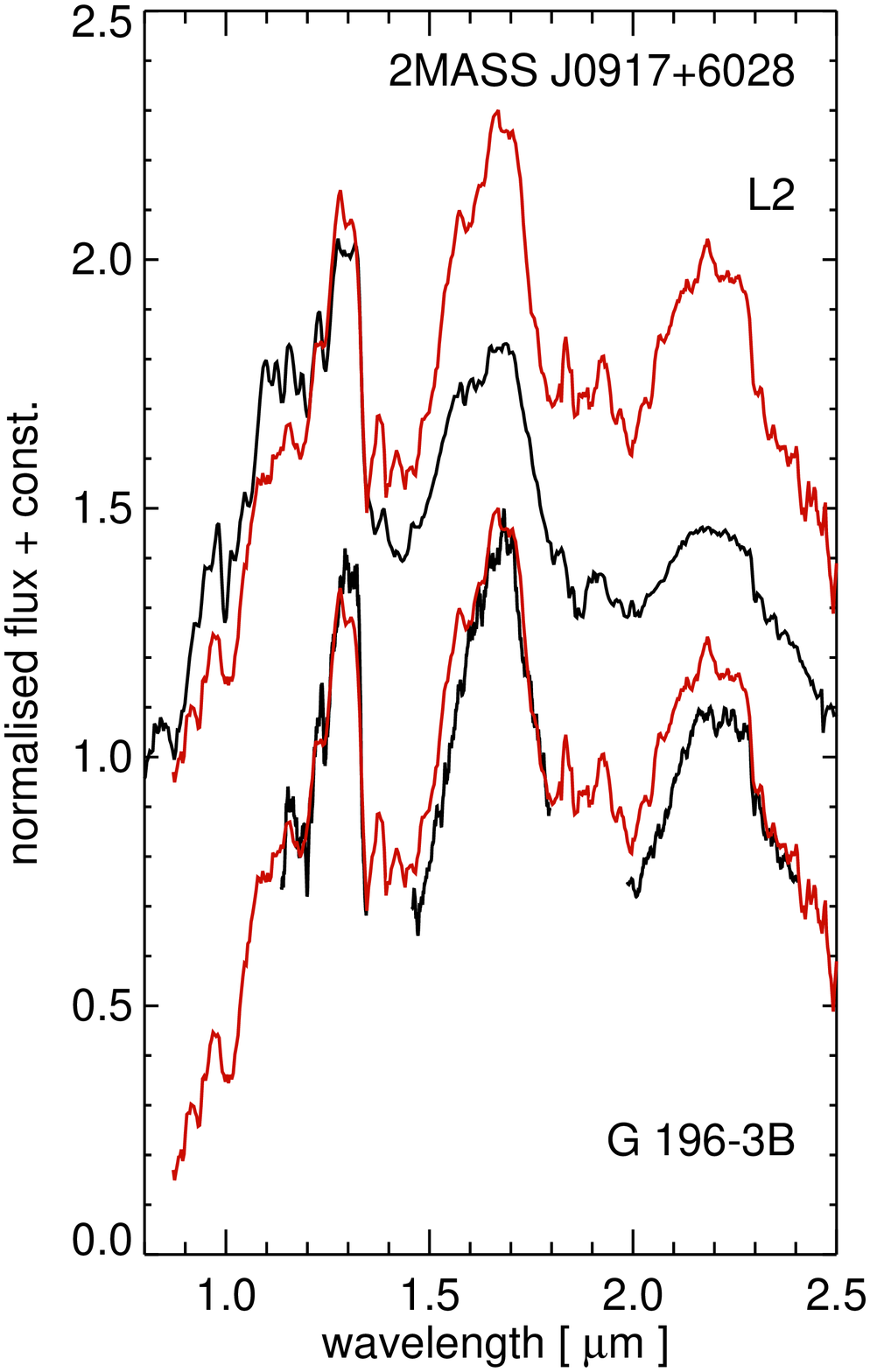}
\includegraphics[scale=0.25]{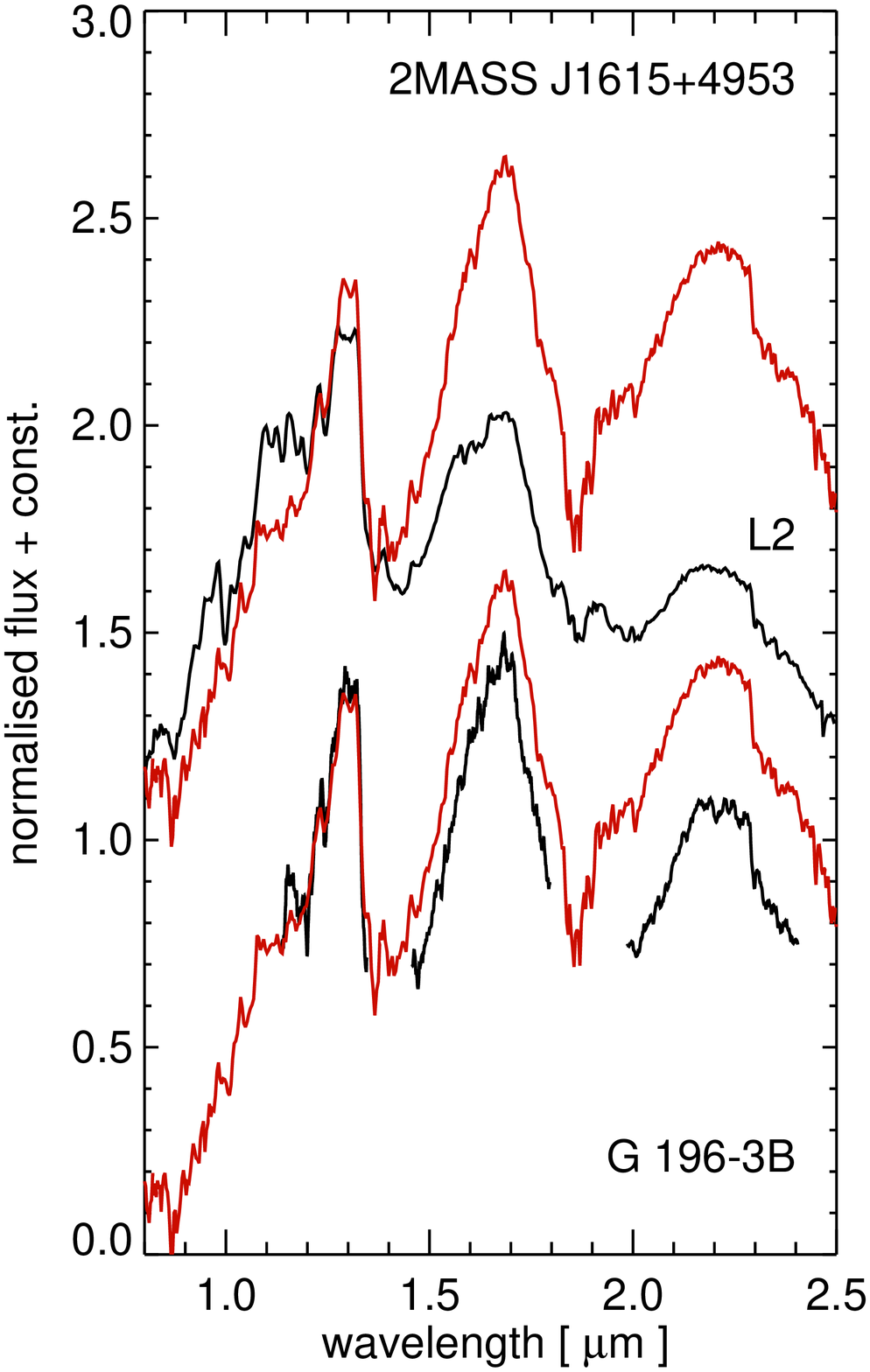}
\end{figure}
\begin{figure}
\caption{\label{fig:red} Comparison of 2MASS J0917+6028 and 2MASS J1615+4953 to the red L0 
dwarf 2MASS J0141$-$4633 \citep{kirkpatrick2006}. All spectra have been normalized to the 
average flux in the 1.15\,-\,1.3\,$\mu$m region.}
\center
\includegraphics[scale=0.3,angle=90]{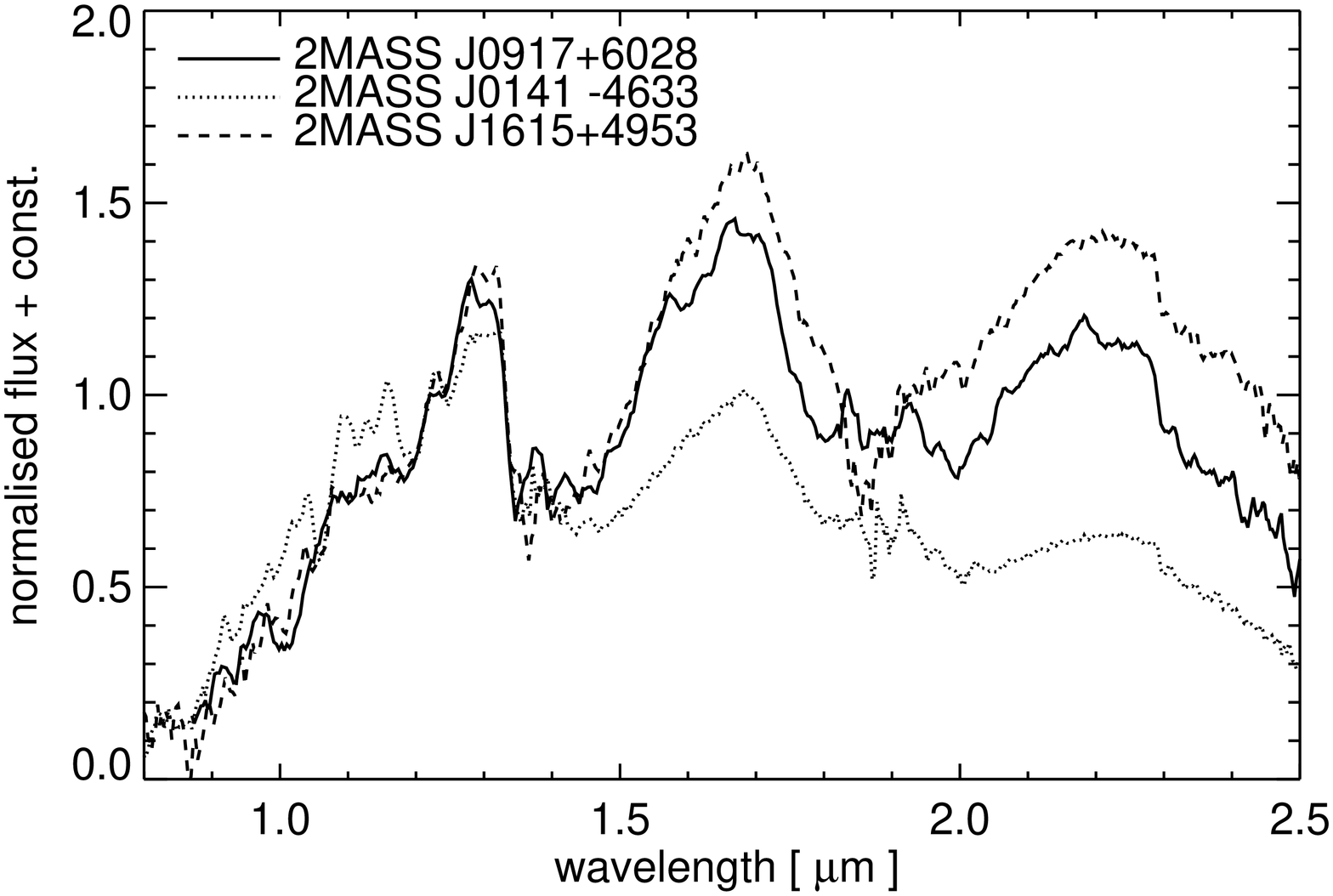}
\end{figure}
\begin{figure}
\caption{\label{fig:color}2MASS colors vs.\ synthetic colors calculated from the SpeX 
spectra. }
\center
\includegraphics[angle=90,scale=0.31]{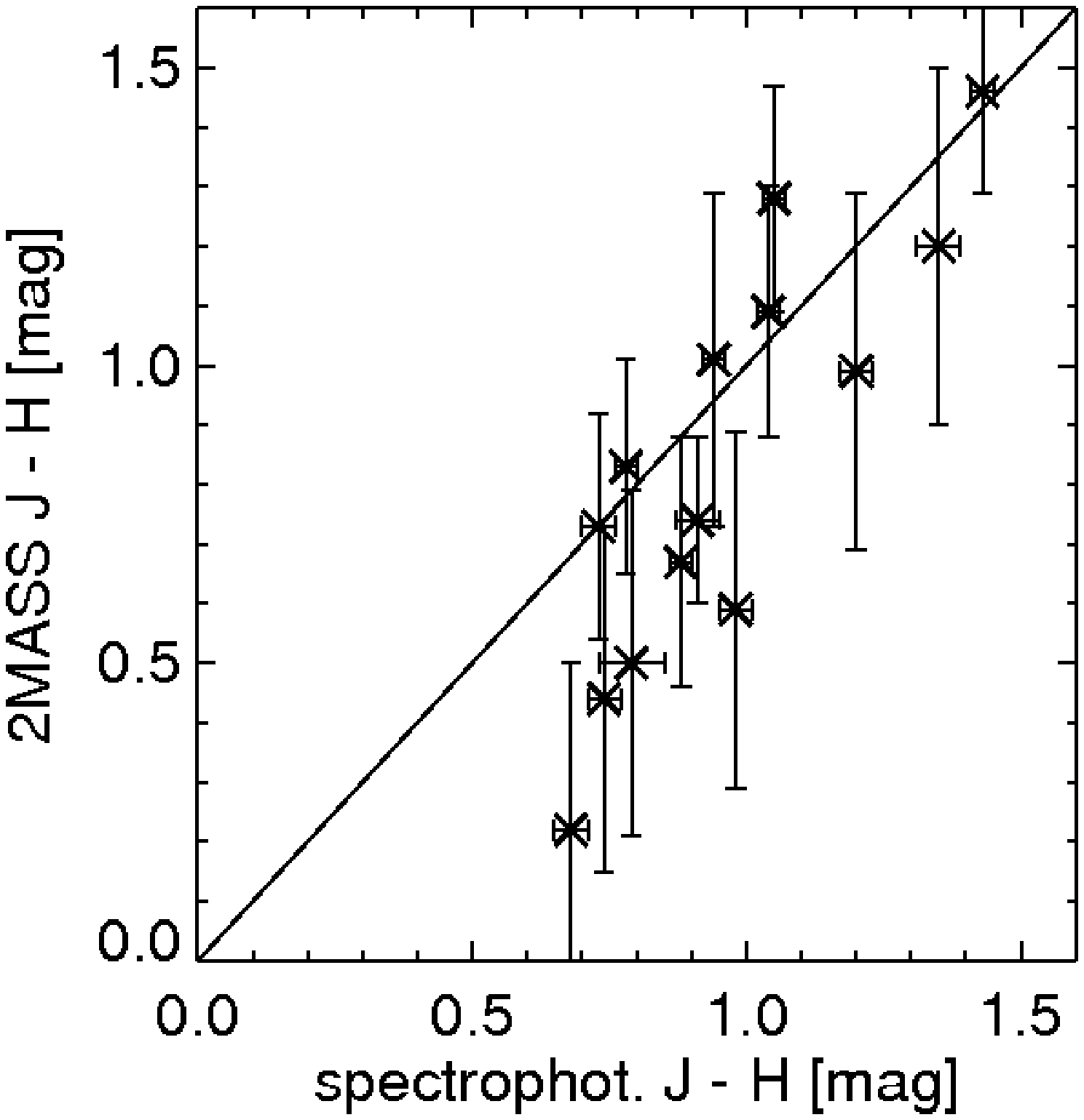}
\includegraphics[angle=90,scale=0.31]{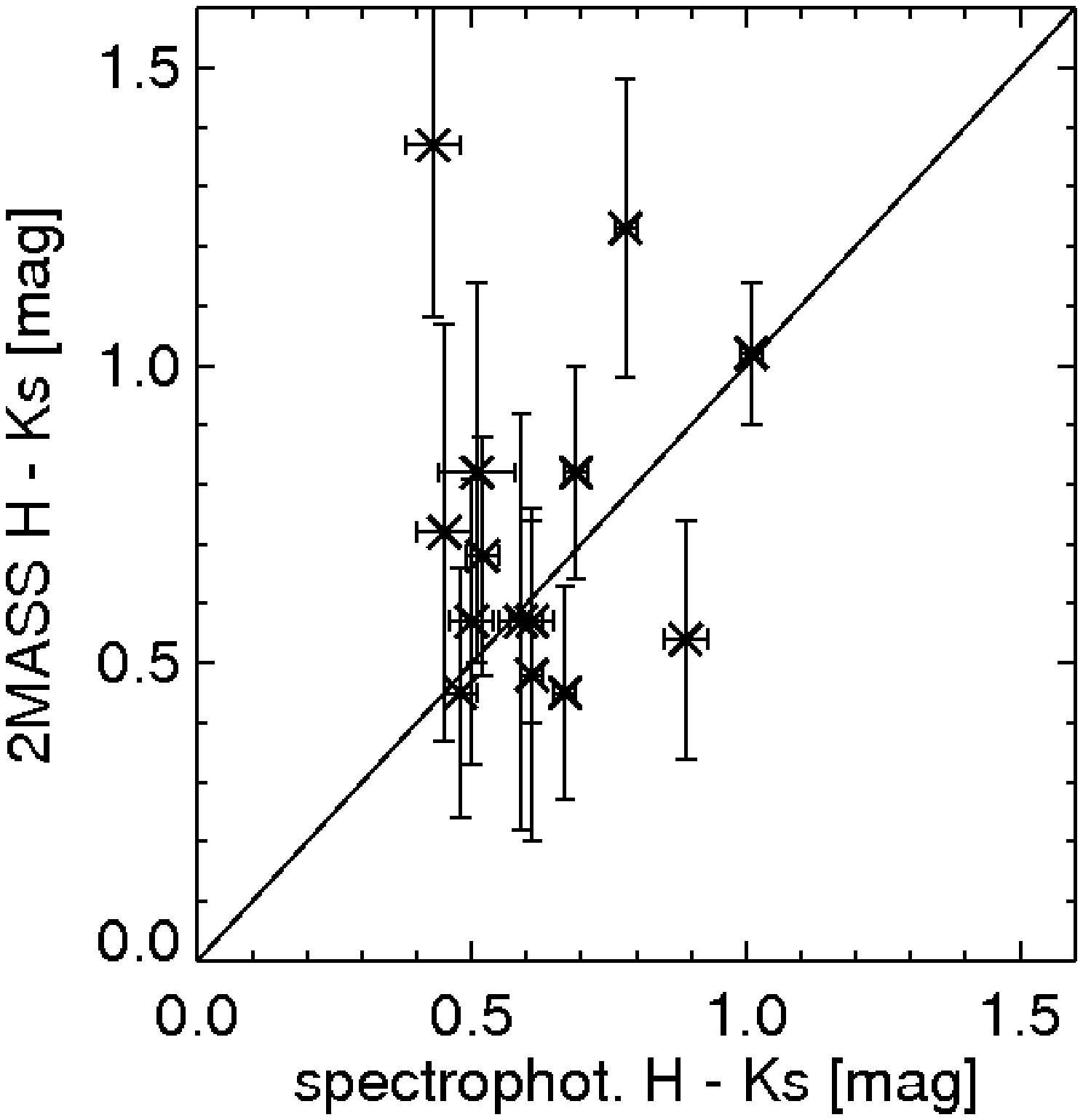}
\includegraphics[angle=90,scale=0.31]{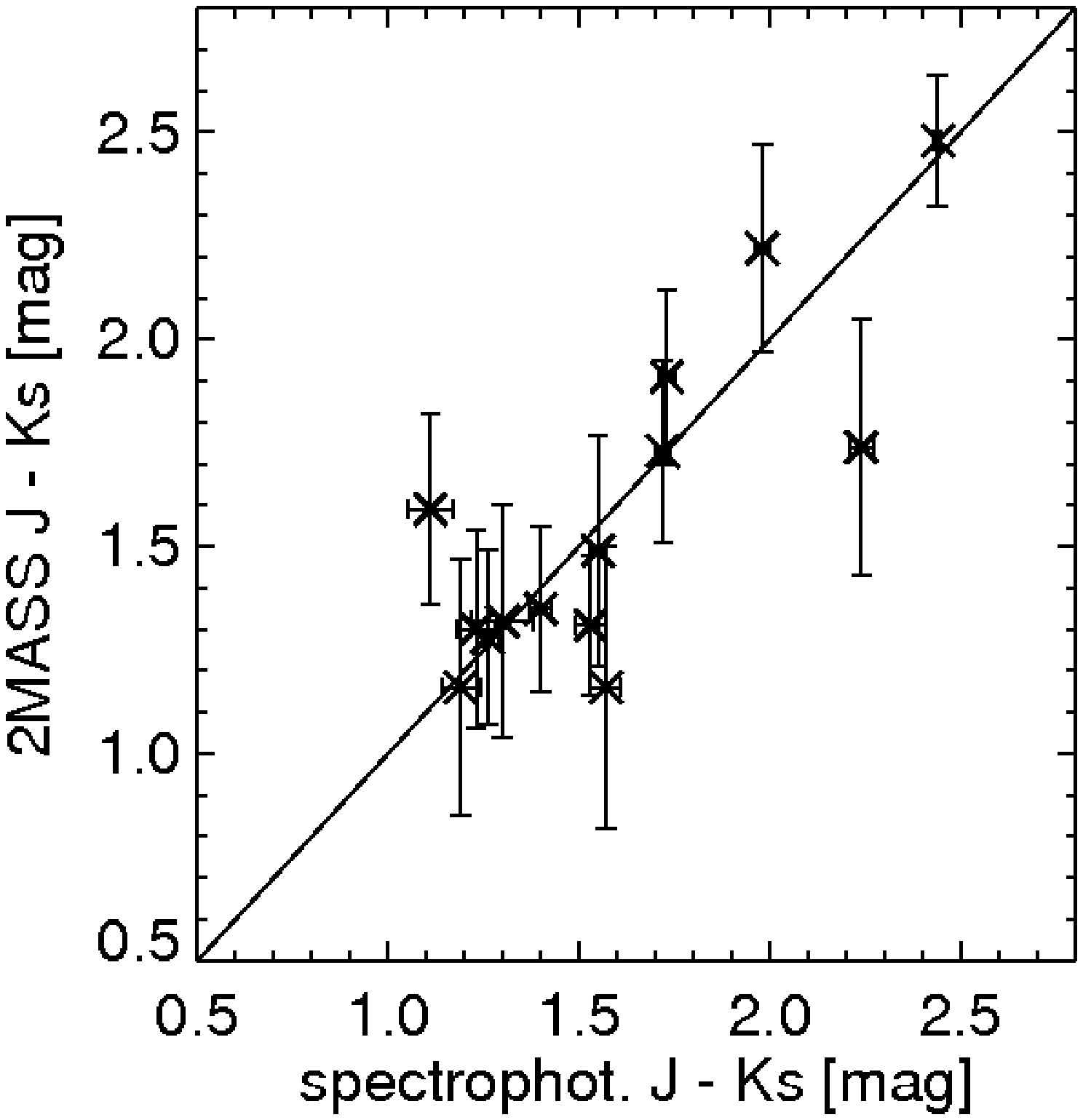}
\end{figure}


\begin{deluxetable}{ccccccc}
\tabletypesize{\footnotesize}
\tablewidth{0pt}
\tablecaption{\label{tab:col} L Dwarf Candidate Sample with SDSS and 
2MASS Photometry\tablenotemark{\dag}.}
\tablehead{\colhead{2MASS ID} & \colhead{$i-z$} & \colhead{$z-J$} & 
	\colhead{$J-H$} & \colhead{$H-K_{S}$} & \colhead{$J-K_{S}$} & \colhead{$J$} \\
	\colhead{(J2000)} & \colhead{[mag]} & \colhead{[mag]} & 
	\colhead{[mag]} & \colhead{[mag]} & \colhead{[mag]} & \colhead{[mag]}}
\startdata 
02292794$-$0053282 & 2.16$\pm$0.13 & 2.91$\pm$0.12 & 0.74$\pm$0.14 & 0.57$\pm$0.17 & 1.31$\pm$0.17 & 16.49$\pm$0.10\\
07354882$+$2720167 & 1.73$\pm$0.16 & 3.03$\pm$0.17 & 0.83$\pm$0.18 & 0.45$\pm$0.21 & 1.28$\pm$0.21 & 16.94$\pm$0.13\\
09175418$+$6028065 & $>$2.36       & 3.48$\pm$0.32 & 1.20$\pm$0.30 & 0.54$\pm$0.20 & 1.74$\pm$0.31 & 17.16$\pm$0.27\\
09264992$+$5230435 & 1.50$\pm$0.27 & 2.88$\pm$0.28 & $<$ 1.19      &  \nodata      & $<$ 1.57      & 16.77$\pm$0.14\\
11191046$+$0552484 & 2.11$\pm$0.13 & 2.88$\pm$0.17 & 1.28$\pm$0.19 & 0.45$\pm$0.18 & 1.73$\pm$0.22 & 16.76$\pm$0.16\\
12172372$-$0237369 & 2.20$\pm$0.21 & 2.99$\pm$0.19 & 1.09$\pm$0.21 & 0.82$\pm$0.18 & 1.91$\pm$0.21 & 16.90$\pm$0.16\\
13081228$+$6103486 & 2.00$\pm$0.14 & 2.75$\pm$0.19 & 0.51$\pm$0.26 & $<$0.67       & $<$1.18       & 16.67$\pm$0.15\\
14140586$+$0107102 & 2.14$\pm$0.17 & 2.86$\pm$0.22 & 1.01$\pm$0.28 & 0.48$\pm$0.28 & 1.49$\pm$0.28 & 16.74$\pm$0.20\\
14232186$+$6154005 & 2.17$\pm$1.25 & 2.93$\pm$0.19 & 0.67$\pm$0.21 & 0.68$\pm$0.20 & 1.35$\pm$0.20 & 16.63$\pm$0.15\\
15341068$+$0426410 & 1.79$\pm$0.11 & 2.86$\pm$0.18 & 0.50$\pm$0.29 & 0.82$\pm$0.32 & 1.32$\pm$0.28 & 16.92$\pm$0.17\\
15422494$+$5522451 & 1.92$\pm$0.30 & $<$3.4        & $>$1.18       & $<$0.76       & \nodata       & $>$17.13\\
15423630$-$0045452 & 2.41$\pm$0.15 & 2.75$\pm$0.14 & 0.73$\pm$0.19 & 0.57$\pm$0.24 & 1.30$\pm$0.24 & 16.71$\pm$0.13\\
15513546$+$0151129 & 1.78$\pm$0.15 & 2.77$\pm$0.18 & 0.22$\pm$0.28 & 1.37$\pm$0.29 & 1.59$\pm$0.23 & 16.85$\pm$0.15\\
16154255$+$4953211 & 2.34$\pm$0.16 & 2.90$\pm$0.16 & 1.46$\pm$0.17 & 1.02$\pm$0.12 & 2.48$\pm$0.16 & 16.79$\pm$0.14\\
17164260$+$2945536 & 1.91$\pm$0.16 & 3.00$\pm$0.22 & 0.59$\pm$0.30 & 0.57$\pm$0.35 & 1.16$\pm$0.34 & 17.06$\pm$0.20\\
17373467$+$5953434 & 2.42$\pm$0.39 & 3.38$\pm$0.21 & 0.44$\pm$0.29 & 0.72$\pm$0.35 & 1.16$\pm$0.31 & 16.88$\pm$0.16\\
21163374$-$0729200 & 2.11$\pm$0.21 & 2.89$\pm$0.25 & 0.99$\pm$0.30 & 1.23$\pm$0.25 & 2.22$\pm$0.25 & 17.20$\pm$0.21\\
\enddata
\tablenotetext{\dag}{SDSS $i$ and $z$ magnitudes are on the AB sinh system 
\citep{fukugita1996}, while 2MASS JH$K_s$ magnitudes are on the Vega system.}
\end{deluxetable}

\begin{deluxetable}{lccccc}
\tabletypesize{\footnotesize}
\tablewidth{0pt}
\tablecaption{Observations of Candidate L Dwarfs with IRTF/SpeX in Prism Mode 
\label{tab:spectr_obs}}
\tablehead{\colhead{2MASS ID} & \colhead{Date} & \colhead{$J$} & 
	\colhead{Slit Width} & \colhead{Exposure} & \colhead{A0 Calibrator}\\
	\colhead{(J2000)} & \colhead{(UT)} & \colhead{(mag)} & 
	\colhead{(arcsec)} & \colhead{(min)} & }
\startdata
02292794$-$0053282 & 2007 Nov 24 & 16.5 & 0.5 & 36 & HD 18571 \\
07354882$+$2720167 & 2007 Nov 24 & 16.9 & 0.8 & 51 & HD 72982 \\
09175418$+$6028065 & 2007 Nov 24 & 17.2 & 0.8 & 24 & HD 88132 \\
09264992$+$5230435 & 2007 Nov 24 & 16.8 & 0.8 & 27 & HD 88132 \\
\nodata\tablenotemark{\dag} & 2008 Mar 24 & 16.8 & 0.8 & 24 & HD 93946 \\
11191046$+$0552484 & 2008 Mar 26 & 16.8 & 0.8 & 36 & HD 107174 \\
12172372$-$0237369 & 2008 Mar 25 & 16.9 & 0.8 & 60 & HD 109969 \\
13081228$+$6103486 & 2008 Mar 26 & 16.7 & 0.8 & 69 & HD 120828 \\
14140586$+$0107102 & 2008 Mar 25 & 16.7 & 0.8 & 60 & HD 132660 \\
14232186$+$6154005 & 2008 Mar 26 & 16.6 & 0.8 & 39 & HD 120828 \\
15341068$+$0426410 & 2008 Mar 27 & 16.9 & 0.8 & 72 & HD 152115 \\
15422494$+$5522451 & 2008 Mar 24 & \nodata\tablenotemark{\ddag} & 0.8 
	& 60 & HD 155838 \\
15423630$-$0045452 & 2008 Mar 27 & 16.7 & 0.8 & 42 & HD 152115 \\
15513546$+$0151129 & 2008 Mar 25 & 16.9 & 0.8 & 27 & HD 132660 \\
16154255$+$4953211 & 2007 Aug 26 & 16.8 & 0.5 & 60 & HD 160883 \\
17164260$+$2945536 & 2007 Aug 26 & 17.1 & 0.5 & 48 & HD 160883 \\
17373467$+$5953434 & 2008 Mar 26 & 16.9 & 0.8 & 42 & HD 176893 \\
21163374$-$0729200 & 2007 Aug 26 & 17.2 & 0.5 & 42 & HD 210781 \\
\enddata
\tablenotetext{\dag}{Repeat observation of the same object, combined with the previous data.}
\tablenotetext{\ddag}{No 2MASS $J$ detection.  $H=16.0$~mag.}
\end{deluxetable}

\clearpage

\begin{deluxetable}{lcccccc}
\tabletypesize{\footnotesize}
\tablewidth{0pt}
\tablecaption{\label{tab:standards} Adopted M, L, and T dwarf spectroscopic standards.}
\tablehead{\colhead{Object} & \colhead{Spectral} & \colhead{Ref.} & \colhead{$J-K_s$} & 
\colhead{$\langle J-K_s\rangle_{\rm SpT}$\tablenotemark{\dag}} & 
\colhead{$\langle z-J\rangle_{\rm SpT}$\tablenotemark{\dag}}  & \colhead{Notes} \\
\colhead{} & \colhead{Type} & \colhead{} & \colhead{(mag)} & \colhead{(mag)} &
 \colhead{(mag)} & \colhead{} }
\startdata
\multicolumn{7}{c}{\bf Optical spectral standards} \\
VB 8 				& M7 V & 1, 2   & $0.96\pm0.04$  & 1.07$\pm$0.27 & 1.92$\pm$0.17 & \\
VB 10 				& M8 V & 1, 3   & $1.14\pm0.03$  & 1.10$\pm$0.20 & 2.07$\pm$0.40 & \\
LHS 2924 			& M9 V & 1, 4   & $1.25\pm0.03$  & 1.15$\pm$0.19 & 2.13$\pm$0.16 & \\
2MASP J0345432$+$254023 	& L0   & 5, 4   & $1.34\pm0.04$  & 1.33$\pm$0.24 & 2.31$\pm$0.14 & \\
2MASSW J1439284$+$192915	& L1   & 5, 6   & $1.21\pm0.03$  & 1.34$\pm$0.24 & 2.46$\pm$0.24 & \\
Kelu--1				& L2   & 5, 7   & $1.67\pm0.03$  & 1.49$\pm$0.24 & 2.55$\pm$0.18 & 
	binary\tablenotemark{a}\\
2MASSW J1146345$+$223053	& L3   & 5, 8   & $1.58\pm0.04$  & 1.60$\pm$0.25 & 2.68$\pm$0.24 &
	binary\tablenotemark{b}\\
2MASSI J1104012$+$195921	& L4   & 3      & $1.43\pm0.04$  & 1.63$\pm$0.38 & 2.64$\pm$0.43 & \\
DENIS-P J1228.2$-$1547		& L5   & 5, 8   & $1.61\pm0.04$  & 1.73$\pm$0.36 & 2.64$\pm$0.41 &
	binary\tablenotemark{c}\\
2MASSI J1010148$-$040649 	& L6   & 9     & $1.89\pm0.07$   & 1.72$\pm$0.29 & 2.63$\pm$0.34 & \\
DENIS-P J0205.4$-$1159		& L7   & 5, 9  & $1.59\pm0.04$   & 1.60$\pm$0.73 & 2.64$\pm$0.23 &
	binary\tablenotemark{d}\\
2MASSW J1632291$+$190441	& L8   & 5, 6   & $1.86\pm0.08$  & 1.72$\pm$0.32 & 2.78$\pm$0.41 & \\
\multicolumn{7}{c}{\bf Near-IR spectral standards} \\
2MASSW J0310599$+$164816	& L9   & 10, 11 & $1.71\pm0.11$  & 1.65$\pm$0.29 & 2.75$\pm$0.16 &  \\
SDSS J120747.17$+$024424.8 	& T0   & 10, 12 & $1.59\pm0.10$  & 1.53$\pm$0.33 & 2.81$\pm$0.42 &  \\
SDSS J015141.69$+$124429.6 	& T1   & 10, 3  & $1.38\pm0.23$  & 1.37$\pm$0.26 & 2.65$\pm$0.78 &  \\
SDSSp J125453.90$-$012247.4 	& T2   & 10, 3  & $1.05\pm0.06$  & 1.03$\pm$0.48 & 3.02$\pm$0.20 &  \\
2MASS J12095613$-$1004008	& T3   & 10, 3  & $0.85\pm0.16$  & 0.97$\pm$0.39 & 2.99$\pm$0.13 &  \\
2MASSI J2254188$+$312349	& T4   & 10, 3  & $0.36\pm0.15$  & 0.60$\pm$0.23 & 3.25$\pm$0.27 &  \\
2MASS J15031961$+$2525196	& T5   & 10, 3  & $-0.03\pm0.06$ & 0.22$\pm$0.43 & 3.34$\pm$0.18 & \\
SDSSp J162414.37$+$002915.6	& T6   & 10, 13 & $<-0.02$       & 0.14$\pm$0.38 & 3.43$\pm$0.11 &  \\
2MASSI J0727182$+$171001	& T7   & 10, 13 & $0.04\pm0.20$  & 0.05$\pm$0.44 & 3.36$\pm$0.15 & \\
2MASSI J0415195$-$093506	& T8   & 10, 3  & $0.27\pm0.21$  & $-0.1\pm$0.51 & \nodata & \\
\enddata
\tablenotetext{\dag}{Mean $J-K_s$ and $z-J$ colors and their standard deviations for 
all objects of a given spectral type, as compiled from DwarfArchives.org (L and T dwarfs) 
or from the Ultracool Dwarf Catalog 
(M dwarfs; http://www.iac.es/galeria/ege/catalogo\_espectral/). }
\tablenotetext{a}{\citet{liu2005, gelino2006}.}
\tablenotetext{b}{\citet{reid2001b}.}
\tablenotetext{c}{\citet{martin1999, bouy2003}.}
\tablenotetext{d}{\citet{koerner1999, bouy2003}.}
\tablerefs{1--\citet{kirkpatrick1999}, 2--\citet{burgasser2008b}, 3--\citet{burgasser2004}, 
4--\citet{burgasser2006a}, 5--\citet{kirkpatrick1999}, 6--\citet{burgasser2004}, 
7--\citet{burgasser2007b}, 8--\citet{burgasser2010},  
9--\citet{reid2006}, 10--\citet{burgasser2006b}, 11--\citet{burgasser2007a}, 
12--\citet{looper2007}, 13--\citet{burgasser2006d}.}
\end{deluxetable}

\begin{deluxetable}{lccc}
\tabletypesize{\footnotesize}
\tablewidth{0pt}
\tablecaption{\label{tab:spt} Spectral classification of single and candidate 
binary L and T dwarfs, including candidate binaries from Paper I.}
\tablehead{\colhead{Object} & \colhead{0.95--1.35~$\micron$ $\chi^2$ fit} & 
\colhead{0.95--2.35~$\micron$ $\chi^2$ fit} & \colhead{Note} \\
 & \colhead{(adopted if single)} & \colhead{(adopted if binary)} &} 
\startdata
\multicolumn{4}{c}{\bf New L Dwarfs} \\
2MASS J02292794$-$0053282 & L2 & L2    & \\
2MASS J07354882$+$2720167 & L1 & L1+L4 & binary \\
2MASS J09175418$+$6028065 & L5 & L8    & red \\
2MASS J09264992$+$5230435 & M8 & M8     & \\
2MASS J11191046$+$0552484 & L4 & L6    & \\
2MASS J12172372$-$0237369 & L4 & L6    & \\
2MASS J13081228$+$6103486 & L2 & L2    & \\
2MASS J14140586$+$0107102 & L4 & L4    & \\
2MASS J14232186$+$6154005 & L4 & L2+T5 & binary \\
2MASS J15341068$+$0426410 & L0 & L0    & \\
2MASS J15423630$-$0045452 & L7 & L1    & blue \\ 
2MASS J15422494$+$5522451 & L4 & L4    & \\ 
2MASS J15513546$+$0151129 & M8 & M8    & \\ 
2MASS J16154255$+$4953211 & L6 & L8    & red/young \\
2MASS J17164260$+$2945536 & L3 & L3    & \\
2MASS J17373467$+$5953434 & L9 & L5+T5 & blue/binary \\
2MASS J21163374$-$0729200 & L6 & L8    & \\
\multicolumn{4}{c}{\bf Peculiar Dwarfs or Binary Candidates from Paper I\tablenotemark{\dag}} \\
2MASS J00521232$+$0012172 & L5\tablenotemark{\star} & L4+T3 & binary \\
SDSSp J010752.33+004156.1 & L8 & L8 & red \\
2MASS J01262109$+$1428057 & L6\tablenotemark{\ddag} & L8 & red/young \\
SDSS J092615.38$+$584720.9 & T4 & T3+T6 & binary \\ 
SDSS J121440.95$+$631643.4 & T5 & T2+T6 & binary \\
2MASS J13243559$+$6358284 & T3 & L9+T2 &  red/binary \\
SDSS J151603.03$+$025928.9 & T0 & L9+T0 & binary \\
2MASS J17310140$+$5310476 & L5 & L5+L8 & binary \\
\enddata
\tablecomments{The spectral types are determined from minimum $\chi^2$ fitting over the 
0.95--1.35~$\micron$ and the 0.95--2.35~$\micron$ regions.  The former fit yields the 
adopted spectral type for each individual object, listed in the second column.  The latter 
fit yields the adopted spectral type combination for candidate unresolved binary systems.}  
\tablenotetext{\dag}{Our 0.95--1.35~$\micron$ spectral classifications for the previously 
identified L and T dwarfs differ from the ones published or referenced in Paper~I by 
$\leq$2 subtypes, unless noted.}
\tablenotetext{\star}{Classified as an L2 dwarf in Paper I by comparison of the 
0.8--1.3\,$\micron$ region to L dwarf standards from \citet{cushing2005}}
\tablenotetext{\ddag}{Classified as an L2 dwarf in Paper I by direct comparison to the 
peculiarly red L2 dwarf G~196--3B.}
\end{deluxetable}

\begin{deluxetable}{l c c c}
\tabletypesize{\footnotesize}
\tablewidth{0pt}
\tablecaption{\label{tab:fit} List of best-fit composite spectral types for probable binary candidates.  }
\tablehead{\colhead{Object} & \colhead{Primary} & \colhead{Secondary} & \colhead{fraction\tablenotemark{\dag}} \\
\colhead{} & \colhead{} & \colhead{} & \colhead{[\%]} }
\startdata
\multicolumn{4}{c}{\bf New Binaries Presented Here} \\
2MASS J07354882$+$2720167 \T & L1 & L4 & 75.4 \\
		             & L1 & L5 & 22.0 \\
2MASS J14232186$+$6154005    & L2 & T5 & 55.6 \\
                             & L2 & T4 & 41.6 \\
2MASS J17373467$+$5953434    & L5 & T5 & 79.4 \\
                             & L4 & T5 & 19.2 \\
\multicolumn{4}{c}{\bf Binaries from Paper I} \\
2MASS J00521232$+$0012172 \T & L4 & T3 & 53.0 \\
                             & L2 & T3 & 43.3 \\
SDSS J092615.38$+$584720.9   & T3 & T6 & 54.0 \\
                             & T4 & T6 & 46.0 \\
SDSS J121440.95$+$631643.4   & T2 & T6 & 78.9 \\
                             & T3 & T8 & 21.1 \\
2MASS J13243559$+$6358284    & L9 & T2 & 99.3 \\
SDSS J151603.03$+$025928.9   & L9 & T0 & 53.4 \\
                             & L8 & T0 & 46.6 \\
2MASS J17310140$+$5310476    & L5 & L8 & 100. \\
\enddata
\tablenotetext{\dag}{The last column gives the percentage of cases the composite template has been 
returned as the best-fit combination during Monte Carlo simulations (\S~\ref{sec:binarity}).}
\end{deluxetable}




\end{document}